%% file: main.tex
\documentclass[journal]{IEEEtran}

\usepackage[utf8]{inputenc}
\usepackage[T1]{fontenc}
\usepackage{lmodern}
\AtBeginDocument{%
  \DeclareFontShape{T1}{lmr}{bx}{sc}{<->ssub*lmr/bx/n}{}%
  \DeclareFontShape{T1}{lmr}{m}{scit}{<->ssub*lmr/m/sc}{}%
}
\usepackage{graphicx}
\usepackage{cite}
\usepackage{amsmath,amssymb}
\usepackage{booktabs}
\usepackage{xcolor}
\usepackage{hyperref}
\usepackage{algorithm}
\usepackage{algpseudocode}
\usepackage{tabularx}
\usepackage{array}
\usepackage{tikz}
\usetikzlibrary{arrows.meta, positioning, shapes.geometric, calc, fit, backgrounds, shadows.blur}

\usepackage{placeins}              

\tikzset{
  stage/.style={rectangle, rounded corners=2pt, draw=black!55, line width=0.5pt,
    fill=#1, minimum height=8mm, inner sep=4pt, align=center, font=\footnotesize},
  artefact/.style={rectangle, draw=black!40, line width=0.4pt, fill=black!3,
    minimum height=7mm, inner sep=3pt, align=center, font=\scriptsize\itshape},
  flow/.style={-{Stealth[length=2mm]}, draw=black!65, line width=0.6pt},
  flowback/.style={-{Stealth[length=2mm]}, draw=black!45, line width=0.5pt, dashed},
}

\title{Multi-Conditioned Diffusion Synthesis of Sand Boils for
Low-Resource Earthen-Levee Inspection}

\author{%
Padam Jung Thapa\textsuperscript{1},
Abdullah Bin Naeem\textsuperscript{2},
Ayon Dey\textsuperscript{2},
Anav Katwal\textsuperscript{2},
Md Tamjidul Hoque\textsuperscript{2,*}\\[4pt]
{\normalsize
\textsuperscript{1}University of Louisiana at Lafayette, Lafayette, LA, USA\\
\textsuperscript{2}Department of Computer Science, Louisiana State University New Orleans, New Orleans, LA, USA\\[3pt]
Emails: tpadamjung@gmail.com;\quad
\{aneem, adey, akatwal, thoque\}@lsuneworleans.edu\\[1pt]
\textsuperscript{*}Corresponding author: thoque@lsuneworleans.edu}%
}

\begin{document}
\maketitle

\input{sections/abstract}
\input{sections/introduction}
\input{sections/related_work}
\input{sections/methodology}
\input{sections/experiments}
\input{sections/results}
\input{sections/discussion}
\input{sections/conclusion}
\input{sections/code_availability}
\input{sections/acknowledgments}

\FloatBarrier                      
\bibliographystyle{IEEEtran}
\bibliography{references}

\end{document}

%% file: sections/abstract.tex
\begin{abstract}
Sand boils on earthen levees are safety-critical defects, yet
pixel-level detection from inspection imagery is limited by the
scarcity of annotated examples. We address this low-resource setting
with a diffusion-based synthesis pipeline. A Stable Diffusion XL
backbone, fine-tuned with DreamBooth on a small curated reference set
and conditioned by a multi-branch ControlNet stack, generates
synthetic sand-boil imagery. A soft-mask inpainting protocol preserves
the real defect region pixel-for-pixel while re-rendering the
surrounding scene, avoiding the seams and colour casts of the
seamless-cloning compositing used previously. A complementary
mask-conditioned ControlNet generates a fresh boil into a chosen mask,
so the conditioning mask is the segmentation label by construction,
with the \textsc{Convex Hull Annotator} repurposed as a drift-checking
quality gate---though certifying that label at scale is not yet solved
by the available real-trained gate, so we release the soft-mask preset,
not this one, as the default. Text conditioning is supplied by a taxonomy-driven
\emph{Prompt Atlas} that expands a single domain specification into a
stratified, CLIP-validated prompt bank and ports to new defect classes
without code changes. From the real training images the pipeline
produces $1{,}020$ synthetic candidates, of which $815$ pass a CLIP
admissibility filter to form the augmented dataset. We evaluate image
quality with distribution and
fidelity/diversity measures against the real reference set and a
Poisson baseline, and audit the augmented set for out-of-distribution
drift and memorisation. No single preset dominates: the presets trade off
fidelity, diversity, and label reliability, so we release the
label-reliable preset as the default and treat a curated mixture as the
natural augmentation set. We scope our claims to image quality, label
provenance, and diversity; downstream segmentation is left to future work. Code and an artefact manifest are released for
reproducibility.
\end{abstract}


%% file: sections/introduction.tex
\section{Introduction}
\label{sec:intro}

\begin{figure*}[t]
\centering
\includegraphics[width=\textwidth]{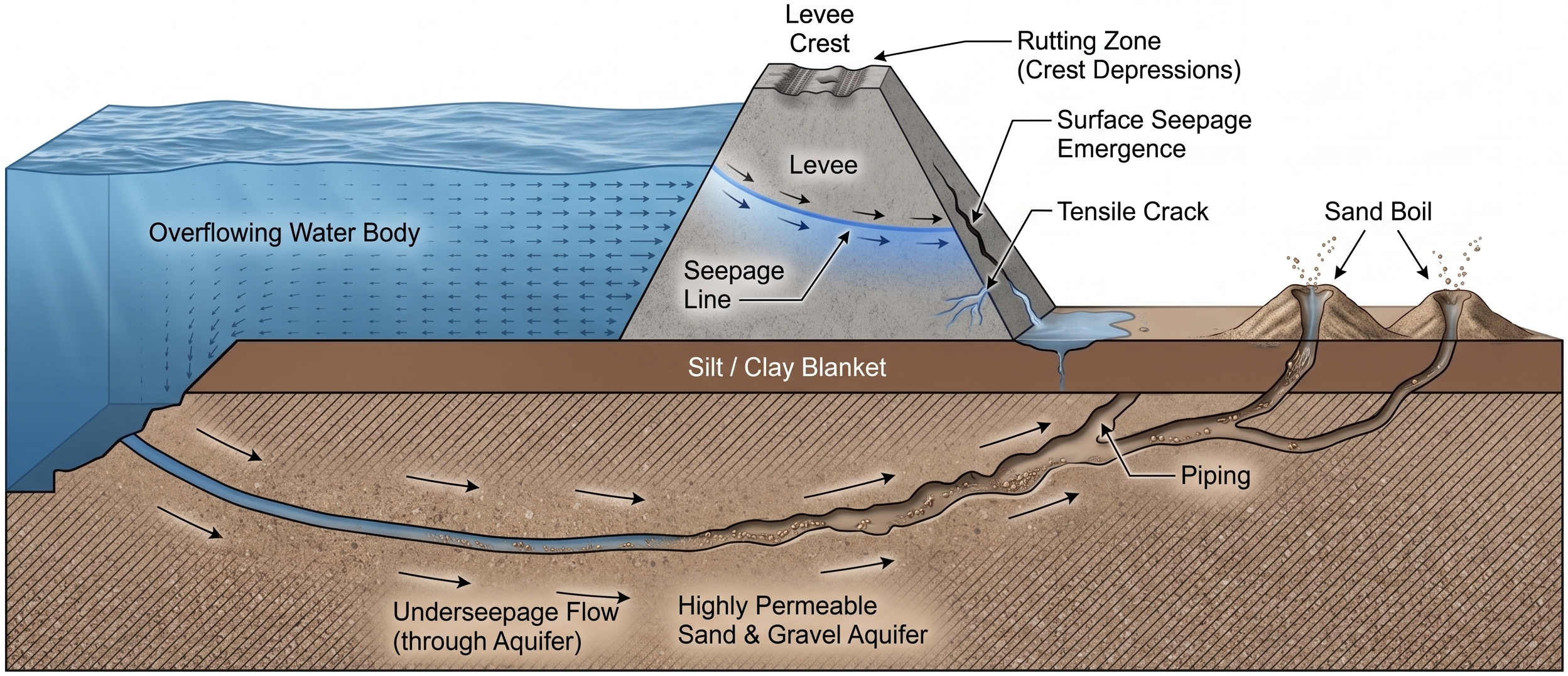}
\caption{Cross-sectional anatomy of a levee system and the surface
defects it produces. High water in the river drives \emph{under-seepage}
through the permeable sand aquifer beneath the silt/clay blanket; where
the flow concentrates, it erodes an upward channel (\emph{piping}) that
breaches the blanket and vents sand and water at the surface as a
\emph{sand boil}. The same hydraulic loading drives a phreatic
\emph{seepage line} through the embankment to a landside \emph{seepage
emergence}, while surface deterioration shows up as \emph{cracks} and
\emph{rutting} on the crest. Sand boils, the early visible stage of
internal erosion, are the focus of this paper, though the generation
pipeline transfers to the other defect classes
(Section~\ref{sec:conclusion}).}
\label{fig:levee_defects}
\end{figure*}

Earthen levees protect large populations and property from flooding,
yet their condition is hard to monitor at scale. The cost of
inadequate inspection became clear during the 2005 Hurricane Katrina
failures, when independent investigators attributed the catastrophic
flooding in New Orleans to a combination of design, construction, and
inspection deficiencies~\cite{ilit2006katrina}. Among the defect
classes that inspectors are trained to recognise, sand boils are
particularly consequential. A sand boil forms when water under hydraulic
pressure carries fine sand upward through a permeable foundation
layer, producing a small saturated dome at the surface
(Figure~\ref{fig:levee_defects}). This is the
early visible stage of internal erosion (also called piping), which,
if left untreated, can progress to a sudden breach during a high-water
event. Detecting and segmenting sand boils at the pixel level from
routine field photographs therefore offers a direct lever for
proactive flood-risk management.

Training a deep segmentation model for this task runs into a stubborn
obstacle: annotated sand boil imagery is rare. The defect appears only
during high water, is spread across long linear levee networks, and
requires hydrologic expertise for accurate pixel-level annotation. The
curated dataset used in this study (whose size is given in
Section~\ref{subsec:dataset}) is, to our knowledge, among the larger
curated sand-boil sets reported. Even at that scale it is too small to
train a modern segmentation network without data augmentation
or pretraining. Conventional spatial and photometric augmentation
(rotation, flips, colour jitter, elastic distortion) inflates the
sample count without broadening the underlying scene distribution. It
cannot synthesise a sand boil under a hydraulic regime, soil type, or
photographic style that the corpus does not already contain. Diffusion-based
generative augmentation~\cite{rombach2022ldm,trabucco2023effective}
promises to fill exactly this gap.

Realising that promise on a niche scientific class raises three
coupled problems, which this paper sets out to solve.

\paragraph{Problem 1: Adapting a general-purpose diffusion model to a
narrow defect class without retraining it.} A stock Stable Diffusion
XL pipeline~\cite{podell2023sdxl} prompted with ``a photo of a sand
boil'' returns images of volcanic vents, coastal surf, or children's
sandboxes; the pretrained base model has never seen civil-infrastructure
defect imagery. Full fine-tuning is impractical at this reference-set scale. DreamBooth fine-tuning with low-rank
adaptation~\cite{ruiz2023dreambooth,hu2022lora} confines the trainable
parameters to a small additive adapter anchored by a rare trigger
token. It fits comfortably on a single accelerator and preserves the
frozen base for unrelated prompts. We adopt this recipe. Our main reason for building on SDXL rather than a
higher-capacity rectified-flow~\cite{liu2023rectflow} successor such as
Stable Diffusion~3.5~\cite{esser2024sd3} is the maturity of its publicly
available ControlNet and IP-Adapter checkpoints, on which our
conditioning stack depends; we also qualitatively observed the
rectified-flow backbone to be more prone to memorise individual training
frames at this adapter rank and data budget.
The backbone choice is therefore deliberate, not a default; we examine it
methodically in Section~\ref{subsec:backbone_choice}.

\paragraph{Problem 2: Constraining the generated scene to a real
field reference while leaving room for textural variety.} Text prompts
alone cannot pin down the outline of a sand boil dome, its
three-dimensional bulge against an otherwise flat ground surface, or
the fine rim-texture transition where saturated sediment meets dry
mud. To this end, we assemble a stack of four ControlNet branches~%
\cite{zhang2023controlnet} on top of the frozen
DreamBooth-fine-tuned denoiser: Canny edges for object outline,
monocular depth~\cite{ranftl2021midas} for three-dimensional layout,
surface normals for dome--rim relief, and holistically-nested edge detection (HED)
soft edges~\cite{xie2015hed} for rim texture. Each
production preset activates a subset; the selected V4 preset combines
HED soft edges, surface normals, and an IP-Adapter style anchor.
To re-render the surrounding scene without disturbing the real defect
pixels, we replace the image-to-image step with a soft-mask
inpainting protocol. The inverted ground-truth mask is eroded and
Gaussian-blurred into a soft gradient, and the denoiser predicts only
inside that gradient at every step. The boundary ring is partially
regenerated throughout the diffusion loop, so that by the time the loop
terminates, the transition has been negotiated against the new lighting
and colour, and the hard-boundary seam of
binary compositing is avoided. This addresses the two failure modes of
the Poisson seamless-cloning step used in our earlier work~%
\cite{thapa2025thesis,perez2003poisson}: the hard mask boundary that
betrayed compositing, and the colour bleed from the new background
into the cloned region.

\paragraph{Problem 3: Producing textual conditioning that exhausts the
scene distribution without manual prompt engineering.} A fine-tuned
generator can only diversify within what the prompts ask for.
Hand-written prompt lists scale poorly across defect classes, offer no
guard against hallucinated or off-distribution prompts, and tie the
augmentation pipeline to the domain expert who wrote them. We replace
the hand-written list with a taxonomy-driven \emph{Prompt Atlas}: a
single JSON specification (concept, trigger token, axis labels,
exemplar references, negative-concept list) is expanded by either
template substitution or a small open-weight language model. It is then
filtered by length, sparse text-similarity de-duplication,
negative-concept screening, and image--text validation against the
reference set with a CLIP similarity score~\cite{radford2021clip}. The
atlas is reproducible from the specification alone and is constructed
to extend to new defect classes (seepage, sinkhole, rutting)
without code changes; the empirical validation in this paper is
restricted to the sand boil class
(Section~\ref{subsec:disc_generalization}).

\paragraph{Contributions.} This paper contributes a generation pipeline
whose components span three thrusts---defect-preserving generation,
label provenance, and reproducible prompt curation---detailed below and
assessed in one empirical study:
\begin{itemize}
\item \textbf{Defect-preserving soft-mask latent blending}
(Section~\ref{subsec:soft_mask}): a per-step composition rule that
pins the defect's image-space appearance verbatim while giving the
diffusion denoiser explicit freedom to negotiate a coherent transition
with the re-rendered surrounding scene, removing the hard-seam and
colour-bleed failure modes of binary inpainting and Poisson seamless
cloning respectively.
\item \textbf{Defect-size-adaptive soft-mask geometry}
(Section~\ref{subsec:soft_mask}): a fully automated size-scaled rule that
sets the soft-mask's erosion radius and Gaussian kernel width
from the defect's bounding-box geometry, so the preservation margin
and transition smoothness scale with the defect's spatial
extent rather than being hand-tuned per image.
\item \textbf{Convex Hull Annotator}
(Section~\ref{subsec:dataset}): a fully automated \emph{shape-aware}
probability-to-annotation pipeline (robust threshold $\to$ connected
components $\to$ resolution-invariant area filter $\to$ shape-aware
geometry adapter routed by form factor $\to$ dual raster/polygon
output) that emits a single-policy training label \emph{and} an
editable polygon annotation. It routes each component between
convex-hull, morphological-closing, and skeleton policies according to
its geometry, so the same code path generalises across
compact, elongated, and thin curvilinear defect classes
(Figure~\ref{fig:cha_pipeline}). The same machinery serves a second
role as a verification gate for the mask-conditioned path below.
\item \textbf{Mask-conditioned generation with exact label provenance,
as a zero-annotation diversity engine}
(Section~\ref{subsec:mask_controlnet}): a binary-mask SDXL ControlNet,
fine-tuned from a public soft-edge checkpoint, that generates
\emph{from} a chosen mask so the conditioning mask is the
ground-truth label by construction. This removes the dependence on an
external segmenter checkpoint that otherwise caps synthetic-label
quality, and repositions \textsc{Convex Hull Annotator} as the
drift-measuring quality-control gate for this path. Because the
conditioning silhouette is decoupled from the real source pool, the same
path renders single-boil scenes from a $39$-mask bank under per-instance
jitter. At zero annotation cost, this contributes exact labels and a variety of shapes, sizes, and placements
that the source-anchored path cannot produce
(Figure~\ref{fig:maskcn_label}). We
report its image quality and the limits of automatically certifying its
labels, and make no downstream-segmentation claim for it here.
\item \textbf{Reference-set similarity quality filtering}
(Section~\ref{subsec:quality_filter}): an automated post-generation
filter that drops out-of-distribution drift \emph{and} memorisation at
the same time, with an admissibility band auto-tuned from the
reference set's leave-one-out CLIP similarity statistics: no human
review, no hand-picked thresholds.
\item \textbf{Taxonomy-driven Prompt Atlas}
(Section~\ref{subsec:prompt_atlas}): a five-stage prompt-curation
pipeline that expands a single JSON domain specification into a
stratified, de-duplicated, image--text-validated prompt bank,
replacing ad-hoc hand-written prompt lists and porting to a new defect
class without code changes.
\item \textbf{Empirical evaluation} of generative image quality:
distribution distance (FID under two independent implementations, KID
with standard deviation, CLIP image--text similarity, LPIPS perceptual diversity)
re-based on the full reference set, \emph{and} a fidelity--diversity
decomposition via improved precision and recall~%
\cite{kynkaanniemi2019prec} and density/coverage
(Section~\ref{subsec:res_manifold}) that separates ``looks like'' from
``covers'' the references; a production-preset comparison
(Section~\ref{subsec:res_quality}); a quantitative comparison against
a Poisson seamless-cloning baseline (Section~\ref{subsec:res_poisson});
empirical validation of the CLIP admissibility filter on the augmented
set ($815$ of $1{,}020$ synthetic candidates admitted), with a
per-bucket diagnostic of where a global threshold is conservative
(Section~\ref{subsec:res_qfilter}); and a CLIP-space
nearest-reference memorisation audit
(Section~\ref{subsec:res_memorisation}). Downstream segmentation use of
the generated data is left to future work.
\end{itemize}

These components are not independent. The generation and labelling
components interlock to give the augmented dataset a single-policy mask
provenance and automatic quality envelopes on both the input and output
sides (Section~\ref{subsec:integration}), and together they are
engineered to make the generated data usable as segmentation
supervision---uniform label policy, on-distribution filtering---a
property we design for but do not test downstream in this paper. The
Prompt Atlas, together with the choice of SDXL over rectified-flow
successors (Section~\ref{subsec:backbone_choice}), is what ports across
defect classes through a single JSON specification; transferring the
full pipeline additionally requires a new DreamBooth adapter and
structural-conditioning sources, so we describe it as designed to
transfer rather than demonstrated on a second class.

\paragraph{Relation to prior work.} This pipeline builds on our earlier
thesis~\cite{thapa2025thesis}, and we delineate the boundary
explicitly. \emph{New in this paper} are the defect-preserving soft-mask
latent-blending compositor and its size-adaptive geometry (which replace
the Poisson seamless-cloning step of the thesis), the mask-conditioned
ControlNet with exact label provenance, the taxonomy-driven Prompt
Atlas, the leave-one-out CLIP admissibility filter, the repositioning of
\textsc{Convex Hull Annotator} as a drift-measuring verification gate,
and an image-quality evaluation re-based on the full reference set under
two independent FID implementations. \emph{Inherited and refined} from
the thesis are the core \textsc{Convex Hull Annotator} annotation
algorithm and the Poisson-composition protocol; the latter appears here
only as an evaluation baseline. This paper claims no contribution to
downstream segmentation, which is left to future work.

The complete pipeline (DreamBooth checkpoint, Prompt Atlas record,
multi-ControlNet inference configuration, soft-mask blending
parameters, per-image seed cascade) is released alongside the paper, so
the reported results can be reproduced.

\paragraph{Scope.} This paper focuses on the \emph{generation} of
synthetic sand boil imagery and on the methodology and metrics for
evaluating that generation. Downstream use of the generated data (the
leak-free cross-validated stacking ensemble and the
confusion-score-mined hard-negative training phase) is left to future
work.

The remainder of the paper is organised as follows.
Section~\ref{sec:related_work} positions the contribution against the
literature on diffusion-based generative augmentation, DreamBooth and
ControlNet, and prompt curation.
Section~\ref{sec:methodology} develops the pipeline.
Section~\ref{sec:experiments} describes the experimental protocol.
Section~\ref{sec:results} reports quantitative and qualitative
results. Section~\ref{sec:discussion} examines the design choices and
the remaining failure modes. Section~\ref{sec:conclusion} concludes.

%% file: sections/related_work.tex
\section{Related Work}
\label{sec:related_work}

The present work sits at the intersection of three research threads:
automated detection of sand boils and related levee defects,
diffusion-based image synthesis as a remedy for data scarcity, and the
emerging practice of systematic prompt construction for
generative-augmentation pipelines. We review each in turn and, in each, mark
precisely where our pipeline departs from established practice.
Segmentation architectures and ensemble learning, which consume the
synthetic data produced here, lie outside this scope and are left to
future work.

\subsection{Sand-boil and levee-defect imagery}
\label{subsec:rw_levee}

Automated inspection of earthen levees began with classical image
processing (edge filters, thresholding, and handcrafted descriptors fed
to shallow classifiers), which degrades badly under the lighting
changes, mud, vegetation, and reflective standing water that dominate
field photographs. Deep learning has since become the norm across
infrastructure distress detection~\cite{guan2023pavement}, and the
sand-boil literature in particular has moved to convolutional
encoder--decoders. Panta \textit{et al.}~%
\cite{panta2023sandboilnet} proposed SandBoilNet, a fully convolutional
encoder--decoder whose partially fine-tuned ResNet-50v2 backbone carries
PCA-based channel-and-spatial attention with Inception blocks on its
skip connections. Its reported intersection-over-union (IoU) of $0.5743$ and
balanced accuracy of $0.8552$ on a held-out real test set are the
reference point for the downstream segmentation in this series. Earlier,
Kuchi \textit{et al.}~\cite{kuchi2019sandboil} were the first to couple a
small synthetic sand-boil set with a CNN, but they classified whole
images rather than segmenting pixels.

The same convolutional recipe recurs across neighbouring
infrastructure defects: culvert-and-sewer segmentation with an enhanced
feature-pyramid network~\cite{alshawi2023efpn}, sinkhole detection with
a depthwise-separable U-Net~\cite{alshawi2024sinkhole}, and
levee-seepage segmentation from the same group~\cite{panta2024seepage}.
Synthetic-aperture-radar monitoring~\cite{james2011sar} probes
subsurface water movement, but at a spatial resolution far too coarse
for the small-scale cones that define a sand boil.

What unites these studies is that none treats diffusion synthesis as a
primary source of training data. Our own earlier prototype~%
\cite{thapa2025thesis} took a first step, namely single-control diffusion
augmentation on Stable Diffusion v1.5 with Poisson seamless
cloning, but the present paper rebuilds it end to end: a stronger
multi-conditioned generator, a soft-mask inpainting protocol that keeps
the labelled defect essentially intact (up to a small, post-processed
rim drift), and a reproducible prompt builder in place of a
hand-written list.

\subsection{Diffusion-based generative augmentation}
\label{subsec:rw_diffusion}

Before diffusion, variational autoencoders~\cite{kingma2014vae} and
generative adversarial networks~\cite{goodfellow2014gan} were the
default synthesisers under data scarcity. Both struggle in our regime:
on the small, visually homogeneous reference sets typical of levee
inspection, StyleGAN variants~\cite{karras2020stylegan2,karras2020stylegan2ada}
are prone to mode collapse and unstable training, and tend to echo a
handful of real images rather than broaden the distribution.

Diffusion models~\cite{ho2020ddpm,song2021ddim} trade this instability
for a many-step denoising process that is markedly easier to train.
Latent diffusion moves the denoising into a compressed latent
space~\cite{rombach2022ldm}, making high-resolution synthesis tractable
on a single GPU, and Stable Diffusion XL (SDXL)~\cite{podell2023sdxl} scales
the backbone and native resolution further. Three extensions turn this
backbone into a controllable, subject-specific generator. DreamBooth~%
\cite{ruiz2023dreambooth} binds a new subject to a rare trigger token
from only a handful of reference images, so later samples render that
subject under novel scenes and lighting. Low-Rank Adaptation~%
\cite{hu2022lora} folds the fine-tuning update into a compact additive
adapter (cutting the trainable parameters from billions to roughly ten
million), so DreamBooth-style training fits on a single accelerator.
ControlNet~\cite{zhang2023controlnet} attaches a parallel branch that
injects a structural signal (a Canny edge map, a monocular depth or
surface-normal estimate, or a holistic-edge map, abbreviated HED) at every skip
connection of the frozen denoiser, enforcing geometry that text prompts
cannot. Orthogonally, the IP-Adapter~\cite{ye2023ipadapter} feeds a
reference-\emph{image} embedding through a decoupled cross-attention
path, supplying a style anchor that we exploit in the V3 and V4 presets
(Section~\ref{subsec:production_presets}).

That diffusion augmentation pays off is by now well documented: in
image classification~\cite{trabucco2023effective,azizi2023synthetic},
across remote-sensing tasks~\cite{liu2024diffrs}, and especially in
data-scarce medical imaging~\cite{kazerouni2023diffusion}, where it has
become a routine response to small annotated cohorts. Generative augmentation of surface and industrial defects is itself an
active area---GAN- and diffusion-based crack, weld, and anomaly
synthesis, including mask-controlled defect
generation~\cite{duan2023dfmgan,hu2024anomalydiffusion}---but pixel-level
earthen-levee defects, and sand boils in particular, have seen almost
none of it. We also note the newer rectified-flow diffusion models,
which follow long compositional prompts more faithfully and render text
better, but demand more memory and lack a mature ecosystem of
structural-conditioning checkpoints. For the structural control our
pipeline depends on, we therefore commit to SDXL, as argued in
Section~\ref{subsec:backbone_choice}.

\paragraph{Composition strategies.} When diffusion is used to augment
defect imagery, the usual recipe is to synthesise a whole image from a
prompt and then either trust the generator's auto-produced mask or
hand-label the result. A second line of work edits inside a fixed
region: RePaint~\cite{lugmayr2022repaint} re-noises and re-denoises
masked pixels in image space, while Blended Diffusion~%
\cite{avrahami2022blended} fuses a text-driven edit with the source
under a region mask. Our soft-mask path takes the complementary region---preserving the
masked defect verbatim and re-rendering only the surrounding scene---but
its contribution over merely complementing an inpainting mask is the
per-step latent composition schedule and the size-adaptive blend band
that keeps the preserved defect aligned with the real source geometry.
We position the method relative to RePaint and Blended Diffusion rather
than benchmark against them: our objective is label-preserving scene
re-rendering, not masked-region editing, so a like-for-like comparison
is left to future work. The segmentation label itself is re-predicted
post hoc rather than inherited (the re-prediction ceiling discussed in
the next paragraph). This soft-mask path sidesteps both
the labelling cost of fully synthetic generation and the hard seams and
colour bleed of the Poisson seamless cloning~\cite{perez2003poisson}
used in our earlier work~\cite{thapa2025thesis}.

\paragraph{The label-provenance gap.} A common thread runs through all
of these recipes: the image is synthesised first and the
\emph{label} is recovered afterwards, whether by trusting an
auto-produced mask, by hand-annotating it, or (as in our soft-mask path) by re-predicting the mask with a segmenter whose accuracy then caps
the label quality. The alternative is to invert the dependency and let a
chosen structural map \emph{be} the label, so the generator is
conditioned to fill it. ControlNet~\cite{zhang2023controlnet} makes this
mechanically possible by accepting an arbitrary structural map as
conditioning. Making the conditioning map the label by construction is the premise of
semantic image synthesis and generate-with-labels
work~\cite{park2019spade,zhang2021datasetgan} and, in the industrial
setting, of mask-controlled anomaly and defect
synthesis~\cite{hu2024anomalydiffusion,duan2023dfmgan}; it is
comparatively under-explored for low-resource civil-infrastructure
defects and, to our knowledge, for sand boils specifically. Our
mask-conditioned path (MaskCN, Section~\ref{subsec:mask_controlnet})
takes this route in that setting and pairs it with the CHA drift gate: it
fine-tunes a binary-mask SDXL ControlNet and generates a fresh defect
into a chosen silhouette, so the mask is the label with no post-hoc
prediction step. Because the conditioning silhouette need no longer come
from a real photograph, the mask bank can be jittered per instance
(flip, rotation, rescale, and repositioning) to obtain a variety
the source-anchored inpainting path cannot produce.
This positions MaskCN as a complementary
diversity engine to the soft-mask path: the latter widens the scene
distribution around real defect geometry; the former widens the
shape, size, and placement distribution of the defect silhouette at zero
annotation cost. We report its image
quality and the limits of automatically verifying its labels with a
real-trained segmenter, and defer any downstream-segmentation claim to
future work.

\subsection{Prompt curation}
\label{subsec:rw_prompts}

Prompt construction is a consequential part of any
generative-augmentation pipeline. Most systems rely on a short,
hand-written prompt list, which scales poorly to new domains and offers
no protection against hallucinated or off-distribution prompts.
Automated prompt construction has been explored mainly for
classification (through chain-of-thought or template expansion), while
the image-synthesis side remains comparatively thin.

We address this with the Prompt Atlas. A single JSON domain
specification (concept, trigger token, axis labels, exemplar
references, and negative concepts) is expanded by template
substitution or a small open-weight language model, and the candidates
are filtered by length, by sparse text-similarity de-duplication,
by negative-concept matching, and by image--text similarity to the reference
set~\cite{radford2021clip,hessel2021clipscore}. The atlas is
reproducible from its specification alone and ports to a new defect
class by editing that single file, with no changes to the builder.

%% file: sections/methodology.tex
\section{Methodology}
\label{sec:methodology}

\begin{figure*}[tp]
\centering
\includegraphics[width=0.98\textwidth]{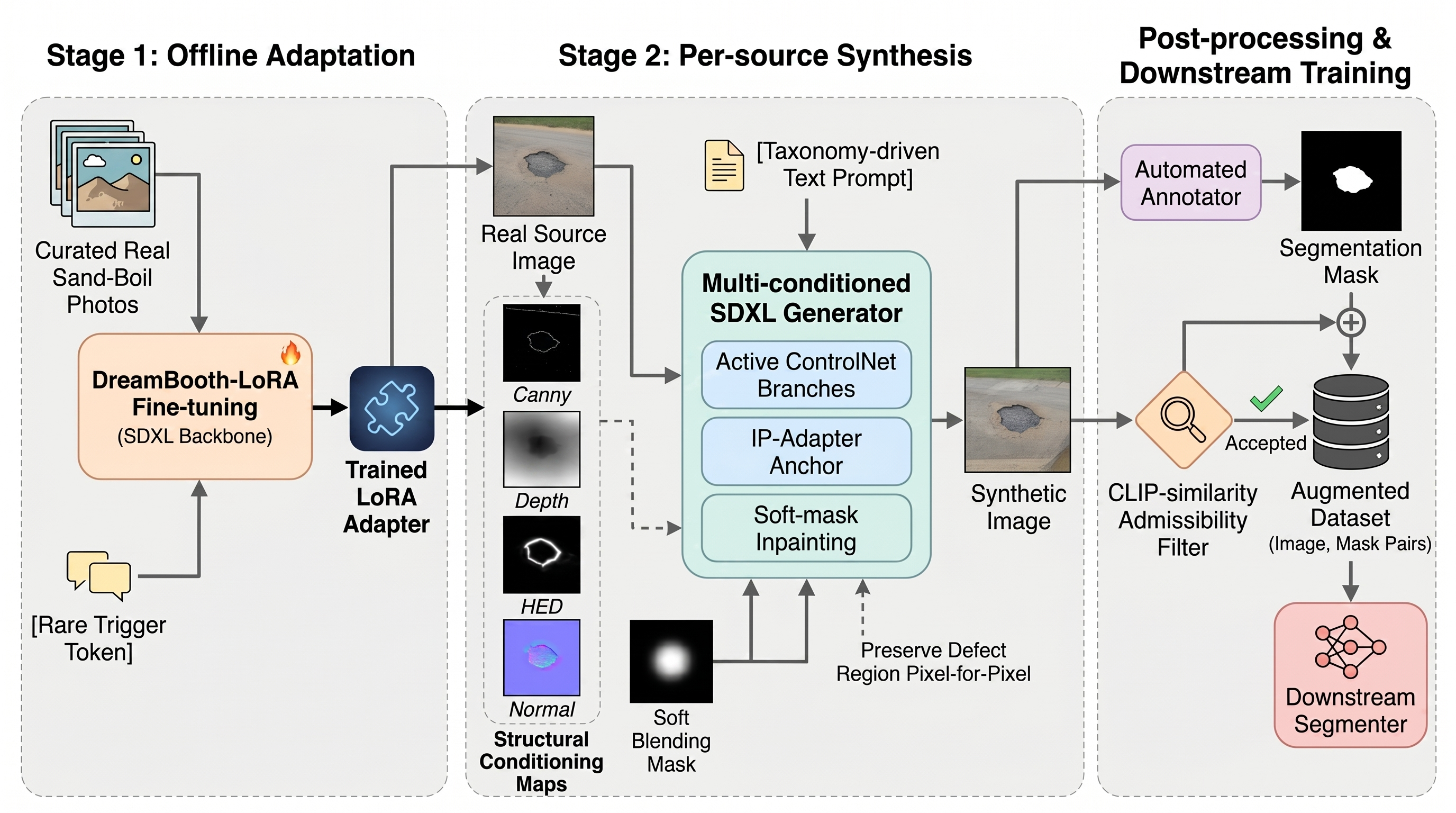}
\caption{Conceptual overview of the synthesis pipeline.
\textbf{Stage~1 (offline adaptation):} a DreamBooth-LoRA adapter is fit
once to the curated reference set, anchored by the rare trigger token
\texttt{sbx}, and loaded into the synthesis stack.
\textbf{Stage~2 (per-source synthesis):} for each real source image, the
four-branch ControlNet stack (Canny, monocular depth, HED soft edges, and
surface normal) and a soft blending mask are extracted; the
multi-conditioned SDXL generator activates a preset-specific subset of
the ControlNet branches, together with an IP-Adapter style anchor and
soft-mask inpainting, and re-renders the surrounding scene while
preserving the real defect region pixel-for-pixel (the selected V4
preset uses HED soft edges and surface normal).
\textbf{Post-processing:} each synthetic image is paired with a mask by
the \textsc{Convex Hull Annotator} and passed through the CLIP
admissibility filter (Section~\ref{subsec:quality_filter}); accepted
(image, mask) pairs enter the augmented dataset used to train the
downstream segmenter.}
\label{fig:pipeline}
\end{figure*}

The generation pipeline has two coupled stages. First, a diffusion-based
generator, fine-tuned on a small curated reference set, produces
structurally faithful synthetic sand boil imagery while preserving the
real defect region pixel-for-pixel. Second, a taxonomy-driven prompt
builder supplies the generator with stratified, validated textual
conditioning. Each synthetic record is paired with a binary mask from a
uniform automated annotator, so every generated image is admissible as
training supervision for downstream segmentation.
Figure~\ref{fig:pipeline} gives a schematic overview, and the
subsections that follow develop each component in turn.

\subsection{Dataset and preprocessing}
\label{subsec:dataset}

The real sand boil dataset is drawn from the U.S.\ Army Corps of
Engineers (USACE) levee-inspection archive~\cite{usace_levee_manual}.
Field inspectors traverse the crest and adjacent areas of an earthen
levee with mobile-phone or DSLR cameras and photograph any
abnormalities they encounter. The archive contains over four thousand
defect photographs. From these, a domain expert curated a subset of
approximately three hundred sand boil images on the basis of
visibility and unambiguous appearance. This subset was then filtered
by automatic intersection-over-union agreement between the manual
annotation and an initial Segment Anything~\cite{kirillov2023sam}
proposal. The final labelled set contains
$N_{\mathrm{train}}{=}199$ training images and
$N_{\mathrm{test}}{=}51$ held-out test images at variable native
resolutions, each paired with a binary pixel-level mask. A high-quality
\emph{subset} of $N_{\mathrm{ref}}{=}39$ images with the highest
agreement score anchors the DreamBooth-LoRA appearance fine-tuning and
supplies the conditioning-mask bank used by mask-conditioned generation (MaskCN, Section~\ref{subsec:mask_controlnet}). All counts in this
paper refer back to these quantities; elsewhere the prose says
``the training set'', ``the held-out test set'', and ``the curated
subset'' to avoid restating the absolute numbers.

\paragraph{Annotation protocols.} Two separate mask-production
protocols are used: one for the real reference images and one for the
synthetic outputs of the generator.

For the \emph{real} images, every mask is hand-drawn at the pixel
level and then cross-checked against a Segment Anything proposal of
the same image~\cite{kirillov2023sam}. The IoU between the two is
recorded, and images with low agreement are returned for
re-annotation rather than admitted to the training set. This yields
masks that preserve concave rim features, and it adds an objective
quality filter that the click-based protocol of our earlier work~%
\cite{thapa2025thesis} lacked.

For the \emph{synthetic} images, we introduce
\textsc{Convex Hull Annotator} (CHA)~%
\cite{thapa2025thesis,thapa_chull_annotator}, a fully automated
\emph{shape-aware} probability-to-annotation pipeline. CHA takes a
sigmoid probability map from a class-specific segmenter
(SandBoilNet~\cite{panta2023sandboilnet} for sand boils) and emits a paired
binary mask plus a VGG-Image-Annotator (VIA)-format polygon
annotation~\cite{dutta2019via}. CHA is the second of two coupled
mask-provenance mechanisms in our pipeline; the first is the
soft-mask inpainting protocol of Section~\ref{subsec:soft_mask}.
Because CHA is applied uniformly across all four production presets,
the augmented dataset has a single, reproducible mask-production
policy. The same shape-aware machinery serves a second role. In the
mask-conditioned generation path of
Section~\ref{subsec:mask_controlnet} the conditioning mask is
already the ground-truth label, so CHA acts instead as a quality-control
gate that measures how far the rendered dome drifted from the
requested silhouette. The shape-aware geometry adapter (stage~4 below)
makes CHA class-agnostic in both roles. The pipeline therefore transfers from
sand boils to seepage, rutting, sinkhole, or crack defects
through a single domain-specification file, with no edits to the
post-processing logic.

\begin{figure*}[tp]
\centering
\includegraphics[width=\textwidth]{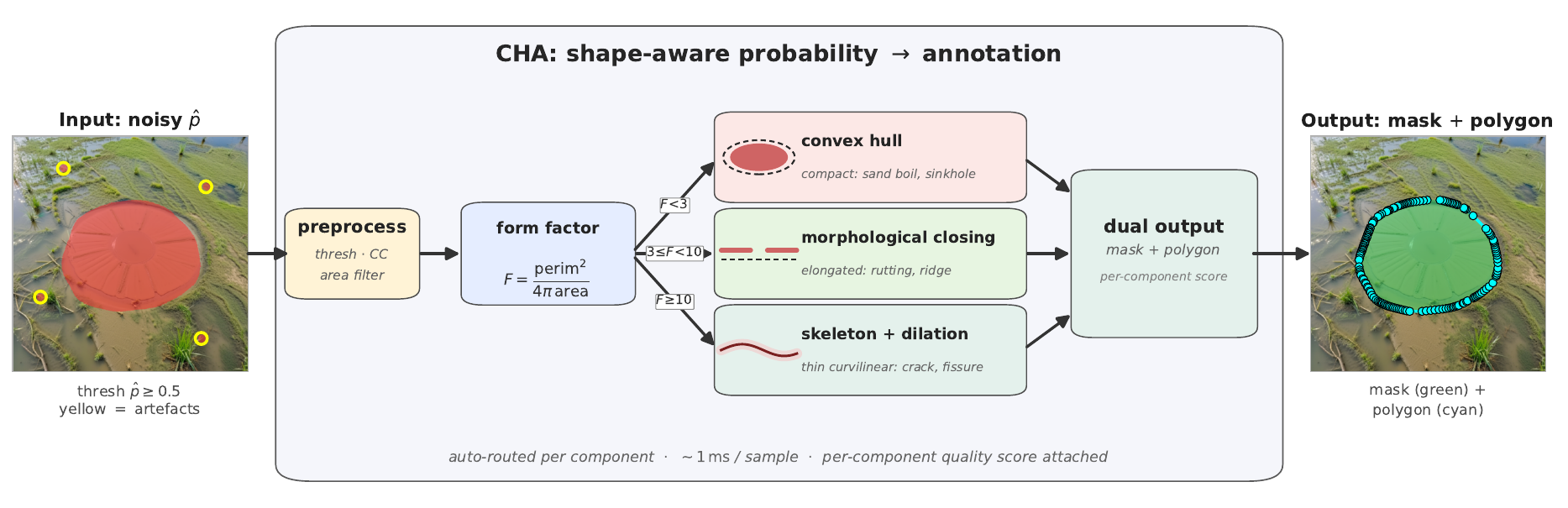}
\caption{\textsc{Convex Hull Annotator} (CHA), the shape-aware
probability-to-annotation pipeline. After thresholding and component
cleanup (connected components, area filter), each surviving component is
routed by its form factor $F=\mathrm{perim}^2/(4\pi\,\mathrm{area})$ to a
convex hull ($F<3$, compact), a morphological closing ($3\le F<10$,
elongated), or a skeleton with controlled dilation ($F\ge10$, thin
curvilinear), yielding a $0/255$ binary raster mask and a VIA-format
polygon with a per-component quality record.}
\label{fig:cha_pipeline}
\end{figure*}

\paragraph{Pipeline.} Given a probability map $\hat p \in [0,1]^{H
\times W \times C}$ from a class-specific segmenter (collapsed
across $C$ output channels by per-pixel max), CHA proceeds in
five stages (Figure~\ref{fig:cha_pipeline}):
\begin{enumerate}\itemsep1pt
\item \textbf{Robust threshold.}~Binarise the probability map at
an image-specific threshold $\tau$. The default policy is
Otsu's method, which selects $\tau$ to maximise the
between-class variance of $\hat p$; alternatives (Li, Yen, mean,
fixed) are exposed in the domain specification. Otsu replaces
the original CHA's mean-of-probability threshold, which is
fragile on near-empty and near-saturated probability maps.
\item \textbf{Connected components.}~Label the binary mask with
eight-connectivity.
\item \textbf{Resolution-invariant area filter.}~Reject
components whose area falls outside $[a_{\min}, a_{\max}]$
expressed as \emph{fractions of image area}; defaults
$a_{\min}{=}0.1\%$ and $a_{\max}{=}85\%$ cover both threshold-noise
suppression and segmenter-blow-up rejection. Expressing the
bounds as fractions removes the original CHA's
$100$-pixel-at-$512^2$ hard-coding, so the same configuration
generalises across image resolutions.
\item \textbf{Shape-aware geometry adapter.}~For each surviving
component, the dimensionless form factor
$F = \mathrm{perimeter}^2 / (4\pi \cdot \mathrm{area})$ is computed
($F{=}1$ is a perfect circle, larger values indicate thinness).
The component is then cleaned with a shape-appropriate policy:
\textbf{convex hull} when $F < 3$ (compact defects: sand boils,
sinkholes), \textbf{morphological closing} when
$3 \leq F < 10$ (elongated-but-thick defects: rutting grooves),
or \textbf{skeleton + controlled dilation} when $F \geq 10$
(thin curvilinear defects: cracks). Auto-routing can be
overridden by an explicit shape policy in the domain spec.
This is the principal generalisation that makes CHA
class-agnostic.
\item \textbf{Dual output with per-component quality scores.}~Union
the cleaned components into a $0/255$ binary raster mask for
downstream segmentation supervision, and emit each component as
a VIA-format polygon~\cite{dutta2019via} carrying a quality
record $(s, \mathrm{solidity}, F)$ where $s = \mathrm{mean}(\hat
p)$ inside the component. The polygons load directly into
standard polygon-annotation tools (VIA, CVAT, LabelMe) for
optional human review; the quality scores let downstream
consumers filter low-confidence components without re-running
the segmenter.
\end{enumerate}
The dual raster-plus-polygon output is what makes CHA an
\emph{annotator} rather than a mask cleaner: the raster path feeds
downstream training, and the polygon path feeds downstream human
review. The shape-aware geometry adapter (stage~4) is what makes CHA
class-agnostic: the same code path handles compact, elongated, and
thin curvilinear defects through automatic form-factor routing, with
no parameter change. This routing is validated on synthetic defects
of all three shape classes in the reference implementation.

The same CHA pipeline is applied uniformly to every synthetic image,
regardless of which production preset produced it (see
Section~\ref{subsec:production_presets}), so mask provenance is
single-policy across the augmented dataset. As
Figure~\ref{fig:cha_pipeline} makes visually concrete, the cleanup
is not cosmetic. The spurious small blobs that survive the naive
$\hat p \geq 0.5$ threshold, if used as training labels directly,
would teach the downstream segmenter to fire on threshold-uncertainty
noise in regions of the synthetic image that contain no defect at all.

For downstream segmentation, images and masks are resampled to
$512 \times 512$ pixels. Image resampling uses cubic interpolation;
mask resampling uses nearest-neighbour interpolation followed by a
binary threshold, so that labels remain strictly $\{0,1\}$ after
resize. After loading, channels are reordered from BGR to RGB and
pixel values are scaled to $[0,1]$. A suite of geometric and photometric
augmentations is drawn from the Albumentations library~%
\cite{buslaev2020albumentations}, applied jointly to image and mask
for spatial transforms and to the image alone for intensity and noise
transforms.

\subsection{DreamBooth fine-tuning with low-rank adaptation}
\label{subsec:lora}

Stable Diffusion XL is a publicly released general-purpose
text-to-image model trained on a large web-scale image corpus by
Stability AI, and it has no prior exposure to civil-infrastructure
defect imagery. A stock SDXL pipeline prompted with ``a photo of a
sand boil'' tends to return unrelated imagery that the word evokes
from web text (volcanic vents, coastal surf, children's sandboxes),
confirming that the base model does not encode the defect class. To
adapt the model without retraining it, we use the DreamBooth-LoRA
recipe of~\cite{ruiz2023dreambooth,hu2022lora}. The pretrained SDXL
weights remain frozen throughout training: no value inside the
released checkpoint is modified. Instead, a small additive adapter
(about $10$ million trainable parameters, against SDXL's ${\sim}2.6$
billion) is fit to the curated USACE reference set and paired with a
rare trigger token \texttt{sbx}. The adapter is therefore the only
component of the generator that has observed real sand boils. At
inference time the frozen base supplies the general photographic world
model (scenes, lighting, materials, geometry), and the adapter steers
those priors toward the defect class whenever the prompt contains
\texttt{sbx}. Unloading the adapter returns the pipeline to plain
pretrained SDXL, and the off-domain behaviour reappears. We
read this as evidence that the sand-boil-specific adaptation is carried
by the adapter rather than the frozen base.

We fine-tune SDXL with a Low-Rank Adaptation~\cite{hu2022lora}
adapter inserted at the four attention projections (query, key,
value, and output) of every transformer block in the denoiser.
For each projection the original weight matrix $W_0$ is frozen
and a trainable update $\Delta W = B A$ is added, with
$B \in \mathbb{R}^{d \times r}$ and $A \in \mathbb{R}^{r \times d}$
and $r{=}16$ in the production SDXL configuration:
\begin{equation}
y = W_0 x + \tfrac{\alpha}{r} B A x ,
\label{eq:lora}
\end{equation}
where $\alpha$ is a fixed scaling constant and the gradient flows
only through $A$ and $B$ (never through $W_0$). For an SDXL
attention matrix of width $d \approx 1280$, each adapter
contributes ${\sim}2rd \approx 40{,}960$ parameters in place of
the original $d^2 \approx 1.6 \times 10^6$, a per-matrix
reduction of roughly $40\times$.

The trigger token is a rare three-letter string, \texttt{sbx},
selected because it does not appear meaningfully in the SDXL
tokeniser. Reference images are paired with the instance prompt
\textit{a photo of sbx sandboil} during training. We train for
$1{,}500$ steps under a cosine learning-rate schedule (peak
$1\mathrm{e}{-4}$), batch size $1$, and gradient accumulation $4$, with
class-image prior preservation ($200$ auto-generated \textit{a photo of
a sandboil} images), using mixed precision (fp16), the 8-bit Adam
optimiser~\cite{dettmers2022bnb}, and gradient checkpointing. (An
initial rank-$8$, $800$-step prototype was used during development; all
results reported here use this production rank-$16$ adapter.) The
trained adapter is serialised as a single safetensors file and loaded
into the SDXL inference pipeline. The full configuration is restated for
reproducibility in Section~\ref{subsec:lora_ablation}.

\subsection{Backbone choice: SDXL versus rectified-flow successors}
\label{subsec:backbone_choice}

Choosing Stable Diffusion XL over a more recent rectified-flow
successor is a deliberate trade-off, because the
two families occupy different operating points. The successors we
considered are Stable Diffusion~3.5~\cite{esser2024sd3} and comparable Multi-Modal
Diffusion Transformer (MMDiT) backbones. Rectified-flow
MMDiT backbones are widely reported to give stronger prompt adherence
on long compositional captions, better in-image text rendering, and a
more cinematic global appearance. They do so at noticeably higher VRAM
cost per generated pixel and with a substantially less mature
ecosystem of pretrained ControlNet, IP-Adapter, and
structural-conditioning weights at the time of writing. SDXL, in
contrast, retains the strongest publicly released ControlNet stack
(Canny, depth, HED, normal, tile, soft-edge), the most mature LoRA
tooling, and the lowest inference latency at $1024 \times 1024$. Its
cost is weaker prompt adherence on unusually long or compositional
prompts and weaker text rendering. For the present application, none
of MMDiT's strengths are on the critical path: our prompts are short
scene descriptions of a single defect on a levee, no in-image text is
required, and the generation is structurally constrained by three
ControlNet branches plus IP-Adapter rather than by the prompt alone.
SDXL's strengths, in turn, are exactly the ones we depend on, so the
trade-off resolves cleanly in favour of SDXL.

Three further observations from our own pipeline experimentation
reinforced this choice. We report them as qualitative findings rather
than as headline numbers. First, the publicly available ControlNet
and IP-Adapter checkpoints for SDXL were directly compatible with our
soft-mask inpainting protocol of Section~\ref{subsec:soft_mask},
whereas equivalent production-quality conditioning checkpoints for the
rectified-flow backbone were not available in a form that
interoperates with the same protocol. Second, at this application's reference-set
scale, a high-capacity MMDiT denoiser fine-tuned
via DreamBooth-LoRA appeared, in qualitative inspection, more prone to
memorising individual training frames: the generator reproduced the
trained dome with little geometric variation across prompts and seeds. This overfitting
failure mode was less pronounced on the lower-capacity SDXL U-Net at
the same adapter rank and data budget. Third, on the H100~NVL hardware
used here, the SDXL inpainting pipeline runs comfortably at
$1024 \times 1024$ with up to three simultaneous ControlNets (or two ControlNets plus an IP-Adapter).
In our informal profiling, the rectified-flow counterpart at the same
conditioning depth left noticeably less headroom for the IP-Adapter
style anchor that the present pipeline relies on for visual variety. We kept the comparison on a fixed structural-augmentation framework
to keep it operationally simple and reproducible. We therefore report SDXL as the
production backbone and treat the rectified-flow alternative as out of
scope.

\subsection{Multi-ControlNet conditioning}
\label{subsec:multicn}

Text prompts alone cannot constrain the geometry of a generated image
to that of a particular field reference. ControlNet~%
\cite{zhang2023controlnet} addresses this by training a parallel
branch of the denoiser to consume a structural map and emit residual
features that are additively combined with the main branch at every
skip connection and at the mid block. The present pipeline makes
four publicly released ControlNet variants available on top of the
DreamBooth-fine-tuned SDXL backbone, so that each generated image can be
constrained from up to four independent structural perspectives; each
production preset activates a subset of these branches
(Section~\ref{subsec:production_presets}).

\begin{enumerate}
\item \textbf{Canny-edge conditioning.} The Canny detector~%
\cite{canny1986} is run on the source image with low and high
thresholds of $30$ and $90$ (a dense setting selected to trace the low-contrast boil rim), producing a binary edge map that
constrains the outline of the sand boil dome together with the
dominant boundaries of the surrounding scene.

\item \textbf{Monocular-depth conditioning.} A pretrained vision
transformer for dense depth prediction~\cite{ranftl2021midas}
produces a per-pixel depth estimate of the source image, which
constrains the three-dimensional layout. A dome bulging out of an
otherwise flat ground surface remains bulging in the synthetic
counterpart.

\item \textbf{Soft-edge (HED) conditioning.} Holistically-Nested Edge
Detection~\cite{xie2015hed} produces a soft probability map rather
than a binary boundary, capturing fine textural transitions (grain
boundaries, wet versus dry patches, the rim where saturated
sediment meets the surrounding mud) that the binary Canny map
cannot represent.

\item \textbf{Surface-normal conditioning.} A surface-normal map
estimated from the source image constrains the local surface
orientation, reinforcing the dome--rim relief so that the bulge of
the boil reads as a coherent three-dimensional form rather than a
flat texture patch. It complements the depth branch, which fixes
coarse layout, by supplying finer per-pixel orientation cues.
\end{enumerate}

\begin{figure}[tp]
\centering
\resizebox{\columnwidth}{!}{%
\begin{tikzpicture}[
    every path/.style={line cap=round, line join=round},
    img/.style={inner sep=0pt, draw=black!50, line width=0.5pt},
    lbl/.style={font=\footnotesize, align=center, inner sep=1.5pt},
    cn/.style={rectangle, rounded corners=3pt, draw=black!55, fill=white,
               line width=0.7pt, align=center, font=\footnotesize\bfseries,
               minimum width=24mm, minimum height=11mm, inner sep=2pt,
               blur shadow={shadow xshift=0.4mm, shadow yshift=-0.4mm,
                            shadow blur radius=0.9mm, shadow opacity=14}},
    nn/.style={rectangle, rounded corners=3pt, draw=black!55, fill=black!7,
               line width=0.7pt, align=center, font=\footnotesize\bfseries,
               minimum width=26mm, minimum height=12mm, inner sep=2.5pt,
               blur shadow={shadow xshift=0.4mm, shadow yshift=-0.4mm,
                            shadow blur radius=0.9mm, shadow opacity=16}},
    mul/.style={rectangle, rounded corners=2pt, draw=black!55, line width=0.5pt,
                fill=black!4, align=center, font=\footnotesize, inner sep=2.5pt},
    sumn/.style={circle, draw=blue!55!black, line width=0.9pt, fill=blue!16,
                 inner sep=0pt, minimum size=9mm, font=\large},
    flow/.style={-{Stealth[length=2.2mm]}, draw=black!55, line width=0.55pt},
    res/.style={-{Stealth[length=2.6mm]}, draw=black!75, line width=0.9pt},
    main/.style={-{Stealth[length=2.8mm]}, draw=black!82, line width=1.1pt},
    rl/.style={font=\footnotesize, inner sep=1pt},
    snowflake/.pic={
      \draw[cyan!60!blue, line width=0.32pt, line cap=round]
        (90:0.95mm)--(270:0.95mm) (30:0.95mm)--(210:0.95mm) (150:0.95mm)--(330:0.95mm);
      \foreach \a in {90,30,150,210,270,330}{
        \draw[cyan!60!blue, line width=0.28pt] (\a:0.95mm)--++(\a+135:0.32mm);
        \draw[cyan!60!blue, line width=0.28pt] (\a:0.95mm)--++(\a-135:0.32mm);
      }},
  ]
\node[img] (src) at (3, 12.7) {\includegraphics[width=21mm]{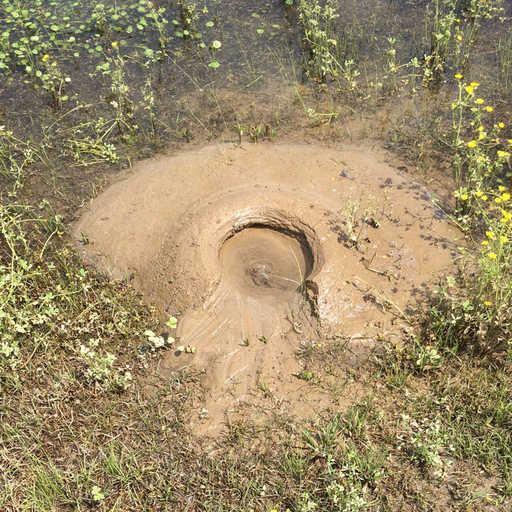}};
\node[lbl, below=0.4mm of src] (lsrc) {Source image $x$};
\node[img] (mc) at (0, 10.0) {\includegraphics[width=18mm]{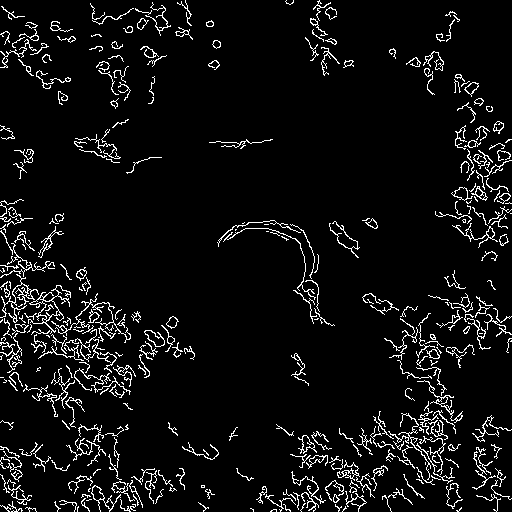}};
\node[img] (md) at (3, 10.0) {\includegraphics[width=18mm]{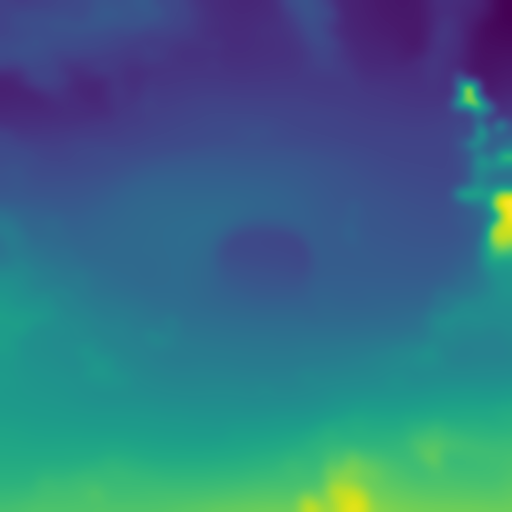}};
\node[img] (mh) at (6, 10.0) {\includegraphics[width=18mm]{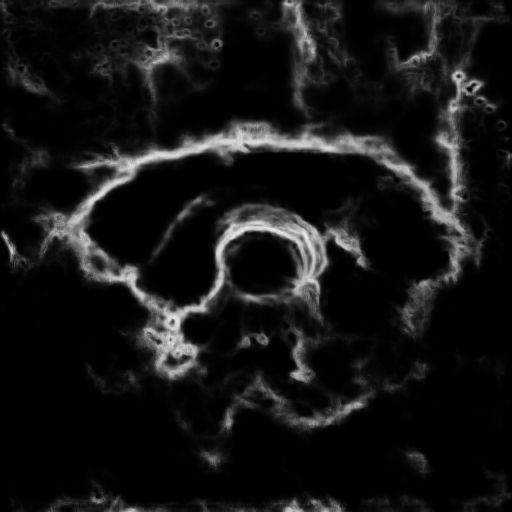}};
\node[lbl, below=0.4mm of mc] (lc) {Canny $c_c$};
\node[lbl, below=0.4mm of md] (ld) {Depth $c_d$};
\node[lbl, below=0.4mm of mh] (lh) {HED $c_h$};
\node[cn] (cc) at (0, 7.6) {ControlNet};
\node[cn] (cd) at (3, 7.6) {ControlNet};
\node[cn] (ch) at (6, 7.6) {ControlNet};
\pic at ([shift={(-2.1mm,-1.7mm)}]cc.north east) {snowflake};
\pic at ([shift={(-2.1mm,-1.7mm)}]cd.north east) {snowflake};
\pic at ([shift={(-2.1mm,-1.7mm)}]ch.north east) {snowflake};
\node[mul] (xc) at (0, 5.9) {$\times\,\lambda_c{=}0.40$};
\node[mul] (xd) at (3, 5.9) {$\times\,\lambda_d{=}0.25$};
\node[mul] (xh) at (6, 5.9) {$\times\,\lambda_h{=}0.40$};
\node[sumn] (sg) at (3, 3.7) {$+$};
\node[nn] (enc) at (-0.8, 3.7) {SDXL Encoder};
\node[nn] (dec) at (6.8, 3.7) {SDXL Decoder};
\draw[flow] (lsrc.south) -- (mc.north);
\draw[flow] (lsrc.south) -- (md.north);
\draw[flow] (lsrc.south) -- (mh.north);
\draw[flow] (lc.south) -- (cc.north);
\draw[flow] (ld.south) -- (cd.north);
\draw[flow] (lh.south) -- (ch.north);
\draw[res] (cc.south) -- node[rl, right] {$r_c$} (xc.north);
\draw[res] (cd.south) -- node[rl, right] {$r_d$} (xd.north);
\draw[res] (ch.south) -- node[rl, right] {$r_h$} (xh.north);
\draw[res] (xc.south) -- node[rl, pos=0.55, left=0.5mm] {$\lambda_c r_c$} (sg.north west);
\draw[res] (xd.south) -- node[rl, right=0.5mm] {$\lambda_d r_d$} (sg.north);
\draw[res] (xh.south) -- node[rl, pos=0.55, right=0.5mm] {$\lambda_h r_h$} (sg.north east);
\draw[main] (enc.east) -- node[above=0.2mm, font=\footnotesize] {$h_i$} (sg.west);
\draw[main] (sg.east)  -- node[above=0.2mm, font=\footnotesize] {$\mathrm{skip}[i]$} (dec.west);
\end{tikzpicture}%
}
\caption{Multi-ControlNet conditioning of SDXL. Canny, depth, and HED
maps from the source $x$ each drive a frozen, pretrained ControlNet; the
residuals, scaled by their weights $\lambda$, are summed with the encoder
feature $h_i$ into $\mathrm{skip}[i]$ (Eq.~\eqref{eq:cn_skip}) before the
decoder. The SDXL backbone is frozen and carries the trained
DreamBooth-LoRA adapter (Section~\ref{subsec:lora}).}
\label{fig:multicn}
\end{figure}
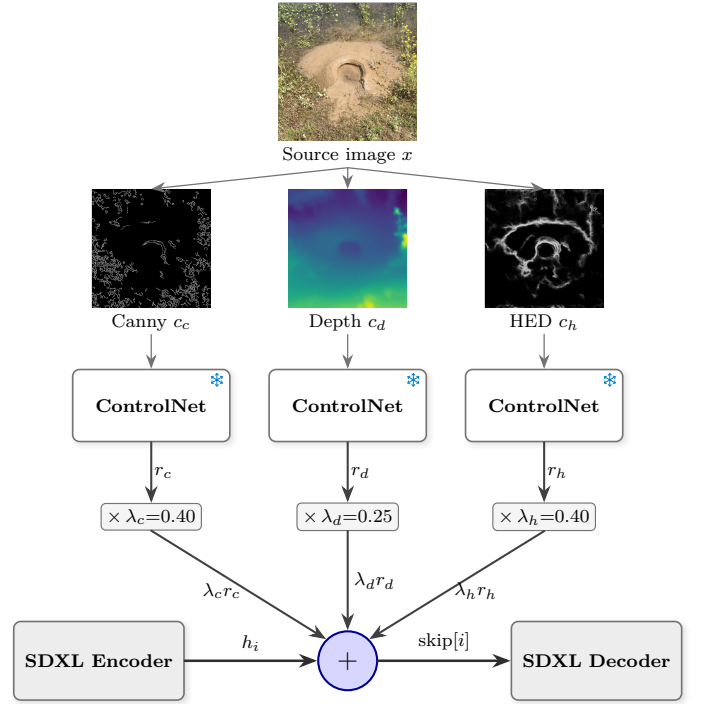

In the forward pass of every diffusion step, the residual at
denoiser skip connection $i$ becomes
\begin{equation}
\mathrm{skip}[i] = h_i + \lambda_c r^{c}_i + \lambda_d r^{d}_i + \lambda_h r^{h}_i + \lambda_n r^{n}_i ,
\label{eq:cn_skip}
\end{equation}
where $h_i$ is the encoder feature, $r^{c}_i, r^{d}_i, r^{h}_i, r^{n}_i$ are
the Canny, depth, soft-edge (HED), and surface-normal branch residuals at
the corresponding resolution, and $(\lambda_c, \lambda_d, \lambda_h,
\lambda_n) = (0.40,\, 0.25,\, 0.40,\, 0.40)$ are the conditioning scales
used throughout. Individual presets activate a subset of the four
branches; the unused branches are simply omitted from the sum. The same
combination is applied at the mid block. The scale budget is itself a
quality lever. Every additional control branch tightens the geometric
constraint, but it also consumes a portion of the denoiser's
representational budget that would otherwise render high-frequency
texture. Loading the stack indiscriminately therefore tends to produce
visibly smoothed outputs even at $1024 \times 1024$ resolution. The
Canny--depth--HED combination is the smallest configuration that
satisfies our core geometric requirements (outline, three-dimensional
layout, rim texture), while the surface-normal branch adds finer
dome--rim relief when a preset calls for it. The ablation of
Section~\ref{subsec:gen_ablation} traces the cost of denser
combinations and the contribution of an IP-Adapter style anchor~%
\cite{ye2023ipadapter}. The anchor imports high-frequency texture from a
real reference photograph and thereby partially restores fidelity in
configurations where one of the structural controls has been dropped.
Figure~\ref{fig:multicn} illustrates the additive-fusion mechanism on
three of these structural controls (Canny, depth, and HED soft edges)
extracted from a real source; the surface-normal branch fuses
identically.

\begin{figure*}[tp]
  \centering
  \includegraphics[width=\textwidth]{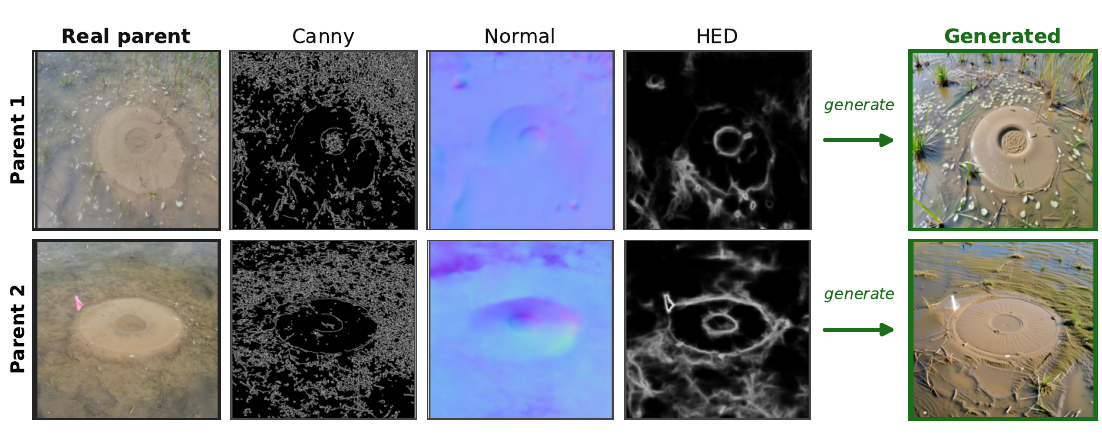}
  \caption{ControlNet conditioning for two real sand boil parents.
  Each row shows the real parent and its three control maps (Canny
  edges, surface normal, and HED soft edges) that jointly
  condition the DreamBooth-LoRA SDXL generator (trigger
  ``\texttt{sbx sandboil}'') to produce the synthetic output (green
  frame).}
  \label{fig:explain_controlnet}
\end{figure*}

Figure~\ref{fig:explain_controlnet} traces, for two real parents, the
three control maps that drive each synthesis (Canny edges,
surface normals, and HED soft edges) through to the generated output.
Qualitatively, the branches capture complementary structure: the normal
map encodes the dome--rim relief and HED the soft boil outline and
sediment texture, while Canny contributes a dense map of high-frequency
scene edges (vegetation, debris, and water lines) and is the weakest
cue for the smooth, low-contrast boil itself. We do not quantify each
branch's contribution; their conjunction is what keeps the synthetic
boil geometrically consistent with its parent while the DreamBooth
trigger repaints faithful texture.

\subsection{Defect-preserving soft-mask latent blending}
\label{subsec:soft_mask}

\emph{Defect-preserving soft-mask latent blending} is a per-step
composition rule that pins the defect's image-space appearance
verbatim while giving the diffusion denoiser explicit freedom to
negotiate a coherent transition with the re-rendered surrounding
scene. This is the first of two coupled
mask-provenance mechanisms in our pipeline; the second is the
\textsc{Convex Hull Annotator} mask producer of
Section~\ref{subsec:dataset}.

\paragraph{The failure modes it replaces.} The image-to-image and
inpainting paths of Stable Diffusion accept a mask that controls
which pixels the denoiser is allowed to modify. The naive choice for
our setting (preserve the sand boil core, re-render the background)
would pass the binary ground-truth mask directly. In practice this
produces a hard, visible seam at the boundary, because the predicted
and original latents change abruptly across a single pixel of the
mask. Our previous work~\cite{thapa2025thesis} used Poisson seamless
cloning~\cite{perez2003poisson} to blend the regenerated background
onto the original sand boil, but this introduced a separate failure
mode: the cloned region tended to inherit the colour cast of the new
background and appeared tinted. Soft-mask latent blending eliminates
both failure modes by construction. The boundary is renegotiated
continuously across the diffusion loop, rather than sharply at a
single mask edge or post hoc in pixel space.

\paragraph{Construction of the soft mask.} The ground-truth sand boil
mask is first inverted, so that $1$ denotes the background region we
wish to regenerate and $0$ denotes the dome core we wish to preserve.
The inverted mask is eroded with an elliptical structuring element
(growing a small protected ring around the dome) and then convolved
with a Gaussian kernel. The result is a smooth gradient mask
$M \in [0,1]^{H \times W}$ that is $0$ in the core, $1$ in the far
background, and varies smoothly across a boundary ring.

\paragraph{Defect-size-adaptive geometry.} The erosion radius and the
Gaussian kernel width control two complementary properties of the
soft mask. The erosion controls the \emph{preservation margin} (how
much of the dome interior is locked verbatim), and the blur controls
the \emph{transition smoothness} (how rapidly the denoiser regains
freedom as we cross the rim). A fixed $(12, 41)$-pixel pair, the
choice in our earlier work~\cite{thapa2025thesis}, over-erodes small
domes while barely affecting large ones. We instead derive both
radii adaptively from the defect's spatial extent. Let
$d = \min(h_{\mathrm{bbox}}, w_{\mathrm{bbox}})$ be the smaller side
of the dome's tight bounding box. We set
\begin{align}
r_{\mathrm{erode}} &= \max\!\left(4,\, \mathrm{round}(0.04\, d)\right), \notag \\
r_{\mathrm{blur}}  &= \max\!\left(15,\,
                       \mathrm{odd}(\mathrm{round}(0.12\, d))\right),
\label{eq:adaptive_radii}
\end{align}
where $\mathrm{odd}(\cdot)$ rounds up to the nearest odd integer
(Gaussian kernel sizes must be odd) and the floors of $4$ and $15$
pixels prevent degenerate operations on extremely small domes. The
coefficients $4\%$ and $12\%$ were selected so that the preservation
margin and the transition ring scale linearly with the dome's size: a
$300$-pixel-wide dome receives a $12$-pixel erosion and a $37$-pixel
blur (close to the fixed $(12, 41)$ values of our prior
work); a
$100$-pixel dome receives a $4$-pixel erosion and a $15$-pixel blur;
and a $600$-pixel dome receives a $24$-pixel erosion and a $73$-pixel
blur. The adaptive policy adds no manual tuning, requires no labelled
data, and is computed in milliseconds from the bounding box of the
binary mask that is already part of the inpainting input.

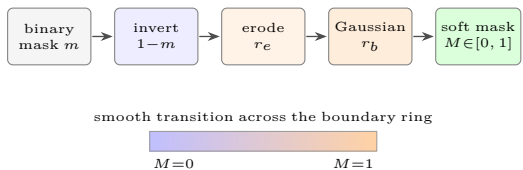
\begin{figure}[!tbp]
\centering
\begin{tikzpicture}[
    node distance=2.5mm and 3mm,
    every path/.style={line cap=round, line join=round},
    smbox/.style={rectangle, rounded corners=2pt, draw=black!50,
                  line width=0.4pt, align=center, font=\tiny,
                  inner sep=2pt, minimum width=11mm, minimum height=7.5mm},
    smarrow/.style={-{Stealth[length=1.6mm]}, draw=black!70, line width=0.5pt,
                    shorten <=0.5pt, shorten >=0.5pt},
  ]
\node[smbox, fill=black!4] (m)   {binary\\mask $m$};
\node[smbox, fill=blue!7,  right=of m]    (inv)  {invert\\$1{-}m$};
\node[smbox, fill=orange!10, right=of inv]  (ero)  {erode\\\textit{$r_e$}};
\node[smbox, fill=orange!14, right=of ero]  (blur) {Gaussian\\\textit{$r_b$}};
\node[smbox, fill=green!14,  right=of blur] (soft) {soft mask\\$M{\in}[0,1]$};
\draw[smarrow] (m)    -- (inv);
\draw[smarrow] (inv)  -- (ero);
\draw[smarrow] (ero)  -- (blur);
\draw[smarrow] (blur) -- (soft);
\node[below=4mm of ero, font=\scriptsize, align=center] (legend)
  {\begin{tikzpicture}[baseline]
    \shade[left color=blue!25, right color=orange!35]
      (0,0) rectangle (30mm, 2.6mm);
    \draw[draw=black!40, line width=0.3pt] (0,0) rectangle (30mm, 2.6mm);
    \node[font=\tiny, below right] at (-0.6mm,0)  {$M{=}0$};
    \node[font=\tiny, below left]  at (30.6mm,0) {$M{=}1$};
    \node[font=\tiny, above]       at (15mm,2.6mm)
      {smooth transition across the boundary ring};
   \end{tikzpicture}};
\end{tikzpicture}
\caption{Soft-mask construction. The binary defect mask $m$ is
inverted, eroded by $r_e$ pixels, and Gaussian-blurred with kernel
$r_b$ into a smooth gradient mask $M$. Both radii are derived
adaptively from the defect bbox minimum side $d$ via
\eqref{eq:adaptive_radii} ($r_e{=}0.04\,d$, $r_b{=}0.12\,d$).
Under per-step blending \eqref{eq:soft_mask}, $M{=}0$ preserves
the dome verbatim, $M{=}1$ regenerates the far background, and
the boundary ring is partially regenerated at every denoising
step.}
\label{fig:softmask}
\end{figure}

\paragraph{Per-step blending.} At each diffusion step $t$ the
inpainting pipeline blends the predicted denoised latent with the
noised source latent under the soft mask:
\begin{equation}
z_t = M \cdot \hat{z}_t + (1 - M) \cdot z^{\mathrm{src}}_t ,
\label{eq:soft_mask}
\end{equation}
where $\hat{z}_t$ is the latent predicted by the denoiser at step
$t$ and $z^{\mathrm{src}}_t$ is the source image latent noised to the
same step. Because $M$ is smooth, the boundary ring is partially
regenerated at every step, so the model can render a coherent
transition between the preserved dome and the re-rendered background.
By the time the diffusion loop terminates, the boundary has been
negotiated against the new lighting and colour at every denoising
step, so the hard-boundary seam of binary compositing is strongly
suppressed. The boundary ring
of $M$ is therefore the exact set of pixels where the denoiser has
freedom to negotiate texture continuity. This property follows
directly from \eqref{eq:soft_mask}, and it is what makes the soft mask
preferable to either binary inpainting (hard seam) or Poisson seamless
cloning (post-hoc colour bleed). Figure~\ref{fig:softmask} traces the
construction of $M$ from the binary defect mask.

\subsection{Mask-conditioned generation: the mask as ground truth}
\label{subsec:mask_controlnet}

The soft-mask path of Section~\ref{subsec:soft_mask} preserves the
dome in image space, but it does not inherit the label: because the
boundary ring is regenerated, the ground-truth mask is
\emph{re-predicted} by \textsc{Convex Hull Annotator} (CHA) from the
public SandBoilNet checkpoint (Section~\ref{subsec:dataset}). Label
quality is therefore upper-bounded by that checkpoint's accuracy on
synthetic images. We remove this ceiling with a second, complementary
provenance path in which the label is exact \emph{by construction}.

\paragraph{Method.} We fine-tune a Stable Diffusion XL ControlNet
whose conditioning input is the binary defect mask itself. Following
standard practice for data-scarce ControlNet training, we do not
train from scratch. We initialise from a publicly released
soft-edge (HED) SDXL ControlNet~\cite{zhang2023controlnet}, whose
edge-to-structure prior is the closest match to a filled
silhouette, and fine-tune on the $199$ curated (image, mask) pairs
with the mask replicated to three channels, at $768 \times 768$ in
full precision with gradient checkpointing (the mixed-precision path
in the public SDXL-ControlNet trainer is numerically unstable for this
configuration). At inference the frozen SDXL backbone, the
DreamBooth-LoRA adapter (Section~\ref{subsec:lora}), and this
mask-ControlNet are composed, and an image is generated
\emph{from} a chosen binary mask in pure text-to-image mode (there is
no source photograph and therefore no denoising-strength, depth, HED,
or Canny conditioning: the silhouette is the only structural signal).
We use classifier-free guidance $8.5$, a DPM++ 2M Karras sampler, and
the same fp16-fix VAE as the rest of the pipeline; to suppress
per-sample artefacts we draw a small candidate pool per prompt and
retain the one with the highest no-reference CLIP-IQA
quality~\cite{wang2023exploring}. Because the generator is conditioned
on that mask, the mask \emph{is} the ground-truth label: there is no
post-hoc prediction step and no dependence on an external segmenter
checkpoint.

\begin{figure*}[tp]
\centering
\includegraphics[width=0.98\textwidth]{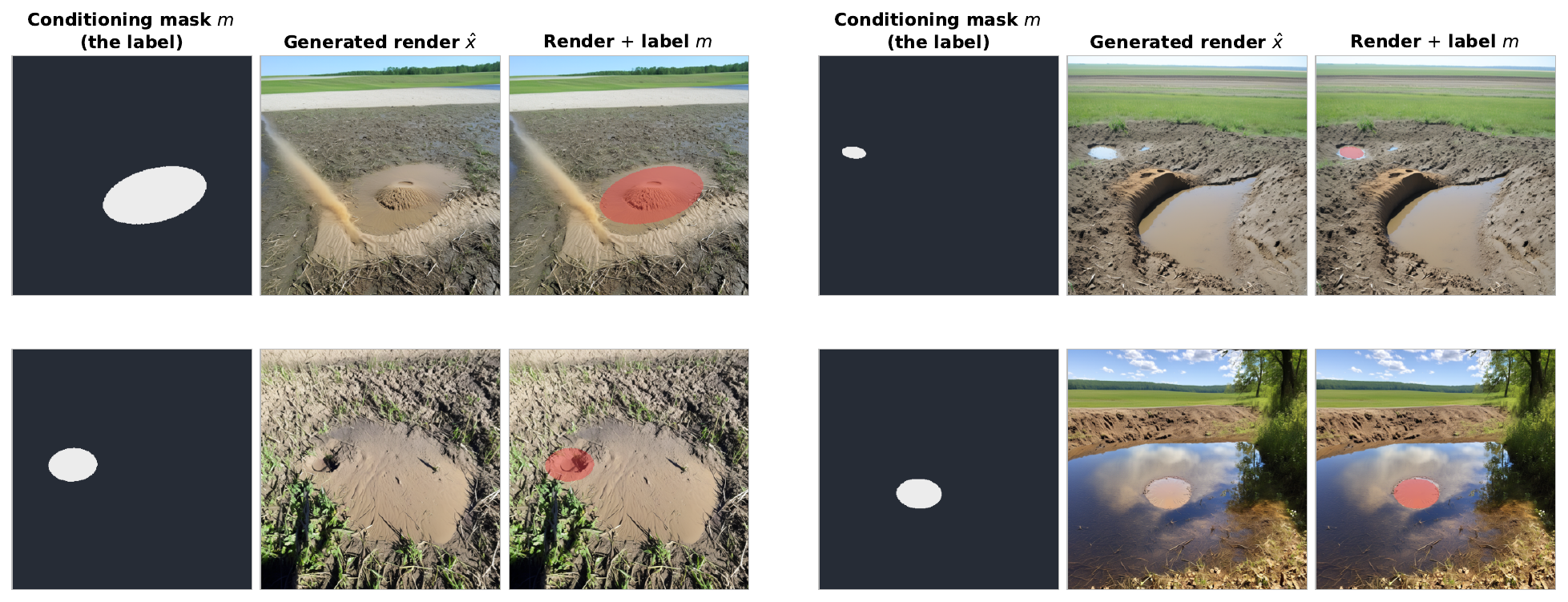}
\caption{Mask-conditioned generation (MaskCN) as a
\emph{label-by-construction} engine: the conditioning mask $m$ is
\emph{exactly} the segmentation label of the image $\hat{x}$ it produces,
so no manual annotation is performed. Each row is a triplet
[conditioning mask $m$ (the label) $\rightarrow$ generated render
$\hat{x}$ $\rightarrow$ render with $m$ overlaid]. Masks are drawn from
the $39$-silhouette bank under per-instance jitter, spanning the
requested-mask size range from small incipient domes to large mature
cones; one mask blob yields one rendered boil. The right column
overlays the conditioning mask on the render to show label registration;
the overlays are shown verbatim, including imperfect cases, and the
quantitative label-drift of these renders under the SandBoilNet-based
gate is analysed separately in Section~\ref{subsec:res_maskcn}. The
distinct sampler (DPM++ 2M Karras) and mask-only conditioning (no
denoising strength, depth, HED, or Canny) distinguish this path from
the image-to-image presets of Section~\ref{subsec:production_presets}.}
\label{fig:maskcn_label}
\end{figure*}

\paragraph{MaskCN conditioning and per-instance diversity.}
Mask-conditioned generation (\emph{MaskCN}) decouples the synthetic
\emph{label geometry} from the catalogue of real source domes, and this
unlocks a second role beyond exact provenance: a controllable
\emph{diversity engine}. The conditioning mask need not come from a real
photograph. Masks are drawn from a bank of $39$ reference silhouettes. The
ControlNet is applied at its default conditioning scale ($1.0$),
yielding one boil per render at a requested size and position. To turn a
fixed bank into a genuine diversity source, each mask is independently
flipped, rotated, rescaled, and repositioned before conditioning
(\emph{per-instance jitter}), so repeated draws of the same source
silhouette yield distinct boil shapes, sizes, and placements rather than
identical copies. The rescale is sampled across a small-to-large size
range that deliberately populates the small-object regime where
pixel-level recall is weakest. Because every perturbation acts on the
conditioning mask itself, the by-construction labelling guarantee is
\emph{exact}: one mask blob reliably yields one rendered boil.
Figure~\ref{fig:maskcn_label} makes this concrete, pairing
each conditioning mask with the render it produces and the mask--render
overlay that shows label registration.

\paragraph{CHA as a verification gate.} This is where CHA's role
shifts rather than disappears. For the diffusion-augmentation presets
(V1--V4), CHA remains the single-policy label producer
(Section~\ref{subsec:dataset}); for the mask-conditioned path it is
repositioned as a \emph{quality-control gate}. CHA runs on the
rendered image, and the intersection-over-union between the
conditioning mask and CHA's prediction measures how faithfully the
generator placed the dome where the mask asked. A sample whose
IoU falls below a threshold is flagged: the rendered dome drifted from
the requested silhouette, so the by-construction label no longer
matches the pixels. The same shape-aware routing (convex hull /
closing / skeleton) that makes CHA a class-agnostic annotator thus
also makes it a class-agnostic verifier, and CHA's standalone library~%
\cite{thapa_chull_annotator} serves both roles unchanged. The gate's
reliability is itself contingent on the SandBoilNet checkpoint
transferring to the synthetic renders; we quantify how tight that
ceiling is in Section~\ref{subsec:res_maskcn}.

\paragraph{Trade-off.} Mask-conditioned generation gives exact,
checkpoint-independent provenance and lets a user draw a mask to obtain a
paired (image, mask) record, at the cost of training one ControlNet per
defect class and curating a bank of conditioning masks. The soft-mask
path, in turn, needs no extra training and inherits the real dome
geometry pixel-for-pixel, at the cost of the CHA re-prediction ceiling.
We release both paths; the soft-mask path with CHA labelling produced the
augmented dataset evaluated in this paper, and the mask-conditioned
ControlNet is released as the checkpoint-independent, diversity-widening
alternative with CHA as its verifier. We are careful to scope MaskCN's
claim. It is a strong \emph{image} and \emph{label-provenance} engine,
but we do not assert a downstream segmentation benefit here; the gate that
would certify its labels at scale is the present bottleneck
(Section~\ref{subsec:res_maskcn}).

\subsection{Prompt Atlas}
\label{subsec:prompt_atlas}

The Prompt Atlas turns a domain into a stratified, validated bank of
text prompts. The builder's input is a JSON domain specification with the following
fields: the \texttt{concept} (here, \textit{sandboil}); the
\texttt{trigger\_token} \texttt{sbx} introduced during DreamBooth
training; a one-sentence \texttt{geometry} description used as system
context when querying a language model; a \texttt{taxonomy} dictionary
mapping axis names to value lists (we use eight axes: geographic
setting, soil type, hydraulic regime, vegetation, season, weather,
lighting, and camera perspective); a list of
\texttt{negative\_concepts} that prompts must not contain (e.g.,
\textit{volcano}, \textit{lava}, \textit{cartoon}); a set of
\texttt{exemplar\_refs} pointing to real images used for image--text
validation; a \texttt{prompt\_template} with placeholders for axis
values; and an explicit list of \texttt{cross\_products} of axes whose
combinations are to be expanded. The same builder accepts any
specification of this shape, so the atlas extends to a new defect
class by writing a new specification file without code changes.

The builder runs five stages.

\paragraph{Stage 1: candidate generation.} In template mode, the
builder enumerates the requested axis cross-products and fills the
template with one value from each axis, drawing the remaining axes
by seeded random sampling. In language-model mode, a small
open-weight model (Qwen2.5-3B~\cite{yang2024qwen2}) is queried once
per axis cell. The query uses a system prompt incorporating the geometry, the
scientific context, and the negative-concept list, and the model returns
three single-sentence variations.

\paragraph{Stage 2: length filter.} Prompts shorter than $12$ words
or longer than $60$ words are discarded.

\paragraph{Stage 3: TF--IDF de-duplication.} Each prompt is
vectorised under a unigram--bigram TF--IDF model and the pairwise
cosine similarity is computed. Whenever a prompt is at or above
$0.85$ cosine similarity with one already retained, it is dropped.

\paragraph{Stage 4: negative-concept screening.} A regular
expression synthesised from the \texttt{negative\_concepts} list is
matched against each prompt; matches are removed.

\paragraph{Stage 5: image--text validation.} Each surviving prompt
is encoded with a CLIP ViT-B/32 text encoder~%
\cite{radford2021clip}; the average image embedding of the
exemplar references is computed with the corresponding CLIP image
encoder. Cosine similarity yields a ranking, and prompts below the
tenth-percentile similarity are dropped.

A final stratified sampling over the cross-product cells yields the
requested number of prompts. The output is a JSON Lines record per
prompt with a stable identifier, the prompt text, the axis bucket,
the CLIP image--text similarity score, and the source flag (template
or language model). The sand boil atlas is built incrementally, and
the production atlas used in this paper is a strict superset of
earlier revisions.

\paragraph{Production atlas statistics.} The v1 sand boil atlas
contains $N\!=\!91$ prompts produced from $102$ template candidates
(target $100$). The per-stage yield is $102 \to 102 \to 102 \to 102
\to 91$: the length, TF--IDF, and negative filters dropped no
candidates, and the CLIP image--text validator removed the bottom
$11$. The CLIP similarity to the exemplar embedding is
$\mu\!=\!0.295 \pm 0.015$ (range $[0.269, 0.331]$); the
tenth-percentile threshold is $0.276$. Prompts average $43.3$ words
(min $31$, max $55$, within the $12$--$60$ filter). Lexical diversity
is low at the unigram level (distinct-$1 = 0.041$) and higher at
the bigram level (distinct-$2 = 0.095$), reflecting a templated style
with substantial axis-driven variation. Of the $44$ taxonomy values
declared across the eight axes, $40$ ($91\%$) appear in at least one
rendered prompt. The four missing values all belong to the
\texttt{season} axis. As a known limitation, the v1 prompt template references seven of the
eight axes and omits the season slot; season values are therefore sampled and recorded in the \texttt{fill} field but
do not reach the rendered prompt text.

\paragraph{Example prompts.} Two representative atlas entries make
the rendered style concrete:
\begin{quote}\footnotesize\itshape
``photo of sbx sandboil at the landside toe of an earthen levee,
saturated foundation sand, steady low-velocity discharge, wetland
reeds and rushes, drizzle with wet ground reflections, raking side
light highlighting the dome relief, wide context shot showing the
levee crest in the background, color-graded raw camera output.''
\end{quote}
\begin{quote}\footnotesize\itshape
``photo of sbx sandboil along a riverbank during receding flood,
saturated foundation sand, intermittent flow with sediment fan
deposits, sparse short grass, foggy early morning, raking side
light highlighting the dome relief, oblique angle inspector survey,
professional documentary photography.''
\end{quote}

\begin{figure*}[tp]
\centering
\includegraphics[width=\textwidth]{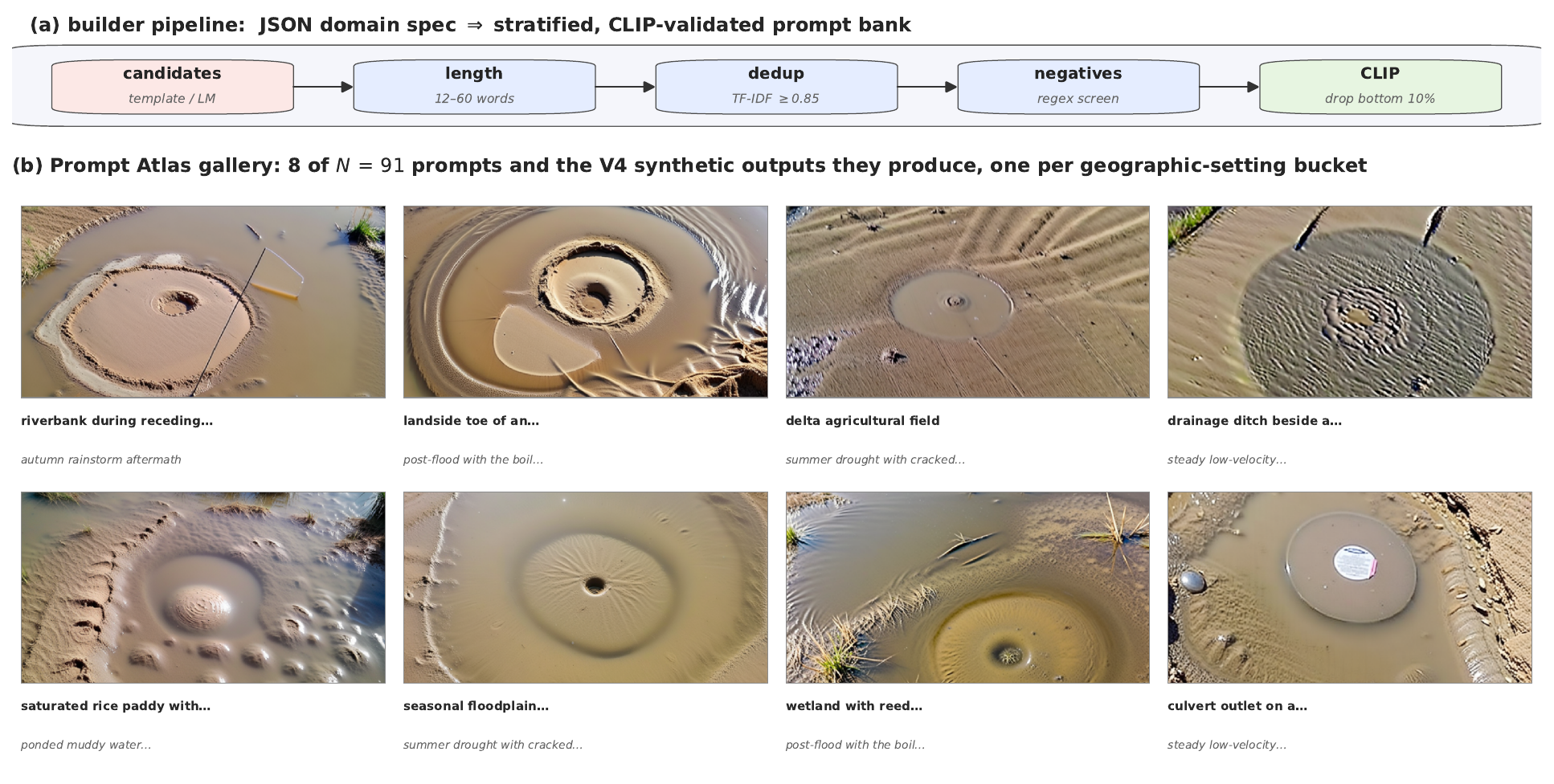}
\caption{The sand boil v1 Prompt Atlas, presented as a visual
catalogue. \textbf{(a)}~\emph{Builder pipeline.} A JSON domain
specification is expanded into candidate prompts (by template
substitution or a small open-weight language model) and then passed
through a length filter, a sparse TF--IDF de-duplication, a
negative-concept regex screen, and a CLIP image--text validation
against the exemplar embedding. The builder is reproducible from
the specification alone and ports to a new defect class by writing
a new specification file with no code change.
\textbf{(b)}~\emph{Atlas gallery.} Eight of the $N\!=\!91$
production prompts paired with the V4 synthetic outputs they
generate (one image per prompt, all from the same generator and
identical scheduler/CFG), one per geographic-setting bucket in the
taxonomy. Each caption gives the
prompt's primary axis value (bold) and one secondary axis value
(italic) so the reader can map an output back to the stratification
cell that produced it. The visual variety is the operational meaning
of the atlas: a fixed defect geometry rendered under controlled
scene variation. The per-prompt CLIP score distribution
($\mu\!=\!0.295 \pm 0.015$, $p_{10}\!=\!0.276$, yield $102\!\to\!91$)
is reported in the text.}
\label{fig:atlas}
\end{figure*}

\noindent Figure~\ref{fig:atlas} summarises the builder and these
statistics.

\subsection{Production presets}
\label{subsec:production_presets}

The five-row ablation of Section~\ref{subsec:gen_ablation} traces
individual design axes (each control branch, the LoRA, the inpainting
path, the soft-edge branch) under fixed seeds and sources. A smaller,
four-cell production-preset comparison is also useful as the practical
distillation of those axes. Each preset bundles a complete recipe
(control set, IP-Adapter use, denoising path) and is the unit a
downstream operator actually picks. All four share the SDXL backbone,
the DreamBooth-LoRA adapter, $45$ diffusion steps at
$1024 \times 1024$, and the conditioning-scale schedule
$(\lambda_c, \lambda_d, \lambda_h, \lambda_n) =
(0.40, 0.25, 0.40, 0.40)$ on whichever branches each preset
activates. IP-Adapter is loaded with scale $0.55$, anchoring style to
a real reference photograph rotated through the source pool.
Classifier-free guidance is $8.5$ for the image-to-image presets and
is per-image jittered for the soft-mask inpainting preset
($\mathrm{CFG} \in [7.5, 9.5]$); strength is jittered for the
inpainting preset in $[0.85, 0.99]$.

The four presets span a single design axis that trades geometric
constraint for style supervision. Two presets load three structural
controls each (Canny plus two of Depth/HED/Normal) but no IP-Adapter,
putting the burden of texture entirely on the denoiser's residual
capacity after the controls have been satisfied. One preset drops
Canny and Depth, retains HED and Normal, and adds an IP-Adapter style
anchor; the anchor injects high-frequency texture (granular sand,
wet-dry transitions, photographic noise) that a structural control
cannot represent. The fourth preset carries the heaviest conditioning
stack (HED, Normal, IP-Adapter, plus the soft-mask inpainting protocol
of Section~\ref{subsec:soft_mask}). This preset preserves the dome
geometry through the diffusion loop, eliminates the Poisson
seam-and-tint failure modes of the prior pipeline~%
\cite{thapa2025thesis,perez2003poisson}, and exposes the IP-Adapter
to a coherent unmasked context for style transfer. The inpainting
preset is selected when geometric fidelity to a specific source-mask
dome is paramount; the style-anchored image-to-image preset is
selected when style-rich sand boil texture without inpainting is
preferred. Without intervention, the inpainting preset's heavy conditioning would
make every output a close variant of the same recipe. To prevent that,
the deployed pipeline
injects seeded per-image jitter on the HED and Normal scales, the
IP-Adapter scale, the soft-mask erode kernel and Gaussian blur kernel,
the sand boil attention boost, the CFG, and the strength. The same
seed-derived RNG drives every jittered axis, so a given (source, seed,
prompt) triple is fully reproducible.

\subsection{Reference-set similarity quality filtering}
\label{subsec:quality_filter}

DreamBooth fine-tuning on a small reference set can fail in two
distinguishable directions. It can drift \emph{off} the target domain
when the prompt pushes the denoiser into a region of the prior the
adapter did not anchor, yielding out-of-distribution outputs. It can
also drift \emph{onto} a memorised training frame when the prompt is
close to one the adapter saw during training, yielding near-duplicates
of the reference set. Both failures degrade downstream training: OOD
samples teach the segmenter spurious correlations, and memorised
samples inflate the effective duplication of a handful of real frames.
The filter introduced here guards against both modes with no human
review.

\paragraph{The filter.} Let
$\Phi: \mathbb{R}^{H \times W \times 3} \to \mathbb{R}^{D}$ denote
the CLIP~\cite{radford2021clip} ViT-B/32 image encoder normalised
to unit length. Compute the centroid embedding of the reference
set,
$\bar{\phi}^{\mathrm{ref}} = \mathrm{normalize}\!\left(
\frac{1}{N_{\mathrm{ref}}} \sum_{x \in \mathcal{R}} \Phi(x)\right)$,
once at pipeline-initialisation time. For each synthetic image
$\hat{x}$, the quality score is the cosine similarity to the
centroid:
\begin{equation}
s(\hat{x}) = \Phi(\hat{x})^{\top}\, \bar{\phi}^{\mathrm{ref}}.
\label{eq:quality_score}
\end{equation}
Samples whose score falls outside the admissibility band
$[s_{\mathrm{lo}}, s_{\mathrm{hi}}]$ are excluded
from the augmented dataset.

\paragraph{Auto-tuned thresholds.} Rather than hand-pick the band, we
compute the per-image leave-one-out CLIP similarity on the reference
set itself: for each $x_k \in \mathcal{R}$,
$s_k = \Phi(x_k)^{\top}
\mathrm{normalize}\!\left(\sum_{j \neq k}\Phi(x_j)/(N_{\mathrm{ref}}-1)\right)$.
The empirical distribution $\{s_k\}$ defines what
``in-distribution'' looks like at the reference-set scale. We set
\begin{equation}
s_{\mathrm{lo}} = \mathrm{median}(\{s_k\}) - 2.5\,\sigma(\{s_k\}),
\quad
s_{\mathrm{hi}} = \max(\{s_k\}) + \epsilon,
\label{eq:band_thresholds}
\end{equation}
with $\sigma$ the median absolute deviation and $\epsilon$ a small
positive offset (we use $0.01$). The lower threshold rejects synthetic
samples that are systematically less reference-like than the most
peripheral real sample. The upper threshold rejects synthetic samples
that are \emph{more} reference-like than any real sample, which under
the small-reference-set regime is the signature of memorisation rather
than legitimate within-distribution generation.

\paragraph{Properties.} The filter is automated (no labelled data, no
manual scoring), reproducible (the band is a deterministic function of
the reference set and a fixed encoder), and content-aware (CLIP's
vision tower compresses both texture and semantic context, so a
sand-boil-styled volcano scene and a near-duplicate sand boil are
rejected for different but principled reasons). The computation is
approximately $4$~ms per sample on a single GPU. The same filter
transfers to any new defect class through its own reference centroid.

\paragraph{Per-bucket diagnostic.} A single global floor implicitly
assumes every scene type sits the same distance from the reference
centroid. To test that assumption we stratify the synthetic set by its
Prompt-Atlas geographic-setting bucket and recompute, within each
bucket, a bucket-local floor
$s_{\mathrm{lo}}^{(b)} = \mathrm{median}(\{s\}_b) - 2.5\,\sigma(\{s\}_b)$
from that bucket's own scores. The global floor
($s_{\mathrm{lo}}=0.839$, set by the tight reference leave-one-out
distribution) proves \emph{stricter} than all bucket-local floors
(which span $0.806$--$0.834$ across buckets): it rejects $205$ of the $1{,}020$
synthetic images, against $121$ when each bucket is judged by its own
floor. The rejections are also far from uniform; they concentrate in
scene types that sit systematically lower in centroid similarity (most
of all the riverbank and delta-field buckets), exactly the cases a
single threshold cannot tell apart from genuine out-of-distribution
drift. We therefore report the per-bucket floors as a diagnostic of
\emph{where} the global filter is conservative, and keep the global
floor in production because it errs toward dropping borderline
samples.

\subsection{Dataset assembly with source tagging}
\label{subsec:dataset_assembly}

Each synthetic image is annotated with a pointer to the real source
image from which it was derived, together with the prompt, seed, and
generator configuration that produced it. The synthetic ground-truth
mask is produced uniformly across all production presets by running
the synthetic image through the \textsc{Convex Hull Annotator}
pipeline of Section~\ref{subsec:dataset}: the public SandBoilNet
checkpoint emits a per-pixel probability map, the
threshold-label-hull cleanup of~%
\cite{thapa2025thesis,thapa_chull_annotator} produces a binary mask,
and that mask becomes the label for the synthetic record. Source
tagging is preserved as an artefact-level discipline: every synthetic
record carries the \texttt{source\_image\_id} of the real parent from
which it was derived, even though the leak-free fold-filtering
machinery that consumes those tags is left to future work.

\subsection{Integration: what the contributions together enable}
\label{subsec:integration}

Three mutually reinforcing claims hold once the
pipeline is assembled end-to-end.

\paragraph{Single-policy mask provenance.} Two coupled mechanisms
guarantee that every synthetic image in the augmented set has a mask
whose origin we can name. Defect-preserving soft-mask latent blending
(Section~\ref{subsec:soft_mask}) ensures the defect \emph{image
pixels} are inherited verbatim from the source.
\textsc{Convex Hull Annotator} (Section~\ref{subsec:dataset})
produces every \emph{mask label} from a single, reproducible
post-processing pipeline applied uniformly across all production
presets. The two mechanisms address different sides of the same
problem (image-side preservation and label-side production) and
together yield an augmented dataset with one mask-production policy
and no per-preset special-casing. The mask-conditioned ControlNet of
Section~\ref{subsec:mask_controlnet} offers a third, stricter
provenance option in which the label is exact by construction and CHA
becomes a verifier; we release it as a checkpoint-independent
alternative to the soft-mask$+$CHA path used for the dataset here.

\paragraph{Automatic quality envelopes at both ends of the
training pipeline.} Defect-size-adaptive soft-mask geometry
(Section~\ref{subsec:soft_mask}) is an \emph{input-side} quality
envelope: it ensures the preservation margin and transition smoothness
are well-scaled to each individual defect rather than tuned to an
average dome. Reference-set similarity quality filtering
(Section~\ref{subsec:quality_filter}) is the \emph{output-side}
envelope: it removes samples that fall outside the in-distribution
band defined by the reference set's leave-one-out similarity
statistics. Both envelopes are automated, both are content-aware, and
both add zero human-annotation effort to the per-defect-class
extension cost.

\paragraph{Portability without code change.} The Prompt Atlas
(Section~\ref{subsec:prompt_atlas}) reduces the per-class textual
specification cost to writing a JSON file. CHA depends only on a
class-specific segmenter checkpoint, the adaptive soft-mask geometry
depends only on the binary mask geometry, and the quality filter
depends only on a reference-set centroid in CLIP space. None of these
depends on the sand boil class itself. Re-instantiating the pipeline
on a seepage, sinkhole, or rutting domain therefore requires
one new domain specification, one new DreamBooth adapter on a small
new reference set, and one new segmenter checkpoint, and nothing
else changes.

These integration properties make the augmented dataset
\emph{admissible as supervision} for a downstream cross-validated stacking
ensemble, and they are the methodological
return on the components developed above.

%% file: sections/experiments.tex
\section{Experiments}
\label{sec:experiments}

This section describes the experimental protocol. It proceeds in
three phases: a recap of the DreamBooth-adapter configuration, a
generator-pipeline ablation, and a production-preset image-quality
comparison against the real reference set.

\subsection{Hardware and software}
\label{subsec:setup}

All experiments run on a single NVIDIA H100 NVL GPU ($95$~GB HBM).
Diffusion-based image generation uses an fp16-mixed-precision
PyTorch stack with the standard Diffusers inference pipelines. Two
decoder settings are critical for image quality. First, the stock
SDXL variational autoencoder produces blotch and colour artefacts when
run in fp16 (a well-documented SDXL issue); we therefore load the
fp16-corrected VAE (\texttt{sdxl-vae-fp16-fix}~\cite{ollin2023vaefix}),
which decodes without the fp16 colour artefacts. Second,
because the $95$~GB HBM removes any memory pressure, we \emph{disable}
VAE tiling and slicing: those split the decode into tiles and leave
visible grid seams (a ``torn'' look) at $1024^2$. The Prompt Atlas
builder runs on CPU only and builds
the sand-boil prompt domain in under twenty seconds. The downstream
segmentation work uses a PyTorch stack
(\textsc{segmentation\_models.pytorch}).

\subsection{DreamBooth adapter configuration}
\label{subsec:lora_ablation}

The adapter used by the production generator is a DreamBooth-LoRA
module trained for
$1{,}500$ steps at rank $16$ and learning rate $1\mathrm{e}{-4}$
under a cosine schedule. Training uses class-image prior
preservation ($200$ auto-generated images for the prompt
\textit{a photo of a sandboil}) on the curated reference set, with
the trigger phrase \texttt{sbx sandboil}. The DreamBooth-LoRA recipe of
Section~\ref{subsec:lora} otherwise applies unchanged.
Section~\ref{subsec:backbone_choice} justifies restricting the
configuration to the SDXL backbone.

\subsection{Generator-pipeline ablation}
\label{subsec:gen_ablation}

To isolate the contribution of each generator component, we run a
five-row ablation, with each row evaluated on the same five source
images, prompts, and seeds.
Table~\ref{tab:gen_ablation} summarises the ablation grid.

\begin{table*}[t]
\caption{Generator-pipeline ablation matrix. Each row produces
five images on the same sources, prompts, and seeds. CN columns
mark the ControlNets active in this ablation (Canny, Depth, HED
soft-edge; the surface-normal branch is exercised only in the
production presets); LoRA marks
whether the DreamBooth adapter is loaded; Path is either img2img
(whole image) or soft-mask inpainting (preserves the dome core).}
\label{tab:gen_ablation}
\centering
\small
\begin{tabular}{@{}cccccll@{}}
\toprule
ID & LoRA & Canny CN & Depth CN & HED soft-edge CN & Path & Notes \\
\midrule
A & \textendash{} & \textendash{} & \textendash{} & \textendash{} & img2img & SDXL baseline (no conditioning) \\
B & \textendash{} & \checkmark & \checkmark & \textendash{} & img2img & Canny + Depth \\
C & \checkmark & \checkmark & \checkmark & \textendash{} & img2img & + LoRA \\
D & \checkmark & \checkmark & \checkmark & \checkmark & img2img & + HED soft-edge \\
E & \checkmark & \checkmark & \checkmark & \checkmark & soft-mask inpaint & + soft-mask inpaint \\
\bottomrule
\end{tabular}
\end{table*}

The last two rows answer two targeted questions. Row D tests
whether adding a HED soft-edge ControlNet on top of Canny and depth
improves visual quality. Row E tests whether the soft-mask
inpainting path yields better seam behaviour than the plain
image-to-image path of row D. The new rows share identical
settings: $45$ diffusion steps, classifier-free guidance scale
$8.5$, denoising strength $0.85$, the DPM-Solver++ (order~3) scheduler~%
\cite{lu2022dpmsolver,lu2022dpmsolverpp}, $1024 \times 1024$ output resolution, and
a full negative prompt that suppresses \textit{volcano},
\textit{lava}, \textit{cartoon}, and other unrelated concepts. We
chose the denoising strength of $0.85$ after a coarse sweep on
five sources. Lower values left the output too similar to the
input, which reduced background diversity; higher values began to
perturb the dome geometry on the non-inpaint rows. The ControlNet
conditioning scales active in this ablation are
$(\lambda_c, \lambda_d, \lambda_h) = (0.40, 0.25, 0.40)$ for Canny,
depth, and HED soft-edge respectively. These are the first three
entries of the full four-way schedule
$(\lambda_c, \lambda_d, \lambda_h, \lambda_n) = (0.40, 0.25, 0.40, 0.40)$;
the surface-normal branch $\lambda_n$ is inactive here. The soft mask in row E uses the fixed $12$-pixel
erosion followed by a $41$-pixel Gaussian blur for this ablation;
the deployed V4 preset instead uses
the adaptive radii of Section~\ref{subsec:soft_mask}. The five source images span varied
environments (rice paddy, levee toe, riverbank, foggy field,
ponded depression). The five prompts are drawn from the
Prompt Atlas (Section~\ref{subsec:prompt_atlas}) in a stratified
manner, so that each source maps to a different geographic-setting cell.

Table~\ref{tab:gen_hparams} consolidates the generation
hyperparameters for reproducibility. The fixed settings are shared
across presets; the two parameters with the largest effect on the
output are the image-to-image denoising strength and the ControlNet
conditioning scale, which trade background diversity against fidelity
to the parent's boil. The diversity-oriented production setting
($\text{strength}=0.85$, scale $0.40$) maximises background variety,
and the selected V4 preset jitters the strength in $[0.85, 0.99]$
around it (Section~\ref{subsec:production_presets}). Because the
synthetic ground-truth mask is produced by the
\textsc{Convex Hull Annotator} from the rendered image itself
(Section~\ref{subsec:dataset}), the label tracks the boil as actually
rendered; out-of-distribution and memorised samples are then removed by
the leave-one-out CLIP admissibility filter
(Section~\ref{subsec:quality_filter}).

\begin{table}[t]
\caption{Generation hyperparameters for the sand-boil synthesiser.
Fixed settings are held across presets. The denoising strength and
ControlNet conditioning scale are the two parameters with the largest
effect on the diversity/fidelity trade-off; the selected V4 preset
jitters the strength in $[0.85, 0.99]$.}
\label{tab:gen_hparams}
\centering
\small
\resizebox{\columnwidth}{!}{%
\begin{tabular}{@{}ll@{}}
\toprule
Parameter & Value \\
\midrule
Backbone & SDXL base 1.0 \\
Domain adapter & DreamBooth-LoRA (rank $16$, $1{,}500$ steps) \\
Adapter weight / trigger & $1.0$ / \texttt{sbx sandboil} \\
ControlNet stack & Canny $+$ Depth $+$ Normal $+$ HED soft-edge \\
\quad preset V1 & Canny $+$ Depth $+$ HED soft-edge \\
\quad preset V2 & Canny $+$ Normal $+$ HED soft-edge \\
Scheduler & DPM-Solver++ (order $3$, Karras) \\
Inference steps & $45$ \\
Guidance scale (CFG) & $8.5$ \\
Resolution & $1024 \times 1024$ \\
Random seed & deterministic per image \\
Prompts & stratified Prompt Atlas \\
\midrule
img2img denoising strength & $0.85$ (V4: $[0.85, 0.99]$) \\
ControlNet conditioning scale & $0.40$ \\
\bottomrule
\end{tabular}%
}
\end{table}

\subsection{Production-preset comparison}
\label{subsec:preset_eval}

The deployed pipeline exposes four production presets, summarised
in Table~\ref{tab:production_presets}. For all four presets, the synthetic ground-truth
mask is produced uniformly post hoc by the
\textsc{Convex Hull Annotator} pipeline of
Section~\ref{subsec:dataset}.

\begin{table*}[t]
\caption{Production presets exposed by the deployed pipeline. All
share the SDXL backbone, the DreamBooth-LoRA adapter, $45$
diffusion steps at $1024 \times 1024$, and conditioning-scale
schedule $(\lambda_c, \lambda_d, \lambda_h, \lambda_n) =
(0.40, 0.25, 0.40, 0.40)$ on whichever branches each preset
activates. ``IPA'' is IP-Adapter~\cite{ye2023ipadapter} at scale
$0.55$ with a real reference rotated through the source pool.
CFG is $8.5$ for V1--V3; V4 jitters CFG in $[7.5, 9.5]$ and
strength in $[0.85, 0.99]$.}
\label{tab:production_presets}
\centering
\small
\begin{tabular}{@{}clccc@{}}
\toprule
ID & Control set & IPA & Path & Notes \\
\midrule
V1 & Canny + Depth + HED soft-edge & \textendash{} & img2img            & Three structural controls (Depth), no style anchor \\
V2 & Canny + Normal + HED soft-edge & \textendash{} & img2img            & Three structural controls (Normal), no style anchor \\
V3 & HED soft-edge + Normal        & \checkmark & img2img            & Style-anchored image-to-image \\
\textbf{V4} & \textbf{HED soft-edge + Normal} & \textbf{\checkmark} & \textbf{soft-mask inpaint} & Style-anchored soft-mask inpainting (selected) \\
\bottomrule
\end{tabular}
\end{table*}

Each preset is evaluated on a fixed batch of $182$ renders (the
per-preset evaluation volume reported in Table~\ref{tab:cost}). We evaluate each preset's samples
under the protocol of Section~\ref{subsec:quality_eval}: Fr\'echet
Inception Distance (FID) and Kernel Inception Distance (KID) against the
real reference set on Inception-v3
features, CLIP score against each sample's own Prompt Atlas prompt,
and Learned Perceptual Image Patch Similarity (LPIPS) diversity within
the per-preset set. The seed cascade,
source rotation, and Prompt Atlas stratification match those of
Section~\ref{subsec:aug_dataset}.

\paragraph{Mask-conditioned generation (MaskCN).} Alongside the four
img2img presets we evaluate the mask-conditioned ControlNet of
Section~\ref{subsec:mask_controlnet} under the same image-quality
protocol. MaskCN is a text-to-image path with mask-only structural
conditioning: it uses no source photograph and therefore has no
denoising-strength, depth, HED, or Canny inputs. It shares the SDXL
backbone, the DreamBooth-LoRA adapter, classifier-free guidance $8.5$,
and the fp16-corrected VAE with the other presets, but swaps the scheduler for
DPM++ 2M Karras. Masks are drawn from a $39$-silhouette bank, yielding
one boil per render; per-instance jitter (flip, rotation, rescale,
reposition) spans boil shape, size, and placement across draws. The
diversity singles folded into the augmentation mixture use the default
conditioning scale ($1.0$), whereas the MaskCN image-quality numbers in
Table~\ref{tab:production_preset_quality} are measured at the best
evaluation configuration selected by the sweep in
Section~\ref{subsec:res_maskcn} (conditioning scale $1.8$, $45$ steps,
DPM++ 2M Karras, with best-of-three retention). This diversity
contribution is also evaluated qualitatively in
Section~\ref{subsec:res_qualitative}; the label-verification gate is
examined separately in Section~\ref{subsec:res_maskcn}.

\subsection{Augmented dataset generation}
\label{subsec:aug_dataset}

The augmented training set is produced by the V4 production preset
(Table~\ref{tab:production_presets}) at $1024 \times 1024$ resolution,
with approximately five synthetic variations per real training image. The
production run yields $1{,}020$ synthetic candidates in total; the
leave-one-out CLIP admissibility filter
(Section~\ref{subsec:quality_filter}) then admits $815$ of them into the
augmented dataset and rejects $205$ as out-of-distribution. Seeds are
assigned deterministically, and prompts are drawn round-robin from
the Prompt Atlas so that the synthetic set is uniformly stratified
across the taxonomy axes. Section~\ref{subsec:reproducibility}
describes the deterministic seed cascade and the full parameter
manifest.

\subsection{Synthetic-image quality evaluation}
\label{subsec:quality_eval}

The image-quality evaluation protocol assesses the augmented
dataset against the full real evaluation set ($N_{\mathrm{eval}}\!=\!199$
images) on two families of measures. We use the full $199$-image set
rather than the smaller curated $39$-image reference because the latter
leaves the $2048$-dimensional Inception covariance rank-deficient and
inflates FID, a known small-sample bias~\cite{chong2020fid}. These $199$ images are the full real
training pool, which also serves as the source set for the
image-to-image presets; FID against it is therefore an
in-distribution reference, and we rely on KID and the
precision/recall decomposition to separate fidelity from diversity.
Because the soft-mask path preserves real dome pixels, its synthetic
images share pixel content with this reference set, which biases its
distribution distances downward and makes them not directly comparable
across families to MaskCN (which uses no source photograph) or the
Poisson baseline; we therefore read these absolute distances within
rather than across the generation families, and rest the selection
argument on the fidelity/diversity decomposition and on label
provenance. A disjoint held-out real set is reserved for downstream
evaluation and is not used for these distribution metrics. The first is distribution distance: FID~%
\cite{heusel2017fid} and KID~\cite{binkowski2018kid} (mean and standard
deviation) on Inception-v3~\cite{szegedy2016inception} features. We report FID with two
implementations (standard \mbox{torchmetrics} and
\mbox{clean-fid}~\cite{parmar2022cleanfid}, which removes resize and
quantisation aliasing) and treat the unbiased KID as the primary
distribution metric at this sample size. We further report CLIP score~%
\cite{radford2021clip,hessel2021clipscore} on the OpenAI CLIP ViT-B/32
model and LPIPS diversity~\cite{zhang2018lpips} (AlexNet variant, at
most $16$ randomly sampled images for tractability). The second is a
fidelity--diversity decomposition that a single FID number cannot
provide: improved precision and recall~\cite{kynkaanniemi2019prec}
and density and coverage~\cite{naeem2020reliable}, computed on the
same $2048$-dimensional Inception features using $k$-nearest-neighbour
manifold radii ($k{=}5$). Precision and density measure how much
synthetic mass falls inside the real manifold (fidelity); recall and
coverage measure how much of the real manifold the synthetic set
spans (diversity). We compute the distribution measures for the
full augmented set and for each production preset, and the
precision/recall/density/coverage decomposition for each production
preset, MaskCN, and the Poisson baseline. The generator ablation
(rows A--E) is instead assessed qualitatively
(Section~\ref{subsec:res_ablation}). Together these separate ``looks
like the references'' from ``covers the references'', which is the
crux of the preset trade-off in Section~\ref{sec:results}. Each
configuration is scored on a single deterministic-seed batch, so KID is
the only measure we report with a spread; differences of a few FID
points or a few thousandths of CLIP or LPIPS should be read as
indicative rather than statistically separated.

\subsection{Reproducibility}
\label{subsec:reproducibility}

Every reported number is reproducible. Each numerical entry in
Section~\ref{sec:results} is
the deterministic product of three identifiers: the trained-adapter
checkpoint, the Prompt Atlas record, and the per-image seed. Each
synthetic record carries a JSON sidecar with the source image
identifier, prompt identifier, seed, ControlNet conditioning
scales, and the production preset. The released code emits a
manifest of every adapter, prompt-bank revision, and seed cascade
used to produce a given augmented set. From the released code and
this manifest, a reader can re-derive every reported number without
manual reconfiguration.

%% file: sections/results.tex
\section{Results and Analysis}
\label{sec:results}

This section reports the generator-pipeline ablation; the
production-preset image-quality comparison; the mask-conditioned
ControlNet (MaskCN) results; a curated two-engine diversity gallery and
source-paired qualitative outputs; a fidelity--diversity decomposition
and a Poisson seamless-cloning baseline; the empirical behaviour of the
CLIP admissibility filter and the memorisation audit; and the wall-clock
cost on a single NVIDIA H100 NVL GPU.

\subsection{Generator-pipeline ablation}
\label{subsec:res_ablation}

\begin{figure*}[tp]
\centering
\includegraphics[width=0.86\textwidth]{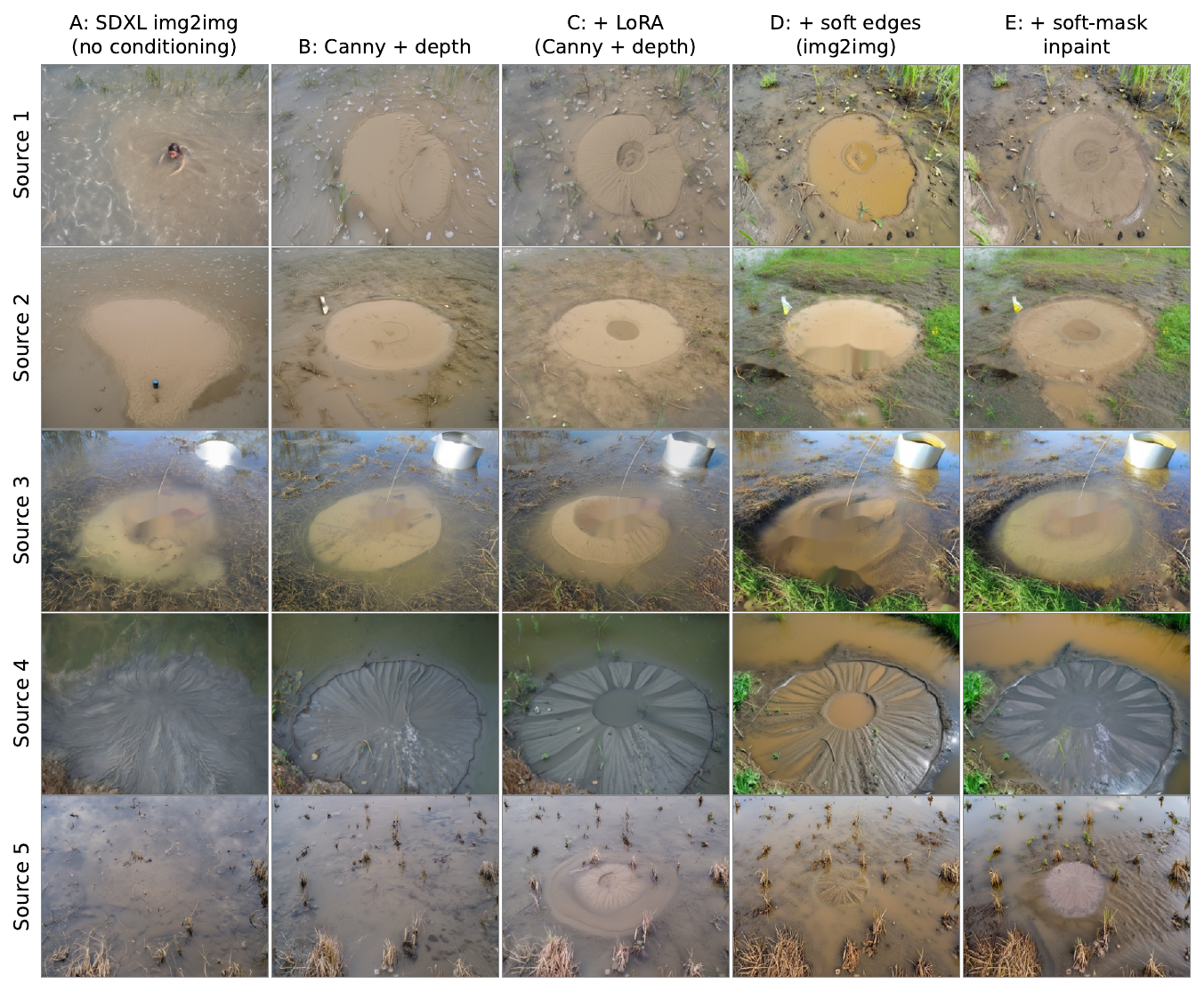}
\caption{Qualitative outputs of the five generator-pipeline
ablation columns on the same five sources, prompts, and seeds.
Columns: \textbf{A} SDXL img2img (no conditioning); \textbf{B}
Canny + depth; \textbf{C} Canny + depth with the
DreamBooth-LoRA adapter; \textbf{D} \textbf{C} plus a soft-edge
(HED) ControlNet; \textbf{E} the soft-mask inpainting path of
Section~\ref{subsec:soft_mask}. From A to E, the dome becomes
progressively better localised and the rim texture sharper; column~E
produces the seam-free, dome-preserving outputs selected for the
augmented dataset.}
\label{fig:gen_ablation_grid}
\end{figure*}

To isolate how each conditioning stage shapes output quality,
Figure~\ref{fig:gen_ablation_grid} shows the qualitative
outputs of the five ablation columns on the same five source images. We
focus on two observations, concerning columns D and E.

\textbf{Column D (img2img with soft edges).} Adding the soft-edge
branch on top of Canny and depth recovers fine-scale rim structure
that the binary Canny map cannot resolve, in particular the
wet--dry transition and the granular sand texture. On sources whose
rim is the most pronounced, the additional branch eliminates the
painterly artefact present in column C. The cost is one extra
ControlNet forward pass per diffusion step, which raises VRAM by
approximately $1.5$~GB and per-step latency by roughly $12\%$.

\textbf{Column E (soft-mask inpainting).} The soft-mask path retains
the source sand-boil region pixel-for-pixel and re-renders only the
surrounding scene. Column D denoises every pixel and may slightly
perturb the dome geometry. In contrast, column E preserves the dome
exactly and produces a seam-free, colour-consistent boundary,
eliminating the two failure modes of the Poisson seamless-cloning
step used in our previous work~\cite{thapa2025thesis}. The dome is
preserved pixel-for-pixel in image space, but the synthetic
ground-truth label is not inherited from the source mask. As
discussed in Section~\ref{subsec:dataset}, the soft-edge blending
region permits a small amount of rim drift, so the label is
re-predicted by the \textsc{Convex Hull Annotator} pipeline on
the rendered synthetic image. This keeps mask provenance uniform
across all production presets.

\subsection{Synthetic-image quality across production presets}
\label{subsec:res_quality}

The image-quality measures, applied to the four image-to-image
production presets of Section~\ref{subsec:production_presets} and to
the mask-conditioned ControlNet of
Section~\ref{subsec:mask_controlnet}, isolate the trade-off between
geometric over-conditioning and style-anchored texture. We read this
table as a \emph{menu of complementary engines}, not a ranking: no
single preset dominates, and minimum distribution distance is
deliberately not the selection criterion. For data destined to augment
a segmenter, label reliability and scene diversity matter more than
proximity to the small reference set---indeed the lowest-FID options
(V1 here, and the degenerate Poisson baseline of
Section~\ref{subsec:res_poisson}) hug the reference distribution rather
than expand it. Accordingly, the released augmented set is generated by
the single \emph{label-reliable} preset (V4,
Section~\ref{subsec:aug_dataset}). A curated \emph{mixture} across
these complementary engines---V1 for manifold coverage, V3/MaskCN for
diversity, V4 for real-structure labels---is the natural augmentation
strategy, to be evaluated in future work. Each preset is
evaluated at $1024 \times 1024$ against the full real reference set
($N_{\mathrm{eval}}\!=\!199$); results are reported in
Table~\ref{tab:production_preset_quality}. We compute FID with two
independent implementations (standard \mbox{torchmetrics} and
\mbox{clean-fid}~\cite{parmar2022cleanfid}), which agree throughout, and
treat the unbiased KID as the primary distribution-match metric at this
sample size.

\begin{table}[t]
\caption{Synthetic-image quality for the four image-to-image production
presets and the mask-conditioned ControlNet (MaskCN), evaluated at
$1024 \times 1024$ against the full real reference set
($N_{\mathrm{eval}}\!=\!199$). FID is reported under two implementations
(\mbox{torchmetrics} at feature dimension $2048$, and
\mbox{clean-fid}~\cite{parmar2022cleanfid}), which agree within three
points. Lower is better for FID/KID, higher for CLIP image--text
similarity and LPIPS within-set diversity; the best value per column is
in bold. V4~($\dagger$) is the preset selected for the augmented dataset
(Section~\ref{subsec:aug_dataset}).}
\label{tab:production_preset_quality}
\centering
\footnotesize
\setlength{\tabcolsep}{3pt}
\begin{tabular}{@{}lccccc@{}}
\toprule
Preset & FID $\downarrow$ & clean-fid $\downarrow$ & KID $\downarrow$ & CLIP $\uparrow$ & LPIPS $\uparrow$ \\
\midrule
V1 & \textbf{169.1} & \textbf{169.4} & \textbf{0.048\,\scriptsize{$\pm$0.008}\normalsize} & 0.302 & 0.595 \\
V2 & 198.7 & 197.0 & 0.073\,\scriptsize{$\pm$0.007}\normalsize & 0.295 & 0.588 \\
V3 & 193.9 & 192.8 & 0.074\,\scriptsize{$\pm$0.007}\normalsize & 0.311 & 0.644 \\
V4$^{\dagger}$ & 201.8 & 201.2 & 0.091\,\scriptsize{$\pm$0.010}\normalsize & 0.302 & 0.633 \\
MaskCN & 189.4 & 190.1 & \textbf{0.048\,\scriptsize{$\pm$0.006}\normalsize} & \textbf{0.317} & \textbf{0.646} \\
\bottomrule
\end{tabular}
\end{table}

The synthetic ground-truth mask is produced uniformly across all
four image-to-image production presets V1--V4 by the
\textsc{Convex Hull Annotator} pipeline of
Section~\ref{subsec:dataset}. This pipeline runs the public
SandBoilNet checkpoint over each synthetic image and convex-hull
cleans the prediction before it is admitted to the training set.
The policy holds even for V4. Soft-mask inpainting preserves the
dome geometry of the source mask more faithfully than the
image-to-image presets V1--V3, yet we deliberately do not inherit
the source mask as the label. As a result, every synthetic sample
has the same mask provenance, and the dataset carries a single,
reproducible label-production policy. The mask-conditioned ControlNet
is the one exception to this policy (it is designed to make the
conditioning mask the label by construction), and we examine how well
that design holds in Section~\ref{subsec:res_maskcn}.

\subsection{Mask-conditioned ControlNet and the limits of CHA as a gate}
\label{subsec:res_maskcn}

The mask-conditioned ControlNet (MaskCN, Section~\ref{subsec:mask_controlnet})
attains the strongest KID, CLIP, and LPIPS scores of any preset in
Table~\ref{tab:production_preset_quality}: it ties V1 for the lowest KID
($0.048$), attains the highest CLIP image--text similarity ($0.317$) and the
highest LPIPS within-set diversity ($0.646$), and its FID ($189.4$, clean-fid
$190.1$) is second only to V1. But, as we show next, its labels cannot
yet be certified by the available gate, and it is therefore \emph{not}
adopted as the production preset. By generating from a chosen
binary mask rather than from a source photograph, it promises an
exact label by construction: the conditioning mask \emph{is} the
ground-truth label.

\begin{figure}[!tbp]
\centering
\includegraphics[width=0.99\columnwidth]{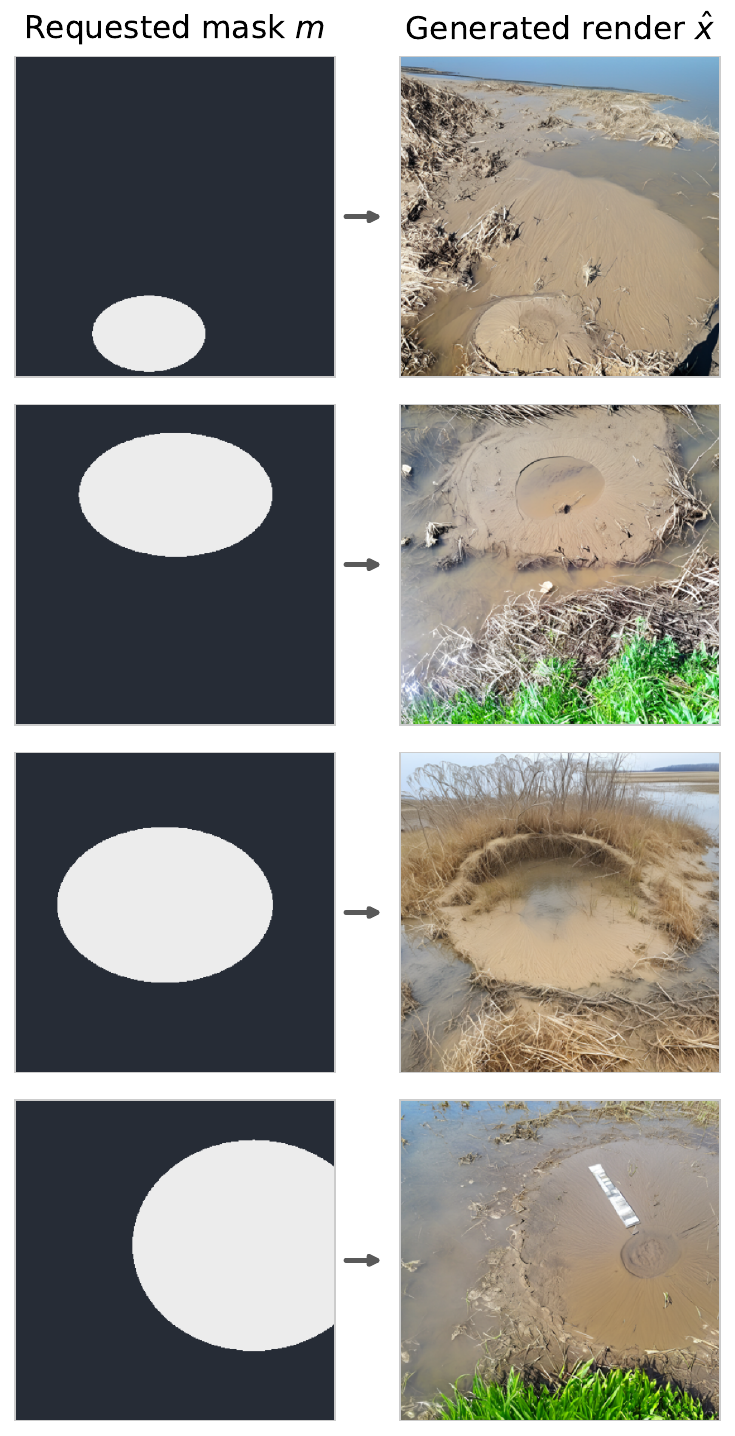}
\caption{Mask-conditioned ControlNet: controllable generation. Each row
pairs a requested binary mask $m$ (left) with the realistic sand boil
the model generates at that location (right), across four examples
spanning the mask-size range (not cherry-picked). Because the image is
generated \emph{from} the mask, the mask is the ground-truth label by
construction. Whether the automated \textsc{Convex Hull Annotator}
gate can \emph{certify} that label on these synthetic renders (where it
under- and over-segments, yielding a median mask--render IoU of only $0.15$, i.e.\ large drift) is taken up
in the surrounding text.}
\label{fig:maskcn_qualitative}
\end{figure}

We test that promise with the \textsc{Convex Hull Annotator} (CHA)
repurposed as a quality-control gate: for each render we run the public
SandBoilNet checkpoint and measure the IoU between the conditioning mask
and the predicted dome (the ``drift''). The result is negative but
informative. Across the retained set the drift IoU has median
$0.15$ and mean $0.17$, and no sample clears the
IoU $\geq 0.5$ acceptance gate. As Figure~\ref{fig:maskcn_qualitative}
shows, the renders are realistic and placed at the requested mask, so
the cause is not principally generator error but a \emph{transfer}
failure of the gate. SandBoilNet, trained on real photographs, alternately
under- and over-segments the synthetic renders, so the IoU it reports is
an unreliable proxy for true mask adherence on this new generation path.

To separate the two effects we swept, on a held-out subset, the
generation controls that govern adherence: the ControlNet conditioning scale
($\{1.0,1.8\}$), the number of inference steps ($\{30,45\}$), and the
sampler (DPM++ 2M Karras vs.\ the stock scheduler). We scored
each configuration by drift IoU and by FID against the
$199$-image reference. Raising the conditioning scale improved the drift
IoU only marginally ($0.137\!\to\!0.143$), and the sampler and step
count made no measurable difference. This confirms that the residual
drift the gate reports is dominated by the segmenter's transfer
behaviour rather than by a generation setting we can tune away. At the
best configuration (conditioning scale $1.8$, $45$ steps, DPM++ 2M
Karras) we additionally applied a best-of-three retention step, keeping
the candidate with the highest CLIP-IQA no-reference
quality~\cite{wang2023exploring} per prompt; this lifted the mean drift
IoU to $0.17$ and produced the image-quality numbers reported above.

The practical consequence is that MaskCN is an excellent \emph{image}
generator whose \emph{label} reliability cannot yet be certified by the
available gate. We therefore retain V4 (soft-mask inpainting), whose dome
is a real preserved structure, as the production preset for the augmented
dataset, and report MaskCN as a strong complementary path whose
label-verification gate is the current bottleneck rather than its image
quality.

Two properties nonetheless make MaskCN a first-class \emph{diversity}
engine alongside the img2img presets, independent of the unresolved gate.
First, its labels are exact \emph{by construction}
(Section~\ref{subsec:mask_controlnet}); the drift IoU above measures only
how well the available SandBoilNet-based gate can \emph{verify} that
label on this new render distribution, not a defect in the label itself.
Second, by decoupling the synthetic label geometry from the catalogue of
real source domes, MaskCN can place a boil of a chosen shape, size, and
position into a scene, independent of where a real dome happened to sit in
any single source image. Figure~\ref{fig:maskcn_label} makes the
label-by-construction guarantee concrete; the resulting cross-engine
diversity is shown in Section~\ref{subsec:res_qualitative}. We make no claim here that adding MaskCN
samples improves downstream segmentation IoU; that evaluation is
ongoing and out of scope for this paper. We claim only that MaskCN contributes
exact zero-annotation labels and a per-instance variety of boil shape, size, and
placement that the img2img production preset cannot produce by design.

\subsection{Qualitative diversity across both generation engines}
\label{subsec:res_qualitative}

\begin{figure*}[tp]
\centering
\includegraphics[width=0.92\textwidth]{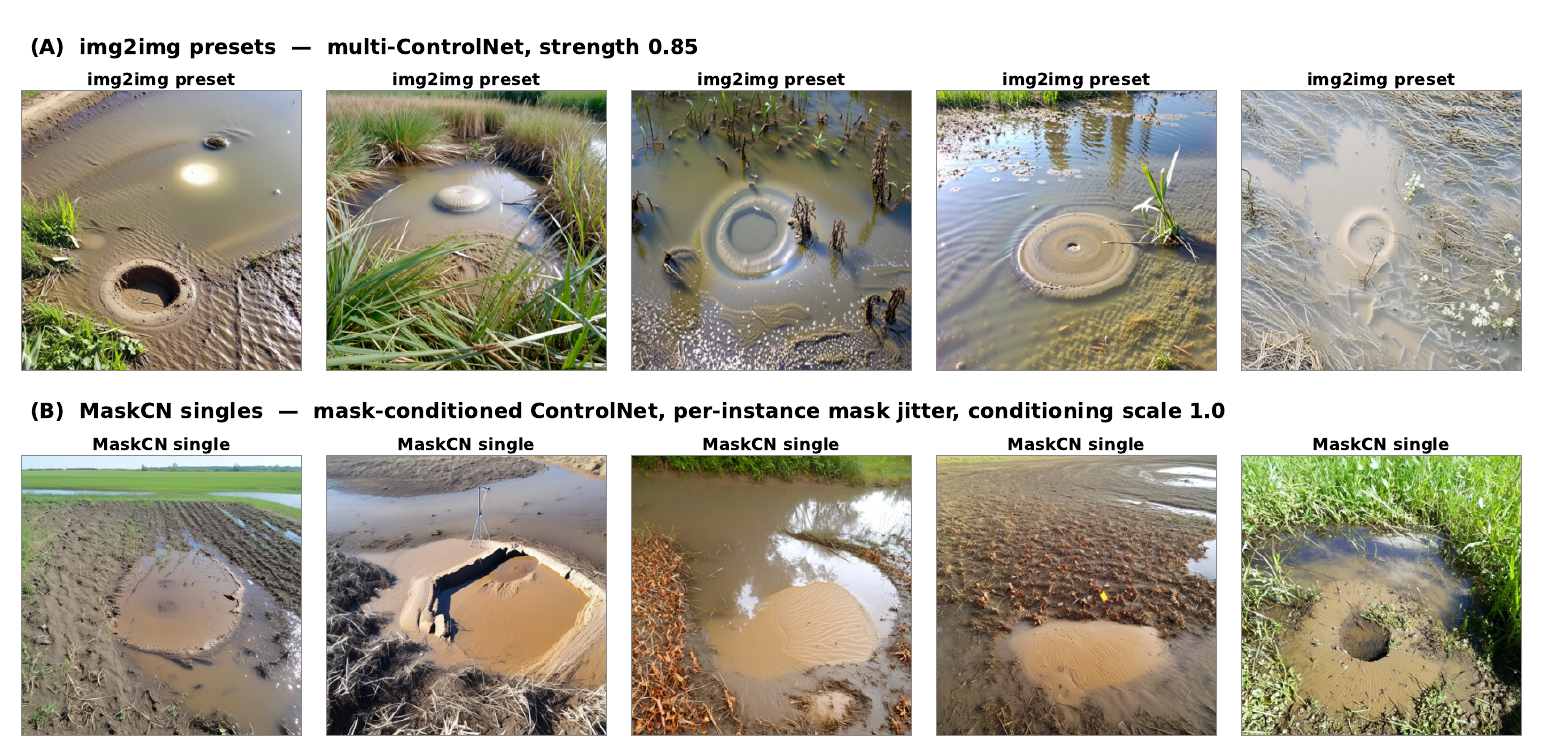}
\caption{Curated best-diverse gallery of synthetic sand boils, arranged as
two labelled rows. \textbf{(A)~Img2img presets} (multi-ControlNet on the
DreamBooth-LoRA SDXL backbone, denoising strength $0.85$; samples
drawn from the V2-curated and V3 production presets of
Section~\ref{subsec:production_presets}). \textbf{(B)~MaskCN singles}
(mask-conditioned ControlNet, DPM++ 2M Karras, per-instance mask jitter,
conditioning scale $1.0$). Tiles are hand-curated for variety, not ranked by
any metric; the panel illustrates image and scene variety only and makes no
segmentation-quality claim. The img2img path widens the \emph{scene} around a
fixed real dome, while MaskCN places a boil of a chosen shape, size, and
position via its conditioning mask, both within the \emph{natural-levee /
floodplain domain}: the strong DreamBooth-LoRA and mask-ControlNet priors mean
prompts for urban, night, or snow collapse back to a levee scene.}
\label{fig:diversity_gallery}
\end{figure*}

Figure~\ref{fig:diversity_gallery} presents the two generation engines
side by side under a single curation policy. Row~(A) shows img2img
preset outputs: each synthetic image retains the dome geometry of its
real source under multi-ControlNet structural conditioning and acquires
a varied surrounding
scene from the Prompt Atlas, spanning soil type, lighting, vegetation,
and viewpoint. Row~(B) shows the mask-conditioned engine, whose
\emph{singles} render one boil into a chosen silhouette at an arbitrary
size and placement, with the conditioning mask serving as the exact label.
The two engines are therefore complementary diversity sources: img2img
widens the \emph{scene} distribution around real defect geometry, while
MaskCN supplies exact zero-annotation labels and places boils at sizes and
positions not tied to any single real source.
The gallery is curated for scene variety, not ranked by
any quality metric, and accordingly implies no claim about downstream
segmentation accuracy.

Within the MaskCN row, per-instance mask jitter further broadens the
engine's output: perturbing each mask's boil shape, size, and position
turns a fixed set of conditioning masks into a wide spread of
boil sizes and off-centre placements rather than collapsing to a few
repeated layouts.

Beyond boil morphology, the mask-conditioned generator also varies the
surrounding \emph{scene} (season, weather, and ground cover) through
the Prompt Atlas. Such
background variety is, in principle, what keeps a downstream segmenter from
keying on a single ``muddy levee'' context, though we make no segmentation
claim here. We bound this claim conservatively, however: the variation stays within the
natural-levee / floodplain domain, because the DreamBooth-LoRA and
mask-ControlNet priors are strong enough that prompts for urban, night, or
snow scenes collapse back to a levee setting---a limit we flag rather than overstate, since the generator does
not produce that wider scene range.

\begin{figure}[!tbp]
\centering
\includegraphics[width=0.99\columnwidth]{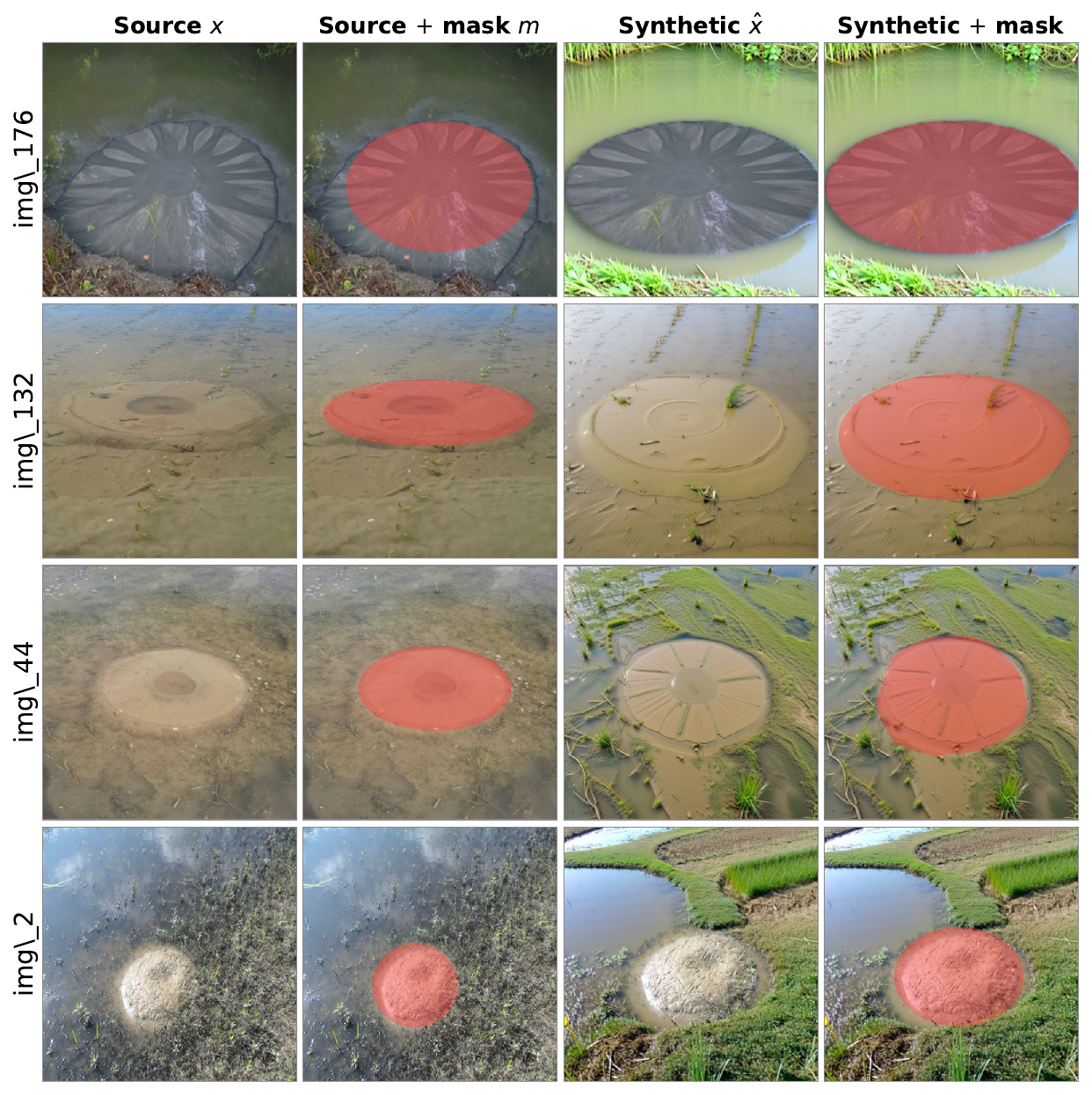}
\caption{Source-paired synthetic outputs (one source per row).
Columns: real source $x$; $x$ with ground-truth mask $m$ overlaid;
synthetic $\hat{x}$ produced under the soft-mask inpainting preset;
$\hat{x}$ with its synthetic mask overlaid. The dome geometry is
preserved between $x$ and $\hat{x}$ by the soft-mask path
(Section~\ref{subsec:soft_mask}); the surrounding scene varies
under Prompt Atlas conditioning. Synthetic masks shown are
human-verified corrections; \textsc{Convex Hull Annotator} is
the default label for uncorrected samples.}
\label{fig:qualitative_samples}
\end{figure}

Figure~\ref{fig:qualitative_samples} complements the diversity gallery
by placing each img2img synthetic image directly beside its real source,
so that source-to-synthetic dome preservation is legible row by row. The
synthetic images preserve the dome geometry of the source through the
soft-mask inpainting path (up to a small rim drift in the soft-blend
region) and acquire varied surrounding scenes from the Prompt Atlas,
including different soil types, lighting, vegetation, and viewpoint. The
synthetic ground-truth masks track the rendered dome.

\subsection{Comparison against a Poisson seamless-cloning baseline}
\label{subsec:res_poisson}

A natural non-generative reference point is the Poisson
seamless-cloning composition strategy~\cite{perez2003poisson}
adopted in our earlier work~\cite{thapa2025thesis}: cut the masked
sand-boil region from one real reference image and paste it into a
different real reference photograph under Poisson colour blending,
with no diffusion model in the loop. We reproduce that baseline
(five Poisson composites per source over the curated foregrounds,
$1024 \times 1024$) and evaluate it against the same full
$199$-image reference under the same FID/KID/LPIPS measure
(Table~\ref{tab:poisson_baseline}).

\paragraph{Why FID alone favours the baseline, and why the
baseline is nonetheless not the better source of training data.}
The Poisson baseline still scores lower on FID
($135.0$ against $201.8$ for V4) and KID
($0.021$\,\scriptsize{$\pm$0.006}\normalsize against $0.091$ for V4),
though the margin is far narrower against the full reference than
against the curated $39$-image set, where the same baseline scored an
artificially low FID of $101$. This is the
expected behaviour, not evidence in favour of the baseline. FID and
KID measure the distance between feature distributions. The Poisson
baseline's outputs are built by pasting \emph{real} sand-boil
foregrounds onto \emph{real} reference photographs, so its output
distribution is a near-subset of the real reference distribution
and the distance to the reference is artificially small. Three
properties separate the two methods qualitatively despite the FID
gap. First, the Poisson baseline cannot vary the defect geometry
beyond the $39$ real-foreground templates: every synthetic sand
boil is one of the $39$ real foregrounds, at most shifted in the
frame, so the effective foreground catalogue is fixed and small.
Second, the Poisson colour-blend step has known seam and
colour-cast failure modes~\cite{perez2003poisson,thapa2025thesis},
visible on most composites in side-by-side qualitative inspection.
Third, LPIPS within-set diversity is close between the two methods
(V4 $=0.633$ vs.\ baseline $=0.620$), so the Poisson baseline's
headline FID advantage does not translate into scene-level
variation. We conclude that FID against a small reference set is a
necessary but not sufficient quality measure for an augmented
training set. The soft-mask inpainting preset is preferred because
it produces (i) seam-free boundaries, (ii) controlled re-rendering
of the surrounding scene under Prompt Atlas conditioning, and (iii)
a synthetic dataset that is not trivially derivable from the
reference set itself.

\begin{table}[t]
\caption{Poisson seamless-cloning baseline versus the proposed
soft-mask inpainting preset (V4), both evaluated against the full
$199$-image real reference and the same number of synthesised outputs
per source. Lower is better for FID/KID, higher for LPIPS diversity.
The baseline reuses real foregrounds (no generative model), so its FID
against the reference is favourable while its scene diversity is bounded
by the reference set itself; clean-fid agrees with the FID shown to
within three points. Best value per column is in bold; the proposed
method's row label is bold.}
\label{tab:poisson_baseline}
\centering
\footnotesize
\setlength{\tabcolsep}{3pt}
\begin{tabular}{@{}p{0.40\columnwidth}ccc@{}}
\toprule
Method & FID $\downarrow$ & KID $\downarrow$ & LPIPS $\uparrow$ \\
\midrule
Poisson seamless-cloning baseline & \textbf{135.0} & \textbf{0.021\,\scriptsize{$\pm$0.006}\normalsize} & 0.620 \\
\textbf{V4 (proposed, soft-mask inpaint)} & 201.8 & 0.091\,\scriptsize{$\pm$0.010}\normalsize & \textbf{0.633} \\
\bottomrule
\end{tabular}
\end{table}

\subsection{Decomposing fidelity from diversity}
\label{subsec:res_manifold}

FID and KID collapse fidelity and diversity into one scalar, which
is why V1 and V4 are hard to separate on those measures alone. The
Improved Precision and Recall~\cite{kynkaanniemi2019prec} and
Density and Coverage~\cite{naeem2020reliable} decomposition, computed on
the same Inception features, pulls the two apart
(Table~\ref{tab:manifold}).

\begin{table}[t]
\caption{Fidelity--diversity decomposition against the full
$199$-image reference set (the same reference as
Table~\ref{tab:production_preset_quality}). \emph{Precision} and
\emph{density} measure synthetic mass inside the real manifold
(fidelity); \emph{recall} and \emph{coverage} measure how much of the
real manifold the synthetic set spans (diversity). All on $2048$-d
Inception features with $k\!=\!5$ nearest-neighbour manifold radii. V4~($\dagger$) is the selected preset; the best value in each column among
the generative presets is in bold.}
\label{tab:manifold}
\centering
\footnotesize
\setlength{\tabcolsep}{4pt}
\begin{tabular}{@{}lcccc@{}}
\toprule
Preset & Precision $\uparrow$ & Recall $\uparrow$ & Density $\uparrow$ & Coverage $\uparrow$ \\
\midrule
V1 & 0.656 & \textbf{0.497} & 0.577 & \textbf{0.673} \\
V2 & 0.478 & 0.307 & 0.309 & 0.432 \\
V3 & 0.517 & 0.171 & 0.454 & 0.482 \\
V4$^{\dagger}$ & 0.315 & 0.226 & 0.149 & 0.322 \\
MaskCN & \textbf{0.829} & 0.236 & \textbf{1.249} & 0.663 \\
\midrule
Poisson baseline & 0.915 & 0.397 & 1.266 & 0.804 \\
\bottomrule
\end{tabular}
\end{table}

Three findings follow. First, among the image-to-image presets the
decomposition tracks the FID ordering: V1 sits closest to and best
spans the real manifold (precision $0.656$, the highest recall $0.497$),
while V4 has the lowest precision ($0.315$) and density---the signature
of a preset that ventures furthest from the tight real manifold, which
is consistent with its high LPIPS within-set diversity in
Table~\ref{tab:production_preset_quality} ($0.633$, the second-highest
among the image-to-image presets, after V3). V4 is thus \emph{not} selected
for manifold fidelity, on which V1 dominates it, but for its single-policy
soft-mask label provenance and its scene-level diversity. Second, MaskCN
has the highest precision ($0.829$) and density ($1.249$) of any
generative preset: it sits very close to the real manifold, consistent
with its low FID/KID, at modest recall ($0.236$): a high-fidelity,
moderate-diversity profile. Third, the Poisson baseline posts the highest
precision ($0.915$) and density ($1.266$), high recall ($0.397$), and
the highest coverage ($0.804$) in the table: because it pastes real
defect foregrounds onto real backgrounds, its samples lie \emph{inside}
the real manifold by construction. Its strong recall and coverage
reflect this reuse of real content rather than genuinely new
scenes---the manifold coverage a degenerate copy-paste route attains
without adding novelty. This is the classic degenerate-augmentation
signature, and the same effect that flatters its FID in
Section~\ref{subsec:res_poisson}. The decomposition therefore confirms
that distribution-matching scores must be read alongside the provenance
of the samples being scored, and that V4's lower fidelity numbers
reflect real scene synthesis rather than a defect.

\subsection{Empirical demonstration of the CLIP quality filter}
\label{subsec:res_qfilter}

Section~\ref{subsec:quality_filter} introduces a
post-generation CLIP admissibility filter whose lower and upper
band thresholds are computed from leave-one-out reference-set
statistics. We apply the filter to the full $1{,}020$-image V4
augmented set without modification.

\begin{itemize}
  \item Reference-set leave-one-out similarity to the cohort
    centroid: median 0.889, MAD 0.020.
  \item Admissibility band:
    $[s_{\mathrm{lo}}, s_{\mathrm{hi}}] = [0.839, 0.963]$.
  \item Augmented-set scores: mean 0.868, range
    [0.627, 0.946].
  \item Filter verdict: 815 accepted, 205
    below band (out-of-distribution drift), 0 above
    band (memorisation signature).
\end{itemize}

\noindent The filter rejects a non-trivial fraction of generations
without any human review or hand-picked threshold. The rejected
examples fall into one of two failure modes whose causes are
diagnostically different: OOD drift and memorisation; here every
rejection is OOD drift.

\subsection{Memorisation audit}
\label{subsec:res_memorisation}

\begin{figure*}[tp]
\centering
\includegraphics[width=0.96\textwidth]{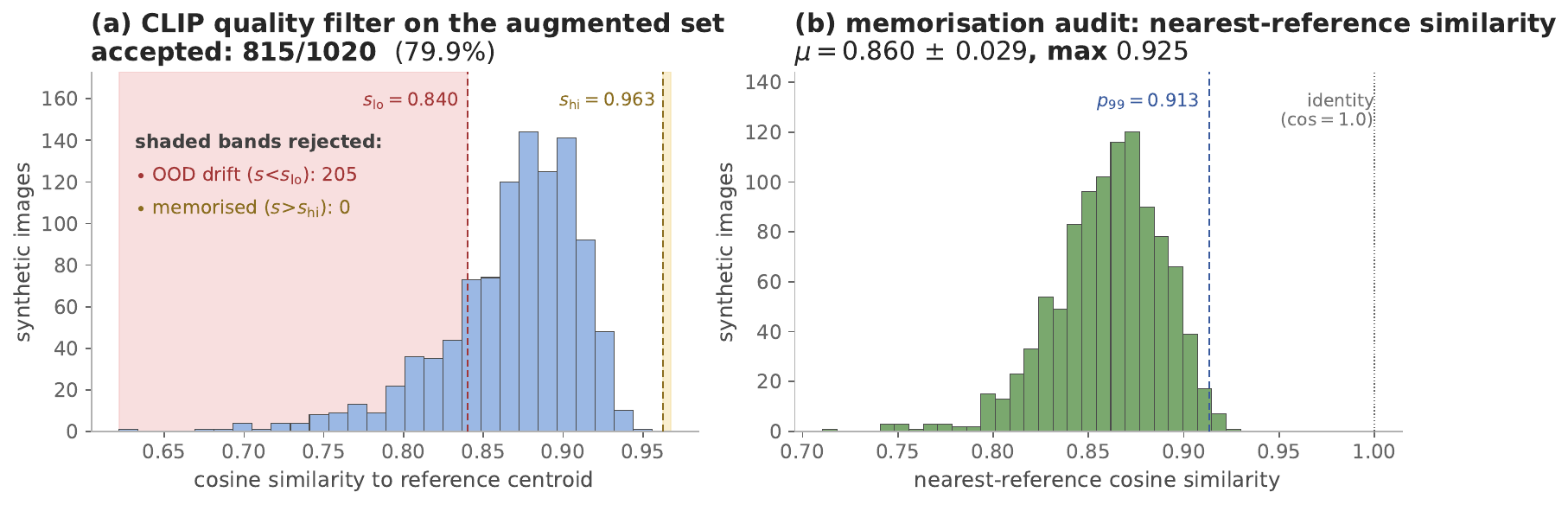}
\caption{Empirical audits over the V4 augmented set
($N_{\mathrm{synth}}\!=\!1{,}020$, $N_{\mathrm{ref}}\!=\!39$).
\textbf{(a)}~Histogram of each synthetic image's cosine similarity
to the reference centroid. The shaded bands are the
leave-one-out admissibility band of
Section~\ref{subsec:quality_filter}: pink rejects out-of-distribution
drift, yellow rejects memorisation-signature outputs. Both
thresholds are deterministic functions of the reference set with
no hand-picked values. \textbf{(b)}~Histogram of nearest-reference
cosine similarity for every synthetic image; the $99$th percentile
sits well below the identity-level cosine of $1.0$, so even the most
reference-like samples are far from identity at the small-reference-set
scale.}
\label{fig:audit}
\end{figure*}

A recurring concern with DreamBooth-fine-tuned
diffusion models on small reference sets is overfitting to
individual reference frames. We evaluate this directly against the
$39$-image curated set the adapter was \emph{trained} on (the relevant
reference for memorisation, as distinct from the $199$-image evaluation
set used for FID/KID). For every synthetic image, we compute the cosine
similarity of its CLIP-ViT-B/32 embedding to its \emph{nearest}
reference image. The
nearest-reference similarity distribution over the $1{,}020$-image
V4 augmented set (Figure~\ref{fig:audit}) has mean 0.860 $\pm$ 0.029, with
$99$th-percentile 0.913 and a maximum of 0.925. The
maximum is comfortably below an identity-level value (cosine
$\approx 1.0$), so no synthetic sample is a CLIP-space duplicate of
any reference frame, and even the most reference-like $1\%$ of
synthetic samples ($99$th-percentile similarity $0.913$) stay well below
that identity level.

\subsection{Interpretability of the conditioned generator}
\label{subsec:res_interpret}

Two views make the pipeline legible: \emph{how faithful} the generated
distribution is to the real one, and \emph{where} the class concept is
grounded. (The conditioning signals themselves, and what each ControlNet
branch contributes, are shown in Figure~\ref{fig:explain_controlnet} of
Section~\ref{subsec:multicn}.)

\begin{figure}[!tbp]
  \centering
  \includegraphics[width=\columnwidth]{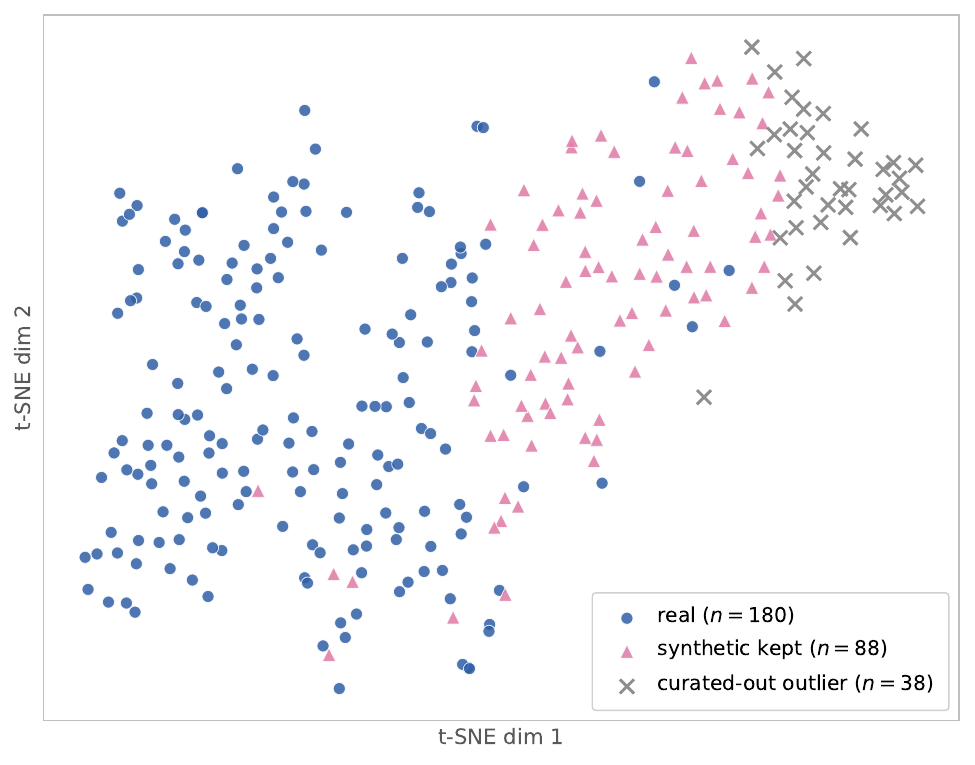}
  \caption{Texture-feature faithfulness of the synthetic pool. Real
  and synthetic images are embedded with PCA followed by t-SNE on the
  $26$-dimensional handcrafted texture/colour descriptor. Synthetic
  samples overlap the real distribution, while the minority farthest
  from their nearest real neighbours form a separable cluster (the
  texture outliers the curation filter removes).}
  \label{fig:explain_embedding}
\end{figure}

\textbf{Synthetic outputs overlap the real texture distribution.}
Figure~\ref{fig:explain_embedding} embeds real and synthetic images in
the $26$-dimensional handcrafted texture/colour space (the descriptor
underlying our curation filter) via PCA followed by
t-distributed stochastic neighbour embedding (t-SNE). On this
2-D t-SNE projection the synthetic samples interleave with the real
points rather than occupying a disjoint region; we note, however, that
t-SNE geometry is only indicative and the quantitative recall and coverage
(Table~\ref{tab:manifold}) qualify this. The overlap is consistent with
the synthetic samples sharing much of the real texture/colour
distribution; a 2-D t-SNE overlap is suggestive rather than conclusive
and does not by itself establish statistical equivalence. We further
note that the selected V4 preset trades manifold fidelity for diversity:
its precision and density are the lowest among the presets
(Table~\ref{tab:manifold}). The
minority of synthetics that sit farthest from their nearest real
neighbours form a \emph{separable} cluster; these are precisely the
texture outliers the curation step removes, which is why curation
sharpens the pool without discarding faithful samples.

\subsection{Computational cost}
\label{subsec:res_cost}

Table~\ref{tab:cost} gives the wall-clock cost of the major phases
of the generation pipeline on a single NVIDIA H100 NVL.

\begin{table}[t]
\caption{Wall-clock cost on a single NVIDIA H100 NVL GPU,
measured from SLURM job-accounting records of the production run.}
\label{tab:cost}
\centering
\footnotesize
\setlength{\tabcolsep}{4pt}
\begin{tabular}{@{}p{0.66\columnwidth}r@{}}
\toprule
Phase & H100 \\
\midrule
Prompt Atlas build (CPU) & 20\,s \\
Augmented dataset generation
   ($1{,}020$ images @ $1024 \times 1024$) & 2\,h 13\,min \\
Per-image generation throughput & $\sim$7.8\,s/img \\
Synthetic-image quality (FID/KID/LPIPS,
   $300$ samples) & $<$2\,min \\
Per-preset evaluation generation
   ($182$ images each) & $\sim$24\,min \\
\bottomrule
\end{tabular}
\end{table}

%% file: sections/discussion.tex
\section{Discussion}
\label{sec:discussion}

The results in Section~\ref{sec:results} show that the
DreamBooth-fine-tuned SDXL generator, combined with multi-ControlNet
conditioning and soft-mask inpainting, produces synthetic sand boil
imagery with controlled stratification, defect-region preservation,
and uniform mask provenance. We restrict the claims of this paper to
the generation pipeline and its image-level quality measures. Whether
training on \emph{real plus synthetic} versus \emph{real only} helps a
downstream segmenter is left to future work. This section explains the main design
choices and reviews the failure modes that remain.

\subsection{What each component contributes}
\label{subsec:disc_components}

The
generator-pipeline ablation of Section~\ref{subsec:res_ablation} makes
a clear case for keeping all three ControlNets. Canny constrains the
broad outline of the sand boil and the major scene boundaries. Depth
constrains the three-dimensional layout. This matters because a sand
boil is a small bulge above an otherwise flat surface, and without
depth conditioning the generator occasionally flattens it. The
soft-edge branch captures the fine textural transition at the boil's
rim, which is exactly where Canny and depth are weakest. Removing any
single control visibly degrades boundaries on at least one of the five
test sources.

The soft-mask inpainting path (row E) is preferred to plain
image-to-image (row D) for augmented-dataset generation, but the
comparison is asymmetric. For visual variety alone, row D is adequate.
For dataset generation, row E is strongly preferred because its
synthetic ground-truth mask aligns pixel-for-pixel with the real mask,
up to the small rim drift that the \textsc{Convex Hull Annotator}
(CHA) step is designed to handle. Under row D the dome shape can drift
slightly during denoising, and the auto-produced ground-truth mask
must be relied upon entirely. Under row E the mask comes from a single
pass over a known-anchored dome region, giving more reliable
boundary labels in our qualitative inspection.

\paragraph{Why V4 is selected despite lower-FID alternatives.}
Table~\ref{tab:production_preset_quality} shows V1 with the lowest FID
among the image-to-image presets ($169.1$) and the mask-conditioned
ControlNet (MaskCN) strongest on CLIP image--text similarity and
within-set diversity, tied with V1 for the lowest KID, while V4
carries a numerically higher FID ($201.8$). We nonetheless select V4 for
the augmented dataset, because the metric leaders each fail a
requirement that FID does not see. MaskCN's labels cannot be certified:
its CHA quality gate reports a median drift IoU of only $0.15$
(Section~\ref{subsec:res_maskcn}), so the mask it is conditioned on is
not a trustworthy ground-truth label for the dome it actually renders.
V1's labels are sounder but still re-predicted, and three further non-FID
reasons favour V4. First, V1 generates whole-image img2img outputs
in which the dome can drift slightly during denoising, so the
synthetic ground-truth mask must be re-predicted by CHA over a dome
that has shifted relative to the source. Under V4's soft-mask
inpainting path the dome pixels are preserved verbatim, so the
mask-production policy is anchored to a known-stable dome region and
yields cleaner boundary labels in qualitative inspection. Second, the
precision/recall decomposition (Table~\ref{tab:manifold}) shows that
V1 leads V4 on manifold fidelity (precision $0.656$ vs $0.315$)
\emph{and} on recall ($0.497$ vs $0.226$). V4's advantage over V1 is
therefore not manifold coverage but its single-policy soft-mask label
provenance, together with a modestly higher LPIPS within-set diversity
($0.633$ versus V1's $0.595$; a small-sample estimate we treat as
suggestive rather than decisive), at equal CLIP image--text similarity
($0.302$). V4 trades manifold fidelity for a trustworthy label and a
wider spread of synthetic scenes\,---\,the properties that matter when the
set is used as segmenter supervision rather than as a sample from the
real distribution. Third, distribution-matching scores reward
\emph{being close to the reference distribution}, which a degenerate
route can achieve. The Poisson seamless-cloning baseline of
Section~\ref{subsec:res_poisson} pastes real foregrounds onto
real backgrounds and shows exactly this: FID~$\approx$~$135$, high
precision and density, and high recall and coverage by construction, yet
no genuinely new content.
The selection criterion is
therefore not minimum FID but the joint requirement of high within-set
diversity (LPIPS spread), adequate fidelity (precision in the
generative-preset range), high CLIP image--text similarity, and
single-policy mask provenance.

We release V4 as the single \emph{label-reliable} production preset: it
is the only one whose synthetic dome is a preserved real structure rather
than a freely re-rendered one. The presets are nonetheless complementary:
V1 attains the broadest real-manifold coverage (Table~\ref{tab:manifold}),
while V3 and the mask-conditioned ControlNet attain the highest within-set
diversity. A curated \emph{mixture} of presets, admitted by the CLIP
quality filter and then sub-sampled for feature-space spread, is therefore
a natural way to widen augmentation diversity without sacrificing
on-distribution quality; quantifying its effect on downstream
segmentation accuracy is left to future work.

\subsection{Mask-conditioned generation as a diversity and
zero-annotation engine}
\label{subsec:disc_maskcn}

The mask-conditioned ControlNet (MaskCN) earns a first-class place in the
pipeline for two reasons that are distinct from (and unaffected by)
the unresolved label-verification gate of
Section~\ref{subsec:res_maskcn}. We state both carefully, because MaskCN's
downstream segmentation benefit is not asserted in this paper.

First, MaskCN is a \emph{zero-annotation} labelling engine. By generating
a fresh boil \emph{into} a chosen binary mask, it makes the conditioning
mask exactly the segmentation label of the image it produces
(Figure~\ref{fig:maskcn_label}), with no manual annotation and no
dependence on a segmenter checkpoint. This removes the
re-prediction ceiling that bounds the soft-mask$+$CHA labels: the
soft-mask path's label quality is capped by how well SandBoilNet
transfers to synthetic renders, whereas MaskCN's label is correct by construction \emph{if} the
generator fills the conditioning silhouette. What remains open is the
\emph{verification} of that adherence at scale: the currently available,
real-trained gate reports a median drift IoU of $0.15$ on these renders,
which we read as largely a transfer artefact of the gate rather than
evidence of a wrong label, though with the present gate we cannot fully
separate gate transfer failure from generator non-adherence
(Section~\ref{subsec:res_maskcn}).

Second, MaskCN is a label-provenance diversity engine that the
source-anchored img2img path structurally cannot match. Because its label
geometry is decoupled from the catalogue of real source domes, MaskCN
draws masks from a bank and jitters each instance with flips, rotations,
rescaling, and repositioning (Figure~\ref{fig:diversity_gallery},
block~B). The boil shape, size, and placement therefore vary independently of
any one real source dome. MaskCN thus widens a different axis of
diversity from the img2img presets: where img2img varies the \emph{scene}
around a fixed real defect geometry, MaskCN varies the \emph{label
geometry itself} at zero
annotation cost. A natural next step is to fold curated
MaskCN singles into the augmentation mixture as a zero-annotation label
source. Whether their by-construction labels hold up under an
independent adherence check and improve downstream segmentation IoU is
left to future work.

\subsection{Backbone choice as a load-bearing decision}
\label{subsec:disc_backbone}

The choice of SDXL over a
more recent rectified-flow alternative is, by design, a load-bearing
decision rather than a default. Section~\ref{subsec:backbone_choice}
details the trade-off. In summary, SDXL pairs with the most mature
publicly available ControlNet and IP-Adapter checkpoints, fits
comfortably on a single H100~NVL at $1024 \times 1024$ with up to
three simultaneous controls (or two controls plus an IP-Adapter), and, in our own qualitative
experiments, was less prone than higher-capacity rectified-flow
alternatives to memorising individual training frames at the
reference-set scale used here. The same pipeline could in principle
be re-instantiated on a rectified-flow backbone once equivalent
structural-conditioning weights mature, with no change to the prompt
curator.

\subsection{Why prompt curation matters for niche datasets}
\label{subsec:disc_prompts}

When real data is scarce, the value of each
synthetic sample depends disproportionately on its prompt. A
synthetic image with an off-topic prompt, such as a sand boil rendered
in a cartoonish style or in an unrelated geographic setting, drags the
augmented distribution away from the real one. Under scarcity, this
contributes far more to model degradation than the same defect would
in a million-image dataset. Hand-written prompt lists offer no guard
against this failure mode and tie the augmentation pipeline to the
domain expert who wrote them. In response, the Prompt Atlas of
Section~\ref{subsec:prompt_atlas} replaces the hand-written list with
a stratified, validated, image--text-anchored bank that is
reproducible from a single domain specification and portable across
defect classes. The CLIP-based image--text validation stage does real
work here: prompts whose CLIP similarity to the reference images falls
below the cohort minimum are dropped, and the resulting bank
provides explicit, reproducible coverage of the taxonomy axes that a
hand-written list of comparable size does not guarantee.

\subsection{Generalisation to other defect classes}
\label{subsec:disc_generalization}

Although the experiments focus on sand boils, the pipeline is
domain-agnostic by construction. Extending to a new defect (rutting,
seepage, sinkhole, and the like) requires only three changes:
writing a new domain specification for the Prompt Atlas with the
appropriate concept, taxonomy axes, and exemplar images; fine-tuning a
new DreamBooth adapter on the new reference set; and supplying the
structural-conditioning sources for the multi-ControlNet stack. The
Prompt Atlas builder accepts any domain specification without code
changes, and the DreamBooth-LoRA recipe reuses the same training
script.

One practical caveat is that the taxonomy axes that matter for one
defect class do not transfer to another. A rutting domain
specification needs road-surface, wheel-path-pattern, and
depth-severity axes, and would not benefit from a hydraulic-regime
axis. Each new domain therefore requires a brief modelling exercise
to choose its axes. This effort is small relative to the cost of
curating real data or hand-writing prompts for every scenario.

\subsection{Failure modes and limitations}
\label{subsec:disc_failures}

Several limitations remain, which we set out below.

\textbf{Style of the reference set.} The DreamBooth adapter is fit
to the curated USACE reference set. The visual style of the
synthetic images is consequently anchored to that of USACE
photography: camera type, colour profile, and ground-level
perspective. A deployed application that receives drone (UAV) or
smartphone imagery from a different source may not generalise as well
as it does to USACE-style imagery. A straightforward remedy is to
augment the adapter's reference set with images from additional
sources.

\textbf{Reflective surfaces.} The downstream segmenter still struggles
with images in which the sand boil is partially obscured by reflective
standing water. The reflective patch breaks the texture cues that the
network relies on. We leave to future work a separate
reflection-masked training condition, or a two-pass approach that
first predicts reflection regions and then segments within the
non-reflective area.

\textbf{Downstream evaluation scope.} This paper reports generative
quality only (FID, KID, CLIP image--text similarity, and LPIPS
diversity) together with the empirical behaviour of the CLIP
admissibility and memorisation audits. A fuller multi-architecture
downstream-segmentation evaluation, with cross-validation and a
leak-free fold filter, is left to future work, which would consume
the augmented dataset
produced here.

\textbf{Cross-domain validation.} The pipeline is constructed to be
class-agnostic: a single JSON specification ports the Prompt Atlas,
and CHA's shape-aware geometry adapter routes by form factor between
compact, elongated, and thin curvilinear cleanup. The empirical
validation in this paper, however, is restricted to the sand boil
class. The class-agnostic claim is exercised by self-tests on
synthetic shape inputs in the released CHA implementation, and the
released code ships domain-specification files for sandboil, sinkhole,
rutting, and crack. An end-to-end empirical demonstration on
a non-sand-boil class is left to a follow-up release once the
corresponding reference sets and DreamBooth adapters are curated.

\textbf{Prompt Atlas template gap.} The v1 prompt template references
seven of the eight taxonomy axes; the \texttt{season} axis is sampled
into the per-prompt \texttt{fill} record but is not rendered into the
prompt text itself. The downstream effect is captured by the 40/44
rendered-coverage figure in
Section~\ref{subsec:prompt_atlas}. Adding the missing template slot
would be a one-line specification edit, but we did not re-run the
production atlas for this paper, in order to preserve a one-to-one
comparison with the previously generated augmented set.

\textbf{Synthetic-mask quality ceiling.} The synthetic ground-truth
mask is produced by CHA from the publicly released SandBoilNet
checkpoint. The quality of the resulting label is therefore
upper-bounded by the prediction quality of that checkpoint on the
synthetic image. For most samples the convex-hull post-processing
absorbs minor prediction noise. On edge cases where the checkpoint
disagrees with the rendered dome by a large margin, however, the
synthetic label is unreliable and the sample must be filtered out. The
deployed pipeline exposes a human-in-the-loop correction interface
that allows an inspector to overwrite the synthetic label per image;
the operational use of that interface is left to future work.

\textbf{CLIP circularity in prompt validation.} The Prompt Atlas filters
candidate prompts by their CLIP image--text similarity to the reference
set, and the V3/V4 presets are in turn conditioned with a CLIP-aligned
style. Validating prompts with the same family of encoder that guides
generation is mildly circular and could bias the atlas toward
CLIP-favoured phrasings. The effect is limited here because field
inspection photography is visually narrow and well-lit, but in a more
heterogeneous domain the validation stage should use an ensemble of
image--text encoders distinct from the generator's conditioning encoder.

\textbf{Un-ablated generation constants.} The soft-mask erosion and
blur coefficients, the per-branch ControlNet conditioning scales, and
the classifier-free guidance value were selected on the five-source
generator ablation rather than swept independently, and the SDXL
backbone choice rests on a qualitative memorisation observation
(Section~\ref{subsec:backbone_choice}) rather than a controlled
backbone benchmark. We report them as fixed, reproducible settings; a
full sensitivity analysis and a rectified-flow backbone comparison are
left to future work and do not affect the conclusions drawn from the
fixed configuration evaluated here.

\subsection{Threats to validity}
\label{subsec:disc_threats}

Two threats to the conclusions drawn here deserve flagging.
First, the generative-quality measures (FID, KID, and LPIPS) computed
against a reference set of the size used here have known sample-size
bias. We report KID with its standard deviation alongside FID to
mitigate the small-sample bias of FID, but a larger reference set
would be needed to make fine-grained between-preset comparisons fully
reliable. Second, the augmented dataset is generated by a single
DreamBooth adapter trained on a single curated reference set. A
residual concern is that any generation-side stylistic bias of that
adapter could narrow the diversity of the augmented set rather than
reflect a general property of diffusion-based augmentation. To address
this directly we trained an ensemble of three DreamBooth-LoRA adapters
on disjoint 13-image partitions of the reference set and
evaluated each on a common prompt/seed set. The three adapters agree
tightly: FID $235.3 \pm 1.9$ (range $4.5$), KID $0.053 \pm 0.005$, and
LPIPS diversity $0.588 \pm 0.007$ across the three independent
partitions. Because adapters fit to disjoint image subsets produce
near-identical quality and diversity, the augmentation behaviour appears
to be driven by the recipe rather than by one particular adapter's
stylistic bias. This rules out single-adapter bias, though not a dataset-
or source-level bias within this single-source reference set. The ensemble-training and evaluation
scripts are released so the control can be reproduced and extended to
disjoint reference sets from other camera sources.

\subsection{Ethical considerations}
\label{subsec:disc_ethics}

The intended use is to support, not to replace,
trained inspectors. Sand boils are safety-critical defects: a false
negative during a high-water event can delay intervention long enough
to permit a breach. Any operational deployment that consumes a
segmenter trained on this augmented data should expose model
uncertainty rather than collapse it to a binary decision, and should
retain a human-in-the-loop correction path for high-stakes
inspections. We also note that the synthetic images produced by this
pipeline are stylistic inheritors of the USACE reference photography.
They are not photorealistic in a forensic sense and should not be
presented as field documentation of actual sand boils.

%% file: sections/conclusion.tex
\section{Conclusion and Future Work}
\label{sec:conclusion}

We presented a generation pipeline for synthetic sand boil imagery.
The pipeline couples a DreamBooth-fine-tuned Stable Diffusion XL
backbone, four independent ControlNet branches (Canny, monocular
depth, surface normal, and HED soft-edge) from which each preset
activates a subset, a soft-mask inpainting protocol
that preserves the real defect region pixel-for-pixel, and a
taxonomy-driven Prompt Atlas. The Prompt Atlas turns a single JSON
domain specification into a stratified, de-duplicated,
image--text-validated prompt bank in under a minute. We chose SDXL
over more recent rectified-flow alternatives deliberately: at the
reference-set scale and under the multi-ControlNet
inpainting regime this application imposes, SDXL offered the best
combination of structural-conditioning maturity, VRAM headroom, and
overfit resistance.

We evaluated the pipeline on standard generative-image-quality
measures (FID under two independent implementations, KID, CLIP
image--text similarity, and LPIPS within-set diversity), together with a
precision/recall and density/coverage decomposition that separates
fidelity from diversity. We reported the production preset selected for
the augmented dataset alongside three image-to-image variants, a
mask-conditioned ControlNet, and a Poisson seamless-cloning baseline
(the compositing strategy used in our earlier pipeline). All were evaluated
against the full real reference set, with KID reported alongside its
standard deviation. We find that no single preset dominates: the
configurations are complementary, trading manifold coverage, within-set
diversity, and label reliability. We therefore release the label-reliable
preset as the default and identify a curated mixture across them as the
natural augmentation set. We demonstrated the leave-one-out CLIP
admissibility filter empirically on the augmented set, and used a
CLIP-space nearest-reference audit to estimate the memorisation rate at
the small-reference-set scale. Downstream use of the augmented
dataset for segmentation is left to future work.

The pipeline is designed to transfer to a new defect class through a
single domain specification. The released code emits a complete artefact
manifest (DreamBooth checkpoint, Prompt Atlas record,
multi-ControlNet inference configuration, and per-image seed
cascade), making the reported results reproducible.

Several extensions remain open.

\textbf{Closed-loop diffusion as a segmentation teacher.} The
present augmentation is open-loop: the synthetic dataset is
generated once, and then training proceeds. A closed-loop variant
would use a downstream segmenter's prediction uncertainty to steer
subsequent rounds of synthesis toward the specific image conditions
on which the current model fails. This is structurally close to
active learning, with diffusion-based augmentation taking the role
usually played by oracle labelling.

\textbf{Scaling the mask-conditioned path.} The mask-conditioned
ControlNet of Section~\ref{subsec:mask_controlnet} is released
alongside the soft-mask pipeline. It generates
from a chosen mask so the label is exact by construction, with
CHA as its drift verifier. It already serves as a zero-annotation
diversity engine: single boils are drawn from a curated mask bank under
per-instance jitter. These contribute exact labels and a variety of boil shapes, sizes, and
placements that the source-anchored path cannot produce. Two extensions
remain
open. First, certifying its by-construction labels at scale needs a gate
that transfers to synthetic renders, since the present CHA gate
(built on the real-trained SandBoilNet checkpoint) under- and over-segments them
(Section~\ref{subsec:res_maskcn}). Second, the ControlNet could be scaled from
the $199$-pair fine-tune to a larger multi-source corpus, and a generative prior could be learned over plausible defect masks so that the curated mask bank is extended by a learned mask distribution. This would let the pipeline synthesise paired (image, mask) records
end-to-end with no real source image at all. Whether MaskCN samples
improve downstream segmentation accuracy is a question for future
work, not this paper.

\textbf{Larger curated benchmark.} A publicly available
levee-defect benchmark with paired real and synthetic data, source
tags, and a held-out test set would let the community compare
methods directly. Building such a benchmark on top of the dataset and
reproducibility framework released here, together with a deployed
inspection-workflow prototype, is left to future work.

\textbf{Domain transfer to other defect classes.} The Prompt Atlas
builder ports to a new domain through its JSON specification alone;
re-instantiating the full multi-ControlNet inpainting pipeline on a
seepage, sinkhole, or rutting domain additionally requires a new
DreamBooth adapter and structural-conditioning sources. Such a transfer would
provide external validation of the generation pipeline that the
present paper's single-class focus does not.

%% file: sections/code_availability.tex
\section*{Code and Data Availability}
\label{sec:availability}

The complete generation pipeline (DreamBooth-LoRA training script,
multi-ControlNet inference configuration, soft-mask inpainting
implementation, \textsc{Convex Hull Annotator} reference
implementation, Prompt Atlas builder, and the per-image seed
cascade) is released alongside this paper. The released artefact
includes the trained LoRA safetensors checkpoint
($\sim$240\,MB), the Prompt Atlas JSON Lines record for the sand
boil domain (\texttt{v1}), and the parameter manifest needed to
reproduce every numerical entry in
Section~\ref{sec:results} from a single command. The
\textsc{Convex Hull Annotator} pipeline is additionally
distributed as a standalone, class-agnostic library~%
\cite{thapa_chull_annotator} with its own domain specifications for
sand boil, sinkhole, rutting, and crack defect classes, so the
post-processing component is reusable outside the present pipeline.

On acceptance, the full artefact (training and inference code, the
trained LoRA checkpoint, the Prompt Atlas record, and the
reproducibility manifest) will be published in a public Git
repository and archived under a versioned Zenodo DOI for long-term
citation.

The curated reference set is derived from the U.S.\ Army Corps of
Engineers (USACE) levee-inspection archive~%
\cite{usace_levee_manual}. Its redistribution is subject to the
USACE archive's terms of use, so the underlying photographs are not
redistributed; the pixel-level binary masks produced by the authors
will be released under the Creative Commons Attribution 4.0
(CC~BY~4.0) license alongside the code.

\section*{Conflicts of Interest}

The authors declare no competing financial or non-financial interests.
This work was carried out in an academic research setting and is
derived from the corresponding author's Master's
thesis~\cite{thapa2025thesis}.

%% file: sections/acknowledgments.tex
\section*{Acknowledgments}
\label{sec:acks}

The authors thank the U.S.\ Army Corps of Engineers for making the
levee-inspection imagery archive~%
\cite{usace_levee_manual} available for academic research, and the
maintainers of the open-source diffusion-model ecosystem (Stability
AI, the diffusers library, the ControlNet authors,
and the IP-Adapter authors) whose publicly released
checkpoints made this pipeline tractable on a single accelerator.
This work was conducted on an institutional GPU cluster at
Louisiana State University New Orleans.

%% file: references.bib
@techreport{ilit2006katrina,
  title={Investigation of the Performance of the New Orleans Flood Protection Systems in Hurricane Katrina on August 29, 2005},
  author={Seed, R.B. and Bea, R.G. and Athanasopoulos-Zekkos, A. and others},
  institution={Independent Levee Investigation Team, University of California Berkeley},
  year={2006}
}

@misc{usace_levee_manual,
  author={U.S. Army Corps of Engineers},
  title={Levee Owner's Manual for Non-Federal Flood Control Works},
  howpublished={\url{https://www.mvr.usace.army.mil/Portals/48/docs/EC/LSP/LeveeOwnersManual.pdf}},
  note={Accessed 2025-02-17}
}

@inproceedings{ruiz2023dreambooth,
  title={DreamBooth: Fine Tuning Text-to-Image Diffusion Models for Subject-Driven Generation},
  author={Ruiz, Nataniel and Li, Yuanzhen and Jampani, Varun and Pritch, Yael and Rubinstein, Michael and Aberman, Kfir},
  booktitle={Proceedings of the IEEE/CVF Conference on Computer Vision and Pattern Recognition (CVPR)},
  year={2023}
}

@inproceedings{zhang2023controlnet,
  title={Adding Conditional Control to Text-to-Image Diffusion Models},
  author={Zhang, Lvmin and Rao, Anyi and Agrawala, Maneesh},
  booktitle={Proceedings of the IEEE/CVF International Conference on Computer Vision (ICCV)},
  year={2023}
}

@article{ye2023ipadapter,
  title={{IP-Adapter}: Text Compatible Image Prompt Adapter for Text-to-Image Diffusion Models},
  author={Ye, Hu and Zhang, Jun and Liu, Sibo and Han, Xiao and Yang, Wei},
  journal={arXiv preprint arXiv:2308.06721},
  year={2023}
}

@inproceedings{rombach2022ldm,
  title={High-Resolution Image Synthesis with Latent Diffusion Models},
  author={Rombach, Robin and Blattmann, Andreas and Lorenz, Dominik and Esser, Patrick and Ommer, Bj{\"o}rn},
  booktitle={Proceedings of the IEEE/CVF Conference on Computer Vision and Pattern Recognition (CVPR)},
  year={2022}
}

@article{podell2023sdxl,
  title={{SDXL}: Improving Latent Diffusion Models for High-Resolution Image Synthesis},
  author={Podell, Dustin and English, Zion and Lacey, Kyle and Blattmann, Andreas and Dockhorn, Tim and M{\"u}ller, Jonas and Penna, Joe and Rombach, Robin},
  journal={arXiv preprint arXiv:2307.01952},
  year={2023}
}

@inproceedings{esser2024sd3,
  title={Scaling Rectified Flow Transformers for High-Resolution Image Synthesis},
  author={Esser, Patrick and Kulal, Sumith and Blattmann, Andreas and Entezari, Rahim and M{\"u}ller, Jonas and Saini, Harry and Levi, Yam and Lorenz, Dominik and Sauer, Axel and Boesel, Frederic and Podell, Dustin and Dockhorn, Tim and English, Zion and Rombach, Robin},
  booktitle={Proceedings of the 41st International Conference on Machine Learning (ICML)},
  year={2024}
}

@inproceedings{liu2023rectflow,
  title={Flow Straight and Fast: Learning to Generate and Transfer Data with Rectified Flow},
  author={Liu, Xingchao and Gong, Chengyue and Liu, Qiang},
  booktitle={International Conference on Learning Representations (ICLR)},
  year={2023}
}

@inproceedings{park2019spade,
  title={Semantic Image Synthesis with Spatially-Adaptive Normalization},
  author={Park, Taesung and Liu, Ming-Yu and Wang, Ting-Chun and Zhu, Jun-Yan},
  booktitle={Proceedings of the IEEE/CVF Conference on Computer Vision and Pattern Recognition (CVPR)},
  year={2019}
}

@inproceedings{zhang2021datasetgan,
  title={{DatasetGAN}: Efficient Labeled Data Factory with Minimal Human Effort},
  author={Zhang, Yuxuan and Ling, Huan and Gao, Jun and Yin, Kangxue and Lafleche, Jean-Fran{\c{c}}ois and Barriuso, Adela and Torralba, Antonio and Fidler, Sanja},
  booktitle={Proceedings of the IEEE/CVF Conference on Computer Vision and Pattern Recognition (CVPR)},
  year={2021}
}

@inproceedings{hu2024anomalydiffusion,
  title={{AnomalyDiffusion}: Few-Shot Anomaly Image Generation with Diffusion Model},
  author={Hu, Teng and Zhang, Jiangning and Yi, Ran and Du, Yuzhen and Chen, Xu and Liu, Liang and Wang, Yabiao and Wang, Chengjie},
  booktitle={Proceedings of the AAAI Conference on Artificial Intelligence},
  year={2024}
}

@inproceedings{duan2023dfmgan,
  title={Few-Shot Defect Image Generation via Defect-Aware Feature Manipulation},
  author={Duan, Yuxuan and Hong, Yan and Niu, Li and Zhang, Liqing},
  booktitle={Proceedings of the AAAI Conference on Artificial Intelligence},
  year={2023}
}

@inproceedings{hu2022lora,
  title={{LoRA}: Low-Rank Adaptation of Large Language Models},
  author={Hu, Edward J. and Shen, Yelong and Wallis, Phillip and Allen-Zhu, Zeyuan and Li, Yuanzhi and Wang, Shean and Wang, Lu and Chen, Weizhu},
  booktitle={International Conference on Learning Representations (ICLR)},
  year={2022}
}

@inproceedings{kingma2014vae,
  title={Auto-Encoding Variational Bayes},
  author={Kingma, Diederik P. and Welling, Max},
  booktitle={International Conference on Learning Representations (ICLR)},
  year={2014}
}

@inproceedings{goodfellow2014gan,
  title={Generative Adversarial Nets},
  author={Goodfellow, Ian and Pouget-Abadie, Jean and Mirza, Mehdi and Xu, Bing and Warde-Farley, David and Ozair, Sherjil and Courville, Aaron and Bengio, Yoshua},
  booktitle={Advances in Neural Information Processing Systems (NeurIPS)},
  year={2014}
}

@inproceedings{karras2020stylegan2,
  title={Analyzing and Improving the Image Quality of {StyleGAN}},
  author={Karras, Tero and Laine, Samuli and Aittala, Miika and Hellsten, Janne and Lehtinen, Jaakko and Aila, Timo},
  booktitle={Proceedings of the IEEE/CVF Conference on Computer Vision and Pattern Recognition (CVPR)},
  year={2020}
}

@inproceedings{karras2020stylegan2ada,
  title={Training Generative Adversarial Networks with Limited Data},
  author={Karras, Tero and Aittala, Miika and Hellsten, Janne and Laine, Samuli and Lehtinen, Jaakko and Aila, Timo},
  booktitle={Advances in Neural Information Processing Systems (NeurIPS)},
  year={2020}
}

@article{kuchi2019sandboil,
  title={Machine Learning Applications in Detecting Sand Boils from Images},
  author={Kuchi, Aditi and Hoque, Md Tamjidul and Abdelguerfi, Mahdi and Flanagin, Maik C.},
  journal={Array},
  volume={3}, pages={100012}, year={2019}
}

@article{panta2023sandboilnet,
  title={Deep Learning Approach for Accurate Segmentation of Sand Boils in Levee Systems},
  author={Panta, Manisha and Hoque, Md Tamjidul and Niles, Kendall N. and Tom, Joe and Abdelguerfi, Mahdi and Flanagin, Maik},
  journal={IEEE Access},
  volume={11}, pages={126263--126282}, year={2023}
}

@article{panta2024seepage,
  title={Application of Deep Learning for Segmenting Seepages in Levee Systems},
  author={Panta, Manisha and Thapa, Padam Jung and Hoque, Md Tamjidul and Niles, Kendall N. and Sloan, Steve and Flanagin, Maik and Pathak, Ken and Abdelguerfi, Mahdi},
  journal={Remote Sensing},
  volume={16}, number={13}, pages={2441}, year={2024}
}

@article{alshawi2023efpn,
  title={Imbalance-Aware Culvert-Sewer Defect Segmentation Using an Enhanced Feature Pyramid Network},
  author={Alshawi, Rasha and Ferdaus, Md Meftahul and Abdelguerfi, Mahdi and Niles, Kendall and Pathak, Ken and Sloan, Steve},
  journal={arXiv preprint arXiv:2408.10181},
  year={2024}
}

@article{alshawi2024sinkhole,
  title={A Depth-Wise Separable {U-Net} Architecture with Multiscale Filters to Detect Sinkholes},
  author={Alshawi, Rasha and Hoque, Md Tamjidul and Flanagin, Maik C.},
  journal={Remote Sensing},
  volume={15}, number={5}, pages={1384}, year={2023}
}

@inproceedings{james2011sar,
  title={Earthen Levee Monitoring with Synthetic Aperture Radar},
  author={Aanstoos, James V. and Hasan, Khaled and O'Hara, Charles G. and Prasad, Saurabh and Dabbiru, Lalitha and Mahrooghy, Majid and Gokaraju, Balakrishna and Nobrega, Rodrigo},
  booktitle={2011 IEEE Applied Imagery Pattern Recognition Workshop (AIPR)},
  year={2011}
}

@article{guan2023pavement,
  title={Deep Learning Approaches in Pavement Distress Identification: A Review},
  author={Guan, Sizhe and Liu, Haolan and Pourreza, Hamid R. and Mahyar, Hamidreza},
  journal={arXiv preprint arXiv:2308.00828},
  year={2023}
}

@inproceedings{trabucco2023effective,
  title={Effective Data Augmentation with Diffusion Models},
  author={Trabucco, Brandon and Doherty, Kyle and Gurinas, Max and Salakhutdinov, Ruslan},
  booktitle={International Conference on Learning Representations (ICLR)},
  year={2024}
}

@article{azizi2023synthetic,
  title={Synthetic Data from Diffusion Models Improves {ImageNet} Classification},
  author={Azizi, Shekoofeh and Kornblith, Simon and Saharia, Chitwan and Norouzi, Mohammad and Fleet, David J.},
  journal={Transactions on Machine Learning Research},
  year={2023}
}

@article{liu2024diffrs,
  title={Diffusion Models Meet Remote Sensing: Principles, Methods, and Perspectives},
  author={Liu, Yidan and Yue, Jun and Xia, Shaobo and Ghamisi, Pedram and Xie, Weiying and Fang, Leyuan},
  journal={IEEE Transactions on Geoscience and Remote Sensing},
  volume={62},
  pages={1--22},
  year={2024},
  note={arXiv:2404.08926}
}

@inproceedings{heusel2017fid,
  title={{GANs} Trained by a Two Time-Scale Update Rule Converge to a Local Nash Equilibrium},
  author={Heusel, Martin and Ramsauer, Hubert and Unterthiner, Thomas and Nessler, Bernhard and Hochreiter, Sepp},
  booktitle={Advances in Neural Information Processing Systems (NeurIPS)},
  year={2017}
}

@inproceedings{binkowski2018kid,
  title={Demystifying {MMD} {GANs}},
  author={Bi{\'n}kowski, Miko{\l}aj and Sutherland, Danica J. and Arbel, Michael and Gretton, Arthur},
  booktitle={International Conference on Learning Representations (ICLR)},
  year={2018}
}

@inproceedings{radford2021clip,
  title={Learning Transferable Visual Models from Natural Language Supervision},
  author={Radford, Alec and Kim, Jong Wook and Hallacy, Chris and Ramesh, Aditya and Goh, Gabriel and others},
  booktitle={International Conference on Machine Learning (ICML)},
  year={2021}
}

@inproceedings{hessel2021clipscore,
  title={{CLIPScore}: A Reference-free Evaluation Metric for Image Captioning},
  author={Hessel, Jack and Holtzman, Ari and Forbes, Maxwell and Le Bras, Ronan and Choi, Yejin},
  booktitle={Conference on Empirical Methods in Natural Language Processing (EMNLP)},
  year={2021}
}

@inproceedings{zhang2018lpips,
  title={The Unreasonable Effectiveness of Deep Features as a Perceptual Metric},
  author={Zhang, Richard and Isola, Phillip and Efros, Alexei A. and Shechtman, Eli and Wang, Oliver},
  booktitle={Proceedings of the IEEE Conference on Computer Vision and Pattern Recognition (CVPR)},
  year={2018}
}

@inproceedings{kirillov2023sam,
  title={Segment Anything},
  author={Kirillov, Alexander and Mintun, Eric and Ravi, Nikhila and Mao, Hanzi and Rolland, Chloe and Gustafson, Laura and others},
  booktitle={Proceedings of the IEEE/CVF International Conference on Computer Vision (ICCV)},
  year={2023}
}

@article{canny1986,
  title={A Computational Approach to Edge Detection},
  author={Canny, John},
  journal={IEEE Transactions on Pattern Analysis and Machine Intelligence},
  volume={PAMI-8}, number={6}, pages={679--698}, year={1986}
}

@inproceedings{xie2015hed,
  title={Holistically-Nested Edge Detection},
  author={Xie, Saining and Tu, Zhuowen},
  booktitle={Proceedings of the IEEE International Conference on Computer Vision (ICCV)},
  year={2015}
}

@inproceedings{ranftl2021midas,
  title={Vision Transformers for Dense Prediction},
  author={Ranftl, Ren{\'e} and Bochkovskiy, Alexey and Koltun, Vladlen},
  booktitle={Proceedings of the IEEE/CVF International Conference on Computer Vision (ICCV)},
  year={2021}
}

@article{perez2003poisson,
  title={Poisson Image Editing},
  author={P{\'e}rez, Patrick and Gangnet, Michel and Blake, Andrew},
  journal={ACM Transactions on Graphics (SIGGRAPH)},
  volume={22}, number={3}, pages={313--318}, year={2003}
}

@article{buslaev2020albumentations,
  title={{Albumentations}: Fast and Flexible Image Augmentations},
  author={Buslaev, Alexander and Iglovikov, Vladimir I. and Khvedchenya, Eugene and Parinov, Alex and Druzhinin, Mikhail and Kalinin, Alexandr A.},
  journal={Information},
  volume={11}, number={2}, pages={125}, year={2020}
}

@inproceedings{dettmers2022bnb,
  title={8-bit Optimizers via Block-wise Quantization},
  author={Dettmers, Tim and Lewis, Mike and Shleifer, Sam and Zettlemoyer, Luke},
  booktitle={International Conference on Learning Representations (ICLR)},
  year={2022}
}

@inproceedings{lu2022dpmsolver,
  title={{DPM-Solver}: A Fast {ODE} Solver for Diffusion Probabilistic Model Sampling in Around 10 Steps},
  author={Lu, Cheng and Zhou, Yuhao and Bao, Fan and Chen, Jianfei and Li, Chongxuan and Zhu, Jun},
  booktitle={Advances in Neural Information Processing Systems (NeurIPS)},
  year={2022}
}

@article{lu2022dpmsolverpp,
  title={{DPM-Solver++}: Fast Solver for Guided Sampling of Diffusion Probabilistic Models},
  author={Lu, Cheng and Zhou, Yuhao and Bao, Fan and Chen, Jianfei and Li, Chongxuan and Zhu, Jun},
  journal={arXiv preprint arXiv:2211.01095},
  year={2022}
}

@article{yang2024qwen2,
  title={{Qwen2.5} Technical Report},
  author={Yang, An and others},
  journal={arXiv preprint arXiv:2412.15115},
  year={2024}
}

@inproceedings{szegedy2016inception,
  title={Rethinking the Inception Architecture for Computer Vision},
  author={Szegedy, Christian and Vanhoucke, Vincent and Ioffe, Sergey and Shlens, Jon and Wojna, Zbigniew},
  booktitle={Proceedings of the IEEE Conference on Computer Vision and Pattern Recognition (CVPR)},
  year={2016}
}

@mastersthesis{thapa2025thesis,
  author  = {Thapa, Padam Jung},
  title   = {Toward Robust Semantic Segmentation in Levee Infrastructure Monitoring: Enhancing Accuracy with High-Fidelity Synthetic Data and Ensemble Learning},
  school  = {University of New Orleans},
  year    = {2025},
  address = {New Orleans, LA, USA},
  type    = {{Master's Thesis}},
  note    = {University of New Orleans Theses and Dissertations No.~3231. \url{https://scholarworks.uno.edu/td/3231/}}
}

@misc{thapa_chull_annotator,
  author       = {Thapa, Padam Jung},
  title        = {{Convex\_Hull\_Annotator}: Automated convex-hull mask annotation from a {SandBoilNet} probability map},
  year         = {2024},
  howpublished = {\url{https://github.com/padam56/Convex_Hull_Annotator}},
  note         = {Companion tool to \cite{thapa2025thesis}; applies thresholding, connected-component analysis, and per-component convex hulls to a CNN-predicted sandboil probability map.}
}

@inproceedings{dutta2019via,
  author    = {Dutta, Abhishek and Zisserman, Andrew},
  title     = {The {VIA} Annotation Software for Images, Audio and Video},
  booktitle = {Proceedings of the 27th ACM International Conference on Multimedia},
  series    = {MM '19},
  year      = {2019},
  pages     = {2276--2279},
  publisher = {ACM},
  doi       = {10.1145/3343031.3350535},
}

@inproceedings{kynkaanniemi2019prec,
  author    = {Kynk{\"a}{\"a}nniemi, Tuomas and Karras, Tero and Laine, Samuli and Lehtinen, Jaakko and Aila, Timo},
  title     = {Improved Precision and Recall Metric for Assessing Generative Models},
  booktitle = {Advances in Neural Information Processing Systems (NeurIPS)},
  year      = {2019},
}

@inproceedings{naeem2020reliable,
  author    = {Naeem, Muhammad Ferjad and Oh, Seong Joon and Uh, Youngjung and Choi, Yunjey and Yoo, Jaejun},
  title     = {Reliable Fidelity and Diversity Metrics for Generative Models},
  booktitle = {Proceedings of the 37th International Conference on Machine Learning (ICML)},
  year      = {2020},
}

@inproceedings{parmar2022cleanfid,
  author    = {Parmar, Gaurav and Zhang, Richard and Zhu, Jun-Yan},
  title     = {On Aliased Resizing and Surprising Subtleties in {GAN} Evaluation},
  booktitle = {Proceedings of the IEEE/CVF Conference on Computer Vision and Pattern Recognition (CVPR)},
  year      = {2022},
}

@inproceedings{wang2023exploring,
  author    = {Wang, Jianyi and Chan, Kelvin C.K. and Loy, Chen Change},
  title     = {Exploring {CLIP} for Assessing the Look and Feel of Images},
  booktitle = {Proceedings of the AAAI Conference on Artificial Intelligence},
  year      = {2023},
}

@inproceedings{ho2020ddpm,
  author    = {Ho, Jonathan and Jain, Ajay and Abbeel, Pieter},
  title     = {Denoising Diffusion Probabilistic Models},
  booktitle = {Advances in Neural Information Processing Systems (NeurIPS)},
  year      = {2020},
}

@inproceedings{song2021ddim,
  author    = {Song, Jiaming and Meng, Chenlin and Ermon, Stefano},
  title     = {Denoising Diffusion Implicit Models},
  booktitle = {International Conference on Learning Representations (ICLR)},
  year      = {2021},
}

@inproceedings{lugmayr2022repaint,
  author    = {Lugmayr, Andreas and Danelljan, Martin and Romero, Andr{\'e}s and Yu, Fisher and Timofte, Radu and Van Gool, Luc},
  title     = {{RePaint}: Inpainting Using Denoising Diffusion Probabilistic Models},
  booktitle = {Proceedings of the IEEE/CVF Conference on Computer Vision and Pattern Recognition (CVPR)},
  year      = {2022},
}

@inproceedings{avrahami2022blended,
  author    = {Avrahami, Omri and Lischinski, Dani and Fried, Ohad},
  title     = {Blended Diffusion for Text-Driven Editing of Natural Images},
  booktitle = {Proceedings of the IEEE/CVF Conference on Computer Vision and Pattern Recognition (CVPR)},
  year      = {2022},
}

@article{kazerouni2023diffusion,
  author    = {Kazerouni, Amirhossein and Aghdam, Ehsan Khodapanah and Heidari, Moein and Azad, Reza and Fayyaz, Mohsen and Hacihaliloglu, Ilker and Merhof, Dorit},
  title     = {Diffusion Models in Medical Imaging: A Comprehensive Survey},
  journal   = {Medical Image Analysis},
  volume    = {88},
  pages     = {102846},
  year      = {2023},
}

@inproceedings{chong2020fid,
  author    = {Chong, Min Jin and Forsyth, David},
  title     = {Effectively Unbiased {FID} and Inception Score and Where to Find Them},
  booktitle = {Proceedings of the IEEE/CVF Conference on Computer Vision and Pattern Recognition (CVPR)},
  year      = {2020},
}

@misc{ollin2023vaefix,
  title        = {{SDXL-VAE-FP16-Fix}: a numerically stable fp16 variational autoencoder for {SDXL}},
  author       = {Boer Bohan, Ollin},
  year         = {2023},
  howpublished = {\url{https://huggingface.co/madebyollin/sdxl-vae-fp16-fix}},
  note         = {Hugging Face model repository}
}
